\documentclass[plaindraft, jppdefaultmargins]{jpp-AAS}

\usepackage[utf8]{inputenc}
\usepackage[T1]{fontenc}    
\usepackage{comment}
\usepackage{psfrag}
\usepackage{pstool}         

\usepackage{parskip}      
\usepackage{microtype}     
\usepackage{enumitem}     

\setlist[enumerate,1]{leftmargin=3em, labelsep=0.5em} 

\usepackage[tbtags]{amsmath}    
\usepackage{amsthm}
\usepackage{thmtools}
\usepackage{mathtools}    
\usepackage{mathrsfs}     
\usepackage{scalerel}     
\usepackage{stackengine}  
\usepackage{xparse}       
\usepackage{calc}         

\newcommand{\nn}{\nonumber\\}
\newcommand{\al}{&\phantom{{}={}}\negmedspace}

\usepackage{graphicx}
\usepackage{xcolor}
\usepackage{tikz}

\graphicspath{{Figures/}} 

\usetikzlibrary{arrows.meta} 

\definecolor{myred}{HTML}{d24727}
\definecolor{myorange}{HTML}{E6A91D}
\definecolor{links}{RGB}{219, 70, 50}
\definecolor{mygrey}{RGB}{156, 138, 144}

\usepackage{tabularx}     
\usepackage{siunitx}      

\usepackage{subfiles}     
\usepackage{xr-hyper}     
\usepackage{needspace}

\usepackage[bookmarks=false]{hyperref}
\usepackage[figure]{hypcap}   

\hypersetup{
    colorlinks=true,
    linkcolor=links,
    urlcolor=links,
    citecolor=links
}
\usepackage{orcidlink}

\AtBeginDocument{%
}
\AtEndDocument{%
\ifSubfilesClassLoaded{%
  \bibliographystyle{plainnat}
  \bibliography{resonant}%
}{}}%

\newenvironment{acknowledge}
{%
  \section*{Acknowledgements} 
}
{%
}

\newenvironment{funding}
{%
  \section*{Funding}
}
{%
}

\ifCUPmtlplainloaded \else
  \checkfont{eurm10}
  \iffontfound
    \IfFileExists{upmath.sty}
      {\typeout{^^JFound AMS Euler Roman fonts on the system, using the 'upmath' package.^^J}\usepackage{upmath}}
      {\typeout{^^JFound AMS Euler Roman fonts on the system, but you dont seem to have the 'upmath' package installed.^^J}\providecommand\upi{\pi}}
  \else
    \providecommand\upi{\pi}
  \fi
\fi

\ifCUPmtlplainloaded \else
  \checkfont{msam10}
  \iffontfound
    \IfFileExists{amssymb.sty}
      {\typeout{^^JFound AMS Symbol fonts on the system, using the 'amssymb' package.^^J}\usepackage{amssymb}\let\leq=\leqslant\let\geq=\geqslant}{}
  \fi
\fi

\ifCUPmtlplainloaded \else
  \IfFileExists{amsbsy.sty}
    {\typeout{^^JFound the 'amsbsy' package on the system, using it.^^J}\usepackage{amsbsy}}
    {\providecommand\boldsymbol[1]{\mbox{\boldmath $##1$}}}
\fi

\newcommand{\rmd}{{\rm d}}
\newcommand{\rme}{{\rm e}}
\newcommand{\rmi}{{\rm i}}

\newcommand{\0}{\bb{\cdot}}
\newcommand{\bh}{\hat{\bb{b}}}
\newcommand{\vp}{v_{\parallel}}

\newcommand{\?}{\mkern1.5mu}

\newcommand\bb[1]{\boldsymbol{#1}}

\newcommand{\grad}[1]{{\bb{\nabla}{#1}}}

\newcommand{\con}[1]{{\partial_{#1}\bb{x}}}

\NewDocumentCommand{\mybar}{ O{0.80} O{1pt} m }{%
    \mathrlap{\hspace{#2}\overline{\scalebox{#1}[1]{\phantom{\ensuremath{#3}}}}}\ensuremath{#3}}
\NewDocumentCommand{\mybarup}{ O{0.80} O{1pt} m }{%
    \mathrlap{\hspace{#2}\overline{\scalebox{#1}[1.12]{\phantom{\ensuremath{#3}}}}}\ensuremath{#3}}
\NewDocumentCommand{\mybardown}{ O{0.80} O{1pt} m }{%
    \mathrlap{\hspace{#2}\overline{\scalebox{#1}[0.93]{\phantom{\ensuremath{#3}}}}}\ensuremath{#3}}

\makeatletter
\newcommand{\raisemath}[1]{\mathpalette{\raisem@th{#1}}}
\newcommand{\raisem@th}[3]{\raisebox{#1}{$#2#3$}}
\makeatother

\usepackage[nameinlink,noabbrev]{cleveref}

\crefname{section}{\S}{\S\S}
\Crefname{section}{\S}{\S\S}

\crefformat{section}{#2\S#1#3}
\Crefformat{section}{#2\S#1#3}

\crefrangeformat{section}{#3\S\S#1#4\kern0.05em--\kern0.05em#5#2#6}
\Crefrangeformat{section}{#3\S\S#1#4\kern0.05em--\kern0.05em#5#2#6}

\crefrangeformat{figure}{#3figures~#1#4\kern0.05em--\kern0.05em#5#2#6}
\Crefrangeformat{figure}{#3Figures~#1#4\kern0.05em--\kern0.05em#5#2#6}

\newcommand{\crefand}[2]{\hyperref[#1]{figures~\ref*{#1}} and~\hyperref[#2]{\ref*{#2}}}

\usepackage{xparse}

\NewDocumentCommand{\crefsub}{mm}{\hyperref[#1]{\cref*{#1}\kern0.05em(#2)}}

\NewDocumentCommand{\Crefsub}{mm}{\hyperref[#1]{\Cref*{#1}\kern0.05em(#2)}}

\NewDocumentCommand{\crefsubrange}{mmmm}{%
  \hyperref[#1]{figures~\ref*{#1}\kern0.05em(#2)}--%
  \hyperref[#3]{\ref*{#3}\kern0.05em(#4)}%
}
\NewDocumentCommand{\Crefsubrange}{mmmm}{%
  \hyperref[#1]{Figures~\ref*{#1}\kern0.05em(#2)}--%
  \hyperref[#3]{\ref*{#3}\kern0.05em(#4)}%
}

\NewDocumentCommand{\crefsuband}{mmmm}{%
  \hyperref[#1]{figures~\ref*{#1}\kern0.05em(#2)}%
  \space and~\hyperref[#3]{\ref*{#3}\kern0.05em(#4)}%
}
\NewDocumentCommand{\Crefsuband}{mmmm}{%
  \hyperref[#1]{Figures~\ref*{#1}\kern0.05em(#2)}%
  \space and~\hyperref[#3]{\ref*{#3}\kern0.05em(#4)}%
}

\crefname{definition}{definition}{definitions}   
\Crefname{definition}{Definition}{Definitions}   

\newcommand{\defref}[1]{\hyperref[#1]{definition~\ref*{#1}}}      

\newcommand*\circled[1]{\tikz[baseline=(char.base)]{
    \node[shape=circle,draw,inner sep=0.8pt, scale=0.8] (char) {#1};}}

\renewcommand{\eg}{e.g.~}

\newcommand{\ie}{i.e.~}
\newcommand{\viz}{viz.~}

\makeatletter
\NewDocumentCommand{\mycitep}{O{} O{} m}{%
  \begingroup
  \def\NAT@aysep##1{,~}%
  \ifx\relax#1\relax
    \ifx\relax#2\relax
      (\citealp{#3})%
    \else
      (\citealp{#3}#2)%
    \fi
  \else
    \ifx\relax#2\relax
      (#1\citealp{#3})%
    \else
      (#1\citealp{#3}#2)%
    \fi
  \fi
  \endgroup
}
\makeatother

\makeatletter
\AtBeginDocument{%
  \def\NAT@aysep#1{,~}%
}
\makeatother

\newtheorem{definition}{Definition}[section]

\makeatletter
\newif\iftag@here

\newcommand*{\taghere}[1][0pt]
{\ifmeasuring@\else
  \global\tag@heretrue
  \tikz[remember picture,overlay]{\coordinate (taghere) at (0pt,#1);}%
\fi}

\def\place@tag{%
    \iftagsleft@
      \kern-\tagshift@
      \iftag@here
        \global\tag@herefalse
        \tikz[remember picture,overlay]%
          {\path (taghere) -| node[anchor=base]{\rlap{\boxz@}} (0pt,0pt);}%
      \else
        \if1\shift@tag\row@\relax
            \rlap{\vbox{%
                \normalbaselines
                \boxz@
                \vbox to\lineht@{}%
                \raise@tag
            }}%
        \else
            \rlap{\boxz@}%
        \fi
        \kern\displaywidth@
      \fi
    \else
      \kern-\tagshift@
      \iftag@here
        \global\tag@herefalse
        \tikz[remember picture,overlay]%
          {\path  (taghere) -|  node[anchor=base]{\llap{\boxz@}} (0pt,0pt);}%
      \else
        \if1\shift@tag\row@\relax
            \llap{\vtop{%
                \raise@tag
                \normalbaselines
                \setbox\@ne\null
                \dp\@ne\lineht@
                \box\@ne
                \boxz@
            }}%
        \else \llap{\boxz@}%
        \fi
      \fi
    \fi
}
\makeatother

\DeclareRobustCommand{\hairskip}{\hskip 0.085em\relax}
\def\emdash{---}
\def\d@sh#1#2{\unskip#1\hairskip#2\hairskip\ignorespaces}
\def\Dash{\d@sh\nobreak\emdash}
\def\Ldash{\d@sh\empty{\hbox{\emdash}\nobreak}}
\def\Rdash{\d@sh\nobreak\emdash}

\shorttitle{Energetic-particle orbits near rational flux surfaces: I. Passing particles}
\shortauthor{T.~E.~Foster et al.}

\title{Energetic-particle orbits near rational flux surfaces in stellarators: I. Passing particles}

\author{
    {Thomas~E.~Foster\aff{1,2}%
        \orcidlink{0000-0001-6620-1217}}
        \corresp{\email{thomas.foster@princeton.edu}},
    {Felix~I.~Parra\aff{1,2}%
        \orcidlink{0000-0001-9621-7404}},
    {Roscoe~B.~White\aff{2}%
        \orcidlink{0000-0002-4239-2685}},
    {Jos\'e~Luis~Velasco\aff{3}%
        \orcidlink{0000-0001-8510-1422}},
    {Iv\'an~Calvo\aff{3}%
        \orcidlink{0000-0003-3118-3463}},
    and {Elizabeth~J.~Paul\aff{4}%
        \orcidlink{0000-0002-9355-5595}}
}
    
\affiliation{
    \aff{1}Department of Astrophysical Sciences, Princeton University,
    \\[\affilskip]
    Peyton Hall, Princeton, NJ 08544, USA
    \\[\affilskip]
    \aff{2}Princeton Plasma Physics Laboratory, PO Box 451, Princeton, NJ 08543, USA
    \\[\affilskip]
    \aff{3}Laboratorio Nacional de Fusi\'on, CIEMAT, E-28040 Madrid, Spain
    \\[\affilskip]
    \aff{4}Department of Applied Physics and Applied Mathematics,
    \\[\affilskip]
    Columbia University, New York, NY 10027, USA
}

\begin{document}

\maketitle

\begin{abstract}
Recent simulations \cite[]{White2022, White2022b, White2025} have shown that, even when the magnetic field of a stellarator possesses nested toroidal flux surfaces, the orbits of passing energetic particles can exhibit islands. These `drift islands' arise near rational flux surfaces, where they are likely to enhance alpha-particle transport \Dash flattening the alpha density profile locally \Dash unless they can be avoided by suitable design of the stellarator magnetic field. To investigate how this might be achieved, we derive an equation for the drift-island shape in a general stellarator. This result follows from the solution to a more fundamental problem: that of calculating the orbits of passing particles near a rational flux surface. We show that these orbits are determined by conservation of an adiabatic invariant associated with the closed rational-surface field lines. We use this `transit adiabatic invariant' to prove that there are no drift islands, for all passing particles, if and only if the magnetic field satisfies a weaker version of the \cite{Cary1997} condition for omnigeneity; we call such magnetic fields `cyclometric'. The drift-island width scales as $\sim\!(\rho_\star\delta/s)^{1/2}\,a$ ($\rho_\star$ is the normalized gyroradius, $\delta$ is the deviation from cyclometry, $s$ is the magnetic shear, and $a$ is the minor radius), so large drift islands could arise in low-shear stellarators that are insufficiently cyclometric. To ensure accurate results for very energetic particles, we compute higher-order corrections to the transit invariant. Our calculations agree extremely well with ASCOT5 guiding‑centre and full‑orbit simulations of alpha particles in reactor-scale equilibria, even at $\SI{3.5}{\mega\electronvolt}$. Finally, we show how our results can also be derived using Hamiltonian perturbation theory, which provides a systematic framework for calculating passing-particle orbits on both rational and irrational surfaces. 
\end{abstract}

\section{Introduction}\label{sec:introduction}

In stellarator reactors, deuterium--tritium fusion will produce energetic, $\SI{3.5}{\mega\electronvolt}$ alpha particles. To minimise losses of energy from the plasma and avoid unsustainable damage to the reactor walls, these fusion-born alphas must be well confined by the stellarator magnetic field, at least until their energy is transferred back to the thermal plasma by collisions. This collisional slowing-down process takes $\sim\!\SI{0.1}{\second}$ for typical reactor parameters (densities of $\sim\!\SI{e20}{\per\meter\cubed}$ and temperatures of ${\sim\!\SI{15}{\kilo\electronvolt}}$), which is long enough for the alpha particles, travelling at $\sim\!\SI{e7}{\meter\per\second}$, to complete many orbits through the device. Together with the fact that their $\sim\!\SI{5}{\centi\meter}$ gyroradii are a small fraction of the minor radius of a reactor-scale stellarator, this means fusion-born alphas are commonly modelled using collisionless, guiding-centre equations \mycitep[\eg]{Lotz1992, Bader2021, Velasco2021, Leviness2022, Paul2022, Albert2023, Bindel2023, Figueiredo2024}.

\begin{figure}
\centering
\includegraphics[width=\textwidth]{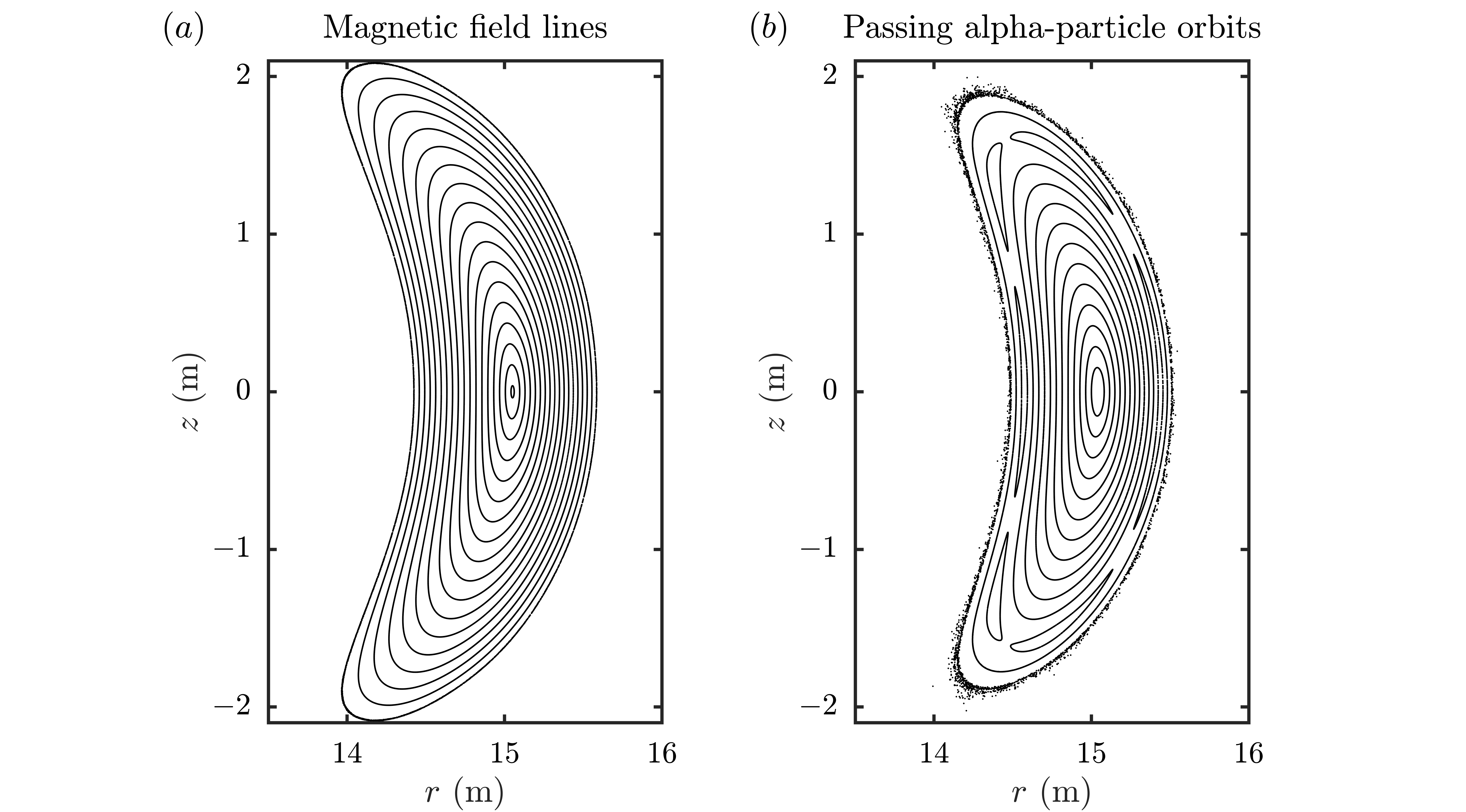}
    \caption{Poincar\'e plots, constructed in the $\phi = 0$ plane using cylindrical coordinates $(r, \phi, z)$, for $(a)$ magnetic field lines and $(b)$ guiding-centre orbits of passing alpha particles, in a W7-X equilibrium scaled up to reactor size. The alpha particles in $(b)$ are co-passing relative to the magnetic field and have energy $\SI{3.5}{\mega\electronvolt}$ and pitch-angle $\mu/\mathcal{E} = \SI{0.143}{\per\tesla}$, where $\mu$ is their magnetic moment and $\mathcal{E}$ is their kinetic energy per unit mass. The chosen value of $\mu/\mathcal{E}$ coincides with the trapped--passing boundary at the last closed flux surface, so the orbits in $(b)$ are barely passing. The field lines and guiding-centre orbits were computed using the ASCOT5 code; more details about the equilibrium and simulation method are provided in \cref{subsec:islandplots}.}
    \label{fig:W7X_cyl}
\end{figure}

Recently, \citet{White2022} simulated the collisionless, guiding-centre motion of alpha particles in stellarators and observed a striking difference between the behaviour of \emph{passing} particles in stellarators and tokamaks \mycitep[see also ]{White2022b, White2025}. It is well known that, in tokamaks, axisymmetry constrains the magnetic field lines to lie on nested toroidal flux surfaces and the particle orbits to lie on nested toroidal drift surfaces. If axisymmetry is broken by a magnetic perturbation, rational flux surfaces break into magnetic islands and nearby drift surfaces break into `drift islands' \mycitep[\eg]{Brambilla1973,Mynick1993, Mynick1993b}. In stellarators \Dash which are inherently nonaxisymmetric \Dash \citet{White2022} found that the orbits of passing particles generally contain drift islands, even when the magnetic field is unperturbed and still possesses a full set of nested toroidal flux
surfaces \mycitep[as shown in \cref{fig:W7X_cyl}; see also ][]{Grad1967, Marcal1982, Wobig2001, White2022b, White2025, Chambliss2024}.

These `equilibrium-induced' drift islands grow with energy, raising concerns that they could be a source of enhanced alpha-particle transport specific to stellarators \cite[]{Paul2022, White2022, White2022b, White2025, Chambliss2024}. Wide islands, in the low-collisionality regimes relevant for fusion-born alphas, allow particles to make large radial excursions through the device. Furthermore, collisions or orbit-class transitions will enable particles to ‘hop’ in and out of these wide orbits, expediting their progress towards the reactor walls. As a result, drift islands are likely to be sites of enhanced neoclassical transport, unless they can be avoided by suitable design of the stellarator magnetic field.

To investigate how this might be achieved, we derive simple equations for the drift-island shape in a general stellarator. In this paper, we focus on passing-particle orbits, while in Paper II \cite[]{Foster2025}, 
we generalize our theory to include trapped particles that complete multiple toroidal transits before bouncing, whose orbits can also form wide drift islands. The question of how drift islands modify \emph{collisional} transport of alpha particles around rational surfaces is left for future work.

The drift-island shape follows from the solution to an even more fundamental theoretical problem: that of calculating the orbits of passing particles near a rational flux surface. We estimate the width of these orbits in \cref{sec:basictheory}. Then, in \cref{sec:adiabaticinvariant}, we show that these orbits are determined by conservation of an adiabatic invariant associated with the closed rational-surface field lines. We derive this `transit adiabatic invariant' by expanding the guiding-centre equations for particles close to the rational surface, then applying an orbit-averaging procedure. Similar adiabatic invariants have been used to describe locally passing particles in stellarators with a small rotational transform per field period or per ripple well \mycitep[\eg]{Dobrott1971, Cary1988, Todoroki1993, Rome1995, Smirnova1997}. These adiabatic invariants described in previous work are not the same as the transit invariant; one difference is that the transit invariant depends explicitly on which low-order rational surface the particles are close to. An adiabatic invariant more similar to the transit invariant was derived by \citet{Hastie1967} \mycitep[see also ][]{Bhattacharjee1992} for passing particles in a magnetic field with either closed field lines or a small rotational transform. Our work generalizes the derivation of \citet{Hastie1967} to the case of particles near low-order rational surfaces in a stellarator with a rotational transform of order unity. Furthermore, we compute higher-order corrections to the transit invariant (see below and \cref{sec:higherorder}) and compare with simulations, conducted using the ASCOT5 code \mycitep{Varje2019}, of particle orbits in various modern stellarator designs.

In \cref{sec:optimisation}, we discuss how the drift-island width depends on properties of the stellarator equilibrium field, such as its number of field periods, aspect ratio, and magnetic shear. In \cref{sec:cyclometry}, we derive a necessary and sufficient condition for the drift-island width to vanish for all passing particles; we refer to magnetic fields that satisfy this condition as `cyclometric'. We also consider nearly cyclometric fields and determine how the drift-island width scales with the deviation from cyclometry.

To model fusion-born alphas in a reactor, it is crucial that our calculations remain valid even at high energies. However, for sufficiently energetic particles, the accuracy of orbit-averaging, which we use to obtain the transit invariant, has been questioned \mycitep[\eg]{Paul2022}. To ensure accurate results even at high energies, in \cref{sec:higherorder} we calculate higher-order corrections to the transit invariant. We derive these corrections by expanding in ${\sqrt{\rho_\star}\ll 1}$, where ${\rho_\star = \rho/L}$ is the energetic-particle gyroradius divided by a typical length scale of the magnetic field. The appropriate expansion parameter is ${\sqrt{\rho_\star}}$ because the orbit width for passing particles near a rational surface scales with $\sqrt{\rho_\star}$, in contrast with the familiar scaling with $\rho_\star$ for particles on irrational surfaces (these estimates are obtained in \cref{sec:basictheory}). When the higher-order corrections are included, our analytical results are in excellent agreement with simulations, indicating that orbit-averaged guiding-centre equations and the transit adiabatic invariant can accurately model passing alpha particles, even at $\SI{3.5}{\mega\electronvolt}$.

Finally, in \cref{sec:irrationalsurfaces}, we show how Hamiltonian perturbation theory can be used to calculate the orbits of passing particles on both rational and irrational flux surfaces in a general stellarator. This framework provides a clear criterion for which flux surfaces are `rational enough' that particle orbits in their vicinity need to be described using the transit invariant.






\section{Drift islands: physical picture and estimates}\label{sec:basictheory}

Drift islands arise in the orbits of passing particles near rational flux surfaces. The drift-island width is, therefore, equal to the orbit width of these particles, by which we mean the typical size of the radial excursions they make from their initial flux surface. In \cref{subsec:orbitwidth}, we estimate this orbit width using a physical picture of the particle motion. It is well known that the orbit width of passing particles typically scales with $\rho_\star$ \mycitep[\eg]{Bishop1966, Galeev1969}. However, we will find that, near low-order rational surfaces, the orbit width is wider, scaling instead with $\sqrt{\rho_\star}$ \mycitep{Grad1967}. In \cref{subsec:magneticislands}, we show that the same estimates can also be obtained using an analogy with magnetic islands.

\subsection{Orbit width of passing particles near rational surfaces}\label{subsec:orbitwidth}

First, we review a standard argument that the orbit width of passing particles typically scales with $\rho_\star$ \mycitep[\eg]{Helander2014} and point out why it does not apply to particles near rational flux surfaces. 

\begin{figure}
\vspace{4mm}
\centering
\includegraphics[width=\textwidth]{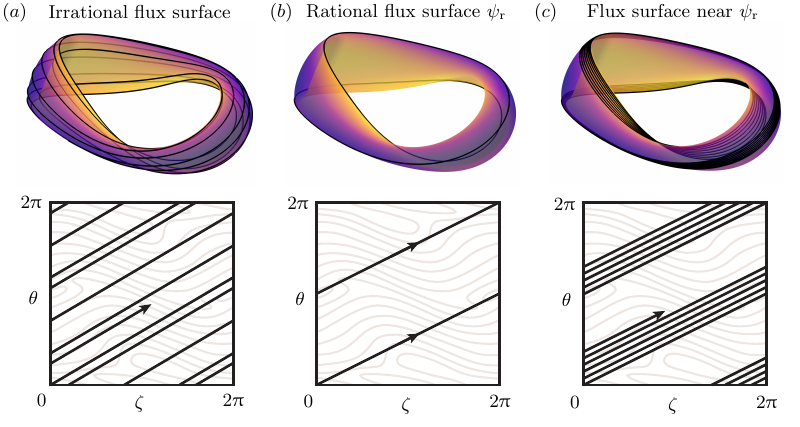}
    \caption{(top row) Illustration of how the rotational transform of a flux surface affects the magnetic field lines in physical space, with colour showing the magnetic field strength. For the flux surface shape, we used a precise-QA equilibrium from \protect{\citet{Landreman2021}}. (bottom row) Cartoons showing how the rotational transform affects the field lines in straight-field-line coordinates for a single-field-period stellarator, with grey contours representing the magnetic field strength. ($a$) On an irrational surface, field lines cover the surface densely and never close. ($b$) On a rational surface $\psi_{\rm{r}}$, field lines close on themselves. ($c$) Near the rational surface $\psi_{\rm{r}}$, field lines require many toroidal turns to explore the entire surface.}
    \label{fig:rationalvsirrational}
\end{figure}

A generic flux surface has an irrational rotational transform. Therefore, each magnetic field line covers the surface densely and a particle streaming along one of these field lines will explore the entire surface (see \cref{fig:rationalvsirrational}(a)). This means that the net radial drift accumulated by the particle over long times is proportional to the flux-surface average of its radial drift velocity divided by its parallel streaming rate, which weights each region of the flux surface according to how long the particle spends there \mycitep[\eg]{Helander2014}. It can be shown that this flux-surface average vanishes, so passing particles do not experience a secular radial drift. Thus, the width $w$ of a generic passing orbit can be estimated using the size of the radial drift velocity, $|\bb{v}_{\rm{d}}|\sim \rho_\star v$ ($v$ is the particle speed), and the typical time required for this drift to average out:
\begin{equation}
    w\sim |\bb{v}_{\rm{d}}|\, \frac{L}{v} \sim \rho_\star L\,.
\end{equation}
This estimate is accurate for most particles but is incorrect for particles close to a low-order rational surface. Near a rational surface, the magnetic field lines almost close on themselves, as shown in \crefsub{fig:rationalvsirrational}{c}, so a particle streaming along one of these field lines will take a very long time to explore the entire surface. In this time, the particle can drift radial distances larger than $\sim\!\rho_\star L$.

%
%

To estimate the orbit width of particles near a rational surface, it is helpful to have in mind the following physical picture. A particle streaming along a closed field line on the rational surface will not sample the entire surface, which means it can experience a net radial drift. This drift will carry the particle onto neighbouring flux surfaces, where the field lines no longer perfectly close on themselves (assuming the magnetic shear is non-zero); compare \crefsuband{fig:rationalvsirrational}{b}{fig:rationalvsirrational}{c}. By streaming along these nearly closed field lines, the particle will migrate tangentially across the rational surface \Dash this is clearest from a picture like \crefsub{fig:rationalvsirrational}{c}. Eventually, its radial drift, which averages to zero over the entire rational surface, must change sign. The particle will then drift back towards the rational surface but will overshoot. By streaming along nearly closed field lines again, the particle will migrate tangentially across the rational surface, in the opposite direction to before, back to its starting point. In a Poincar\'e plot, this cyclic motion traces out a drift island.

To estimate the width of such an orbit, it is convenient to use straight-field-line coordinates $(\psi,\theta,\zeta)$ \mycitep[\eg]{Helander2014}. The radial coordinate $\psi$ is the toroidal magnetic flux contained by the flux surface divided by $2\upi$, and $\theta$ and $\zeta$ are the poloidal and toroidal angles, respectively. Let the rotational transform be $\iota(\psi)$. Suppose the particle is near a low-order rational surface, $\psi=\psi_{\rm{r}}$, with $\iota(\psi_{\rm{r}}) = N/M$, for coprime integers $N$ and $M>0$. The orbit width is the radial distance the particle can drift in the time required for it to migrate across a significant fraction of the rational surface by streaming along nearly closed field lines, such as those depicted in \crefsub{fig:rationalvsirrational}{c}. We denote this orbit width, measured by the variation in the particle's $\psi$ coordinate, by $\Delta\psi$.

%
%
\begin{figure}
\vspace{2mm}
\includegraphics[width=\textwidth]{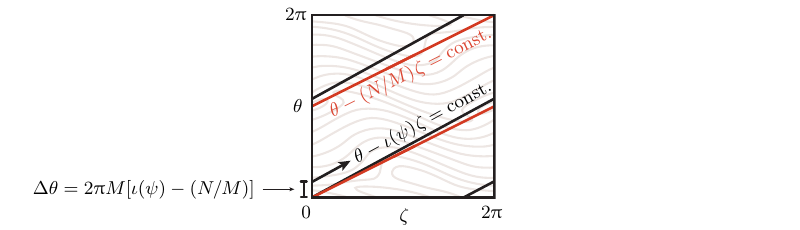}
    \caption{Cartoon showing the difference between a closed curve of constant $\theta - (N/M)\? \zeta$ (shown in red) and a field line (black) in straight-field-line coordinates on a flux surface near the rational surface. In this case, the stellarator has a single field period, $N=1$, and $M=2$. Grey contours represent the magnetic field strength. The change in poloidal angle, $\Delta\theta$, when the field line is followed for one full transit ($\Delta\zeta = 2\upi M$) is indicated.}
    \label{fig:etavsfieldline}
\end{figure}
Consider the motion of the particle from $\zeta = 0$ to $\zeta = 2\upi M$, where the orbit nearly `bites its tail'; throughout the paper, this motion is referred to as a `transit'. In a single transit, $\theta$ changes by $\Delta\theta\simeq 2\upi M[\iota(\psi)-N/M]$, as shown in \cref{fig:etavsfieldline}. Assuming $M\sim N\sim 1$, this means $\Delta\theta\sim\iota_{\rm r}' \Delta\psi$, where $\iota_{\rm r}' \coloneq (\rmd \iota / \rmd \psi)(\psi = \psi_{\rm{r}})$ is related to the global magnetic shear by ${s \coloneq |(\psi/\iota)(\rmd \iota / \rmd \psi)|}$. The number of transits required for the particle to move over a significant portion of the flux surface, by streaming along nearly closed field lines, is $\sim\! 1/\Delta\theta$. Thus, the time required for the particle to migrate over the flux surface is ${\sim\! (1/\iota_{\rm r}' \Delta\psi)(L/v)}$.\footnote{Here, we neglect the tangential drift, which could also be responsible for moving the particle across the flux surface. This can be justified \textit{a posteriori} given the island-width estimate \eqref{eq:scaling}. The change in $\theta$ in one transit due to the tangential drift is $\sim\!\rho_\star$, while \eqref{eq:scaling} implies that the change due to streaming along sheared field lines is $\sim\!(\rho_\star \delta s)^{1/2}$, which usually dominates. If the drift term dominates, the drift islands simply shift to a new radial location, as explained in \cref{subsec:lowshear}.} In this time, the change in $\psi$ due to drifts is 
\begin{equation}\label{eq:driftestimate}
\Delta\psi \sim \biggl(\frac{1}{\iota_{\rm r}'\Delta\psi}\biggr)(\rho_\star\delta)\?\Psi_{\rm{t}}\,,
\end{equation}
where $\Psi_{\rm{t}}$ is a typical value of the toroidal flux through a flux surface, and $\delta$ is a factor that reduces the radial drift rate in nearly omnigeneous stellarators and certain other configurations (\viz cyclometric configurations; we introduce cyclometry and define $\delta$ more precisely in \cref{sec:cyclometry}). Note that $\delta\sim 1$ in an unoptimized device. Solving \eqref{eq:driftestimate} for $\Delta\psi$ gives
\begin{equation}\label{eq:scaling}
\frac{\Delta\psi}{\Psi_{\rm{t}}} \sim \biggl(\frac{\rho_\star\delta}{s}\biggr)^{\!1/2}\,.
\end{equation}
As claimed, the orbit width for particles near rational surfaces, which is equivalent to the drift-island width, scales with $\sqrt{\rho_\star}$ instead of $\rho_\star$. For context, birth alpha particles in a reactor-scale stellarator might have $\rho_\star\sim 0.03$ (using parameters from \citet{Ku2008} or \citet{Drevlak2014}; $\rho_\star$ will be lower in a high-field stellarator). Stellarator designs usually have $s$ between $0.01$ and $1$ \mycitep{Landreman2022, Komori2006} and $\delta$ between $10^{-4}$ and $10^{-1}$ \mycitep{Velasco2023}. Thus, for birth alphas, estimate \eqref{eq:scaling} suggests that the drift-island width can reach $\sim\! 50\%$ of the minor radius. 

Estimate \eqref{eq:scaling} does not capture the dependence of the island width on $N$ and $M$, since we ordered $N\sim M \sim 1$ and disregarded order-unity factors. This dependence will be revealed by the full calculation in \cref{sec:adiabaticinvariant}. Physically, if the flux surface is rational, but with high order (meaning $M$ is large), the particle will sample most of the rational surface before the field lines close, so the radial drift will very nearly average out and the particle will not drift far. We will find that the island width typically decays exponentially with $|N|$ and $|M|$.

\subsection{Drift-island width by analogy with magnetic islands}\label{subsec:magneticislands}

The drift-island width can also be estimated using well-known facts about magnetic islands. This is possible because it can be shown that passing orbits follow the field lines of a modified magnetic field. To demonstrate this, we write the guiding-centre velocity, whose direction determines the trajectory of a particle's guiding centre through space, as%
\footnote{Equation \eqref{eq:guidingcentre} gives the correct velocity along the field lines to order $v$ and across the field lines to order $\rho_\star v$ \mycitep{Hazeltine2018}. In \citet{White1982}, it was pointed out that the right side of this equation could be divided by the near-identity factor $\smash{1+(v_\parallel/\Omega)(\bh\0\bb{\nabla}\bb{\times}\bh)}$ to ensure that Liouville's theorem is satisfied exactly. Since we are only interested in the direction of the guiding-centre velocity, this modification is unnecessary.}
\begin{equation}\label{eq:guidingcentre}
    \frac{\rmd \bb{X}}{\rmd t} = \frac{v_\parallel}{B} \?\biggl(\bb{B} + \bb{\nabla}\bb{\times}\biggl( \frac{v_\parallel\bb{B}}{\Omega} \biggr)\biggr)\,.
\end{equation}
Here, $\bb{X}$ is the guiding-centre position, $t$ is time, $\bb{B}$ is the magnetic field, $B \coloneq |\bb{B}|$ is the magnetic field strength, and $\Omega \coloneq ZeB/mc$ is the gyrofrequency, where $Ze$ and $m$ are the particle charge and mass, respectively, and $c$ is the speed of light (we use CGS units). All functions are evaluated at the guiding-centre position. The parallel velocity in \eqref{eq:guidingcentre} is
\begin{equation}\label{eq:vparallel1}
    v_\parallel \coloneq \sigma\sqrt{2\?(\mathcal{E}-\mu B)}\,,
\end{equation}
where $\mathcal{E}$ is the kinetic energy per unit mass of the particle, $\mu$ is its magnetic moment, and $\sigma \coloneq v_\parallel/|v_\parallel|$, and the curl in \eqref{eq:guidingcentre} is taken at fixed $\mathcal{E}$ and $\mu$. We have ignored the electrostatic potential in \eqref{eq:vparallel1} because it is negligible for sufficiently energetic particles.

According to \eqref{eq:guidingcentre}, the particle does not perfectly follow magnetic field lines. Instead, it moves along the field lines of the divergence-free field \mycitep{Morozov1959, Morozov1966}
\begin{equation}\label{eq:B*}
    \bb{B}^\star \coloneq \bb{B} + \bb{\nabla}\bb{\times}\biggl( \frac{v_\parallel\bb{B}}{\Omega} \biggr)\,,
\end{equation}
where $\mathcal{E}$ and $\mu$ are treated as fixed parameters. For passing particles, this field $\bb{B}^\star$ exists throughout a toroidal region and is a small perturbation to the real magnetic field $\bb{B}$, which we assume has nested toroidal flux surfaces. It is well known that when a magnetic field with nested toroidal flux surfaces is perturbed, rational flux surfaces break up into magnetic islands. Therefore, the field lines of $\bb{B}^\star$, which actually represent guiding-centre trajectories through space, will exhibit islands: these are the drift islands. 

If a magnetic field with nested toroidal flux surfaces is modified by a small perturbation, the width of the resulting magnetic islands scales with the square root of the ratio of the perturbation amplitude to the magnetic shear \mycitep[\eg][]{Boozer2005}. Applying this result to the field in \eqref{eq:B*}, where the size of the perturbation is $\sim vB/\Omega L \sim \rho_\star B$, we find that the drift-island width scales with $(\rho_\star/s)^{1/2}$. 

The scaling with $\smash{\sqrt{\rho_\star}}$ is recovered,\footnote{This is a special case of a general result concerning nearly integrable Hamiltonian systems: perturbations of size $\sim\!\varepsilon$ (relative to the unperturbed Hamiltonian) that remove integrability cause invariant tori in phase space to break up into islands of width $\sim\!\sqrt{\varepsilon}$ around resonances \mycitep[\eg]{Chirikov1979, Lichtenberg2013}.} but what about the factor of $\smash{\sqrt{\delta}}$, which reduces the drift-island width in optimized configurations? Later in the article (in \cref{subsec:omnigeneity}, where we give a precise definition of $\delta$), we will explain how the argument above may be extended to obtain this factor. To summarize, we will show that, in a perfectly cyclometric stellarator, $\bb{B}^\star$ is integrable to higher order in $\rho_\star$. Therefore, in a nearly cyclometric stellarator, the perturbation to $\bb{B}^\star$ that breaks integrability comes from the magnetic drift due to the \emph{non-cyclometric piece} of the field. The size of this perturbation is $\sim\!\rho_\star\delta B$. Using this perturbation size in the magnetic-island-width formula leads to the scaling \eqref{eq:scaling} that we found previously for the drift-island width, including the factor of $\sqrt{\delta}$.

\section{Drift islands and the transit adiabatic invariant}\label{sec:adiabaticinvariant}

In this section, we derive an equation for the drift-island shape by expanding the guiding-centre equations around a rational surface (\crefrange{subsec:coordinates}{subsec:hamiltonianderivation}), which will reveal the existence of an adiabatic invariant (\cref{subsec:invariantphysics}). The shape of the drift islands, as they would appear in a Poincar\'e section, can be visualized by plotting the level curves of this invariant (\cref{subsec:islandplots}).

The calculations in this section rely on three main assumptions. First, we use the guiding-centre equations; in \cref{app:GCvsFO} we compare guiding-centre and full-orbit simulations to verify that the guiding-centre model is accurate in reactor-scale stellarators even for the most energetic alpha particles. Second, we assume that the stellarator magnetic field does not change in time and that it possesses nested toroidal flux surfaces. In fact, this assumption is not necessary; in \cref{app:magneticislands} we explain how the analysis can be generalized to stellarators whose magnetic field contains magnetic islands. Finally, we assume that drift-island overlap does not occur. It is well known that, when the drift islands around two nearby rational surfaces become large enough to overlap, the islands are replaced by a region of stochastic orbits \mycitep{Chirikov1979}, a phenomenon not described by our theory. The assumption of no drift-island overlap is discussed further in \cref{subsec:shear}.

\subsection{Coordinate system}\label{subsec:coordinates}

Near a rational surface, passing particles rapidly execute almost-closed orbits by streaming along almost-closed field lines. Meanwhile, they slowly drift perpendicular to the field lines and they slowly migrate tangentially across the rational surface as described in \cref{subsec:orbitwidth}. It will be useful to choose coordinates that separate the fast and slow timescales.

\begin{figure}
\vspace{3mm}
\centering
\includegraphics[width=\textwidth]{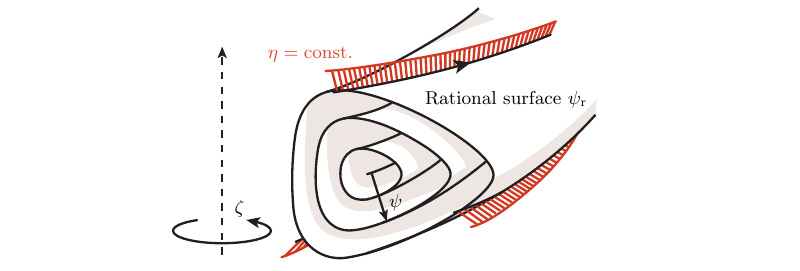}
    \caption{Diagram of the coordinate system that we use. The outermost flux surface displayed here is the rational surface $\psi_{\rm{r}}$, on which a magnetic field line is drawn. The constant-$\eta$ surface that coincides, on $\psi_{\rm{r}}$, with this field line is shown in red. }
    \label{fig:coordinates}
\end{figure}

In \cref{sec:basictheory}, we introduced the straight-field-line coordinates $(\psi,\theta,\zeta)$. We will still use $\zeta$, which varies on the fast timescale of a transit, and $\psi$, which changes on a slower timescale due to drifts. Instead of the poloidal angle $\theta$, we will now employ a combination of angles that changes more slowly,\footnote{An analogous change of coordinates is routinely used in the theory of motion close to a resonance in quasi-integrable Hamiltonian systems. In this context, the resulting coordinate, which is a linear combination of angle coordinates that does not change for particles exactly on-resonance, is called the `slow angle' \mycitep[\eg]{Lichtenberg2013, Hamilton2023}.} namely
\begin{equation}\label{eq:etadef}
\eta \coloneq \theta - \frac{N}{M}\? \zeta\,.
\end{equation}
As above, $N/M$ is the rotational transform of the rational surface $\psi_{\rm{r}}$, where $N$ and $M>0$ are coprime integers. On the rational surface, $\eta$ coincides with the usual field-line label $\alpha \coloneq \theta - \iota(\psi)\? \zeta$. On nearby flux surfaces, however, curves of constant $\eta$ remain closed, with the helicity of the rational-surface field lines, so these curves are not perfectly tangent to the magnetic field. We can think of $\eta$ as an extension of the rational-surface field-line label to nearby flux surfaces, or as a way of telling us which rational-surface field line a particle is close to (see \cref{fig:coordinates}). 

We take $\zeta\in[0,2\upi M)$ to describe position along each constant-$\eta$ curve. Then, each of these curves can be uniquely labelled using ${\eta\in [0,2\upi/M)}$.

%
%

\subsection{Orbit calculation}\label{subsec:hamiltonianderivation}

The coordinates $(\psi(\bb{X}), \eta(\bb{X}), \zeta(\bb{X}))$ of a particle's guiding-centre position $\bb{X}$ evolve according to the guiding-centre equations
\begin{subequations}\label{eq:guidingcentre2}
\begin{align}
\label{eq:dpsidt}\frac{\rmd\psi}{\rmd t} &= \bb{v}_{\rm{d}}\0\bb{\nabla}\psi\,,\\
\label{eq:dalphadt}\frac{\rmd\eta}{\rmd t} &= v_\parallel\bh \0\bb{\nabla}\eta + \bb{v}_{\rm{d}}\0\bb{\nabla}\eta \,,\\
\label{eq:dzetadt}\frac{\rmd\zeta}{\rmd t} &= v_\parallel\bh \0\bb{\nabla}\zeta\,.
\end{align}
Here, $\bh \coloneq \bb{B} /B$ is the unit vector parallel to the magnetic field, $\vp$ is the parallel velocity, defined in \eqref{eq:vparallel1}, and
\begin{equation}
\label{eq:vdrift}\bb{v}_{\rm{d}} \coloneq \frac{v_\parallel}{B}\,\bb{\nabla}\bb{\times}\biggl(\frac{v_\parallel\bb{B} }{\Omega}\biggr)
\end{equation}
\end{subequations}
is the drift velocity, where the curl in \eqref{eq:vdrift} is taken at fixed $\mathcal{E}$ and $\mu$. All functions of position in \eqref{eq:guidingcentre2} are evaluated at the guiding-centre position $\bb{X}$. As a reminder, we did not include the electrostatic potential in the definition \eqref{eq:vparallel1} of the parallel velocity, which means the $\bb{E\times B}$ drift does not appear in \eqref{eq:vdrift}, because potential variations have a negligible effect on the motion of sufficiently energetic particles. Potential variations can easily be incorporated into the following calculations by including the potential in $v_\parallel$.

The physical interpretation of the expressions for $\rmd\psi/\rmd t$ and $\rmd\zeta/\rmd t$ is straightforward: $\psi$ changes due to the radial magnetic drift, while $\zeta$ changes due to parallel streaming along field lines. Meanwhile, the expression for $\rmd\eta/\rmd t$ in \eqref{eq:dalphadt} consists of two terms. The $\bb{v}_{\rm{d}}\0\bb{\nabla}\eta$ `drift term' represents the magnetic drift pushing the particle tangentially across the flux surface. The $\smash{v_\parallel\bh \0\bb{\nabla}\eta}$ `shear term' accounts for changes in $\eta$ due to the slight misalignment between the magnetic field lines, along which the particle rapidly streams, and the curves of constant $\eta$ on the particle's current flux surface (see \cref{fig:etavsfieldline}). This misalignment increases with distance from the rational surface, since the shear term can be expressed as
\begin{equation}\label{eq:shearterm}
    v_\parallel\bh \0\bb{\nabla}\eta = v_\parallel\?\biggl(\iota(\psi) - \frac{N}{M}\biggr)\,\bh \0\bb{\nabla}\zeta\,.
\end{equation}
We now introduce the assumption that the particle is close to the rational surface, in the sense that
\begin{equation}\label{eq:ass}
    \biggl|\iota(\psi) - \frac{N}{M}\biggr|\ll1\,.
\end{equation}
In a low-shear device, this condition may be satisfied everywhere in the plasma. In a higher-shear stellarator, condition \eqref{eq:ass} is equivalent to $|\iota_{\rm r}'(\psi-\psi_{\rm r})|\ll 1$; this defines narrow regions around low-order rational surfaces where our calculations are applicable.

When \eqref{eq:ass} is satisfied, the rate of change of $\eta$ is small, as expected. We refer to $\psi$ and $\eta$ as the `slow coordinates'. To predict the evolution of the slow coordinates over many nearly periodic transits, we average, at fixed $\psi$ and $\eta$, over the transit motion,
\begin{subequations}\label{eq:deltaeqs}
\begin{align}
    \label{eq:deltapsi}\biggl\langle\frac{\rmd\psi}{\rmd t}\biggr\rangle_{\!\!\rm{t}} &= \langle \bb{v}_{\rm{d}}\0\bb{\nabla}\psi \rangle_{\rm{t}}\,, \\
    \label{eq:deltaalpha}\biggl\langle\frac{\rmd\eta}{\rmd t}\biggr\rangle_{\!\!\rm{t}} &= \langle v_\parallel\bh \0\bb{\nabla}\eta \rangle_{\rm{t}} + \langle \bb{v}_{\rm{d}}\0\bb{\nabla}\eta \rangle_{\rm{t}}\,.
\end{align}
\end{subequations}
Here, we have defined the transit average
\begin{subequations}\label{eq:transitaverage}
\begin{equation}
    \langle \? \ldots\?  \rangle_{\rm{t}} \coloneq \frac{1}{\tau_{\rm{t}}}\oint\frac{ (\ldots)}{|v_\parallel|\,\bh \0\bb{\nabla}\zeta}\,\rmd\zeta\,,
\end{equation}
where the transit time is
\begin{equation}
    \tau_{\rm{t}} \coloneq \oint \frac{1}{|v_\parallel|\,\bh \0\bb{\nabla}\zeta}\,\rmd\zeta\,.
\end{equation}
\end{subequations}
Throughout the paper, the closed-contour integral symbol denotes a line integral along a curve of constant $\psi$ and $\eta$, in the direction of the magnetic field (without loss of generality, we assume $\smash{\bh \0\bb{\nabla}\zeta > 0}$ so that the transit time is positive). Therefore, in \eqref{eq:transitaverage}, where the integration variable is the toroidal angle, the integrals are taken from $\zeta = 0$ to $\zeta = 2\upi M$.

We need to evaluate the transit averages appearing in \eqref{eq:deltaeqs}. Using \eqref{eq:vdrift}, the drifts appearing in these equations may be written as
\begin{subequations}\label{eq:driftdiv}
\begin{align}
\label{eq:psidriftdiv}\bb{v}_{\rm{d}}\0\bb{\nabla}\psi &= \frac{v_\parallel}{B}\,\bb{\nabla}\0\Bigl(\frac{v_\parallel}{\Omega}\bb{B} \bb{\times}\bb{\nabla}\psi\Bigr)\,,\\
\label{eq:alphadriftdiv}\bb{v}_{\rm{d}}\0\bb{\nabla}\eta &= \frac{v_\parallel}{B}\,\bb{\nabla}\0\Bigl(\frac{v_\parallel}{\Omega}\bb{B} \bb{\times}\bb{\nabla}\eta\Bigr)\,.
\end{align}
\end{subequations}
The Jacobian for the $(\psi,\eta,\zeta)$ coordinate system is $\smash{((\bb{\nabla}\psi\bb{\times}\bb{\nabla}\eta)\0\bb{\nabla}\zeta)^{-1} = (\bb{B} \0\bb{\nabla}\zeta)^{-1}}$, so evaluating the divergences in \eqref{eq:driftdiv} leads to
\begin{subequations}\label{eq:easyform}
\begin{align}
\bb{v}_{\rm{d}}\0\bb{\nabla}\psi &=  v_\parallel\bh \0\bb{\nabla}\zeta\Bigl[ \partial_\eta\Bigl(\frac{v_\parallel}{\Omega}\? \bb{B} \0\con{\zeta} \Bigr) - \partial_\zeta \Bigl(\frac{v_\parallel}{\Omega}\? \bb{B} \0\con{\eta}  \Bigr) \Bigr]\,,\\
\bb{v}_{\rm{d}}\0\bb{\nabla}\eta &=  v_\parallel\bh \0\bb{\nabla}\zeta\Bigl[ \partial_\zeta \Bigl(\frac{v_\parallel}{\Omega}\? \bb{B} \0 \con{\psi} \Bigr) - \partial_\psi\Bigl(\frac{v_\parallel}{\Omega}\? \bb{B} \0\con{\zeta} \Bigr) \Bigr]\,.
\end{align}
\end{subequations}
Here, $\bb{x}(\psi, \theta,\zeta)$ denotes the point in physical space with coordinates $(\psi, \theta, \zeta)$. Throughout the paper, unless otherwise specified, the partial-derivative symbols $\partial_\psi$, $\partial_\eta$ and $\partial_\zeta$ always represent partial derivatives with respect to one of the coordinates $(\psi,\eta,\zeta)$, with the other two held fixed (and with $\mathcal{E}$ and $\mu$ fixed). Thus, the vectors $\con{\psi} $, $\con{\eta} $ and $\con{\zeta} $ are the contravariant basis vectors for our straight-field-line coordinate system. Using \eqref{eq:easyform}, the transit-averaged drifts are
\begin{subequations}\label{eq:driftaverage}
\begin{align}
\label{eq:psidriftav}\langle\bb{v}_{\rm{d}}\0\bb{\nabla}\psi\rangle_{\rm{t}} &= \frac{mc}{Ze\tau_{\rm{t}}}\,\partial_\eta I\,,\\
    \label{eq:alphadriftav}\langle\bb{v}_{\rm{d}}\0\bb{\nabla}\eta\rangle_{\rm{t}} &= -\frac{mc}{Ze\tau_{\rm{t}}}\,\partial_\psi I\,,
\end{align}
\end{subequations}
where
\begin{equation}\label{eq:Jdef}
    I(\psi,\eta,\mathcal{E},\mu) \coloneq \oint |v_\parallel|\? \bh \0\con{\zeta} \, \rmd \zeta\,.
\end{equation}
Formulas \eqref{eq:driftaverage} are reminiscent of the usual identities for the drift rate of trapped particles \mycitep[see, \eg][, equations (60) and (61)]{Helander2014}. Interestingly, formulas \eqref{eq:driftaverage} are exact in the following sense: their derivation, starting with \eqref{eq:driftdiv}, did not require anywhere that the particle is close to the rational surface $\psi_{\rm{r}}$. We will make use of this fact in \cref{subsec:higherordercalc}. Of course, transit-averaging is only meaningful if the integration contour approximates the actual particle trajectory, which is only the case when ${|\iota-N/M|\ll 1}$.

We also need the transit average of the shear term, which follows from \eqref{eq:shearterm},
\begin{equation}\label{eq:shearaverage}
    \langle v_\parallel\bh \0\bb{\nabla}\eta\rangle_{\rm{t}} = \frac{2\upi M \sigma}{\tau_{\rm{t}}} \biggl(\iota(\psi) - \frac{N}{M}\biggr)\,.
\end{equation}
This equation is physically transparent: in one transit, which takes time $\tau_{\rm{t}}$, the change in $\eta$ due to streaming is $2\upi M\sigma\?  [\iota(\psi)-N/M]$, as shown in \cref{fig:etavsfieldline}.

Finally, using results \eqref{eq:driftaverage} and \eqref{eq:shearaverage} in \eqref{eq:deltaeqs}, we find that the transit-averaged rate of change of the slow coordinates is
\begin{subequations}\label{eq:changeslow}
\begin{align}
    \biggl\langle\frac{\rmd\psi}{\rmd t}\biggr\rangle_{\!\!\rm{t}} &= \frac{\sigma mc}{Ze\tau_{\rm{t}}}\,\partial_\eta \mathcal{I}\,,\\
    \label{eq:avetachange}\biggl\langle\frac{\rmd\eta}{\rmd t}\biggr\rangle_{\!\!\rm{t}} &= -\frac{\sigma mc}{Ze\tau_{\rm{t}}}\,\partial_\psi \mathcal{I}\,.
\end{align}
\end{subequations}
Here, we have introduced the `transit adiabatic invariant'
\begin{equation}\label{eq:invariant}
    \mathcal{I}(\psi,\eta,\mathcal{E},\mu,\sigma) \coloneq \sigma I(\psi,\eta,\mathcal{E},\mu) - \frac{2\upi M Z e}{mc}\!\int_{\psi_{\rm{r}}}^\psi \biggl(\iota(\psi') - \frac{N}{M}\biggr)\,\rmd\psi'\,,
\end{equation}
which, according to \eqref{eq:changeslow}, is conserved on average over multiple transits:
\begin{equation}
    \biggl\langle\frac{\rmd\mathcal{I}}{\rmd t}\biggr\rangle_{\!\!\rm{t}} = \partial_\psi\mathcal{I}\,\biggl\langle\frac{\rmd\psi}{\rmd t}\biggr\rangle_{\!\!\rm{t}} + \partial_\eta\mathcal{I}\,\biggl\langle\frac{\rmd\eta}{\rmd t}\biggr\rangle_{\!\!\rm{t}} = 0\,.
\end{equation}
The existence of the adiabatic invariant \eqref{eq:invariant} for passing particles near a rational flux surface is one of the key results of this paper.

\subsection{Transit adiabatic invariant}\label{subsec:invariantphysics}

We now discuss the physical meaning of the transit adiabatic invariant. Its definition in \eqref{eq:invariant} is a sum of two terms. The first is a `kinetic' term $I(\psi,\eta,\mathcal{E},\mu)$, which can be simplified using the fact that $\smash{\bh} $ is parallel to $\con{\zeta} $ to leading order in $|\iota - N/M|\ll 1$,
\begin{equation}
    \bh  = (\bh \0\bb{\nabla}\zeta)\biggl[\con{\zeta}  + \biggl(\iota(\psi)-\frac{N}{M}\biggr) \con{\eta}  \biggr]\simeq (\bh \0\bb{\nabla}\zeta)\, \con{\zeta} \,.
\end{equation}
Then,
\begin{equation}\label{eq:kineticterm}
    I \simeq \oint |v_\parallel|\,\rmd l\,.
\end{equation}
Here, $l$ denotes arc length along the integration contour, which, as before, is the closed curve obtained by holding the values of $\psi$ and $\eta$ fixed. The kinetic term, in the form \eqref{eq:kineticterm}, is similar to the usual second adiabatic invariant for trapped particles \mycitep[\eg]{Helander2014}. 

The second term in \eqref{eq:invariant} is a `geometric' term that depends only on the distance $\psi - \psi_{\rm{r}}$ from the rational surface; it is independent of $\mathcal{E}$ and $\mu$. This term has a simple physical interpretation: it is proportional to the `helical flux' $\chi_{\rm{r}}(\psi)$ \mycitep{Boozer2016} passing between flux surface $\psi$ and the rational surface $\psi_{\rm{r}}$,
\begin{equation}\label{eq:helicalflux}
\chi_{\rm{r}}(\psi) \coloneq \int_{\psi_{\rm{r}}}^\psi \biggl(\iota(\psi') - \frac{N}{M}\biggr)\,\rmd \psi'\,.
\end{equation}
The reason $\chi_{\rm{r}}(\psi)$ is called the helical flux is as follows. It is straightforward to check that $\bb{B}  = \bb{\nabla}\bb{\times}\bb{A}$ for a vector potential $\bb{A} \coloneq (\psi-\psi_{\rm{r}})\? \bb{\nabla}\eta - \chi_{\rm{r}}(\psi)\?  \bb{\nabla}\zeta$. Then, we have
\begin{equation}\label{eq:lineintegral}
\oint \bb{A}\0\rmd\bb{x} = -2\upi M \chi_{\rm{r}}(\psi)\,.
\end{equation}
The name `helical flux' is appropriate because we can apply Stokes' Theorem to the line integral on the left side of \eqref{eq:lineintegral} to conclude that $-2\upi M \chi_{\rm{r}}(\psi)$ equals the magnetic flux through the two-dimensional surface bounded by the two curves labelled by $(\psi,\eta)$ and $(\psi_{\rm{r}},\eta)$, which will look similar to the red helical strip in \cref{fig:coordinates}.

Using formulas \eqref{eq:Jdef} and \eqref{eq:lineintegral} for the kinetic and geometric terms, respectively, the transit adiabatic invariant can be expressed as
\begin{equation}\label{eq:hamiltonianform}
 \mathcal{I} = \oint \?\biggl( v_\parallel\bh  + \frac{Ze}{mc}\bb{A} \biggr)\0\,\rmd\bb{x}\,.
\end{equation}
Equation \eqref{eq:hamiltonianform} is similar to the standard formula
\begin{equation}\label{eq:adiabaticinvariant}
\mathcal{A} = \oint \bb{p}\0\rmd\bb{q}
\end{equation}
for adiabatic invariants in Hamiltonian systems \mycitep{Hazeltine2018}. Here, $\bb{q}$ is the vector of generalized position coordinates and $\bb{p}$ is the associated canonical momentum. In Cartesian coordinates, the canonical momentum for a charged particle in a magnetic field is $\bb{p} = m\bb{v} + (Ze/c)\bb{A}$. For trapped particles, \eqref{eq:adiabaticinvariant} leads to the usual second adiabatic invariant because the vector potential term in the canonical momentum cancels out upon integrating along both halves of the bounce motion (along the field line and back). However, for passing particles, this vector potential term \emph{does not} vanish, because the integral is taken along a closed loop circulating in one direction around the stellarator.

\subsection{Drift-island shape}\label{subsec:islandplots}

We have shown that passing-particle orbits near a rational surface $\psi_{\rm{r}}$ conserve the transit adiabatic invariant $\mathcal{I}$. Plotting the level sets of $\mathcal{I}$ allows us to visualize these orbits, and any drift islands they trace out, as they would appear in a Poincar\'e plot. For example, plotting the level sets of $\mathcal{I}$ in the $(\eta,\psi)$ plane shows the drift islands as they would appear in a Poincar\'e plot of $\psi$ versus $\theta$ at a fixed toroidal angle.

%
%
\begin{figure}
\vspace{3mm}
\centering
\includegraphics[width=\textwidth]{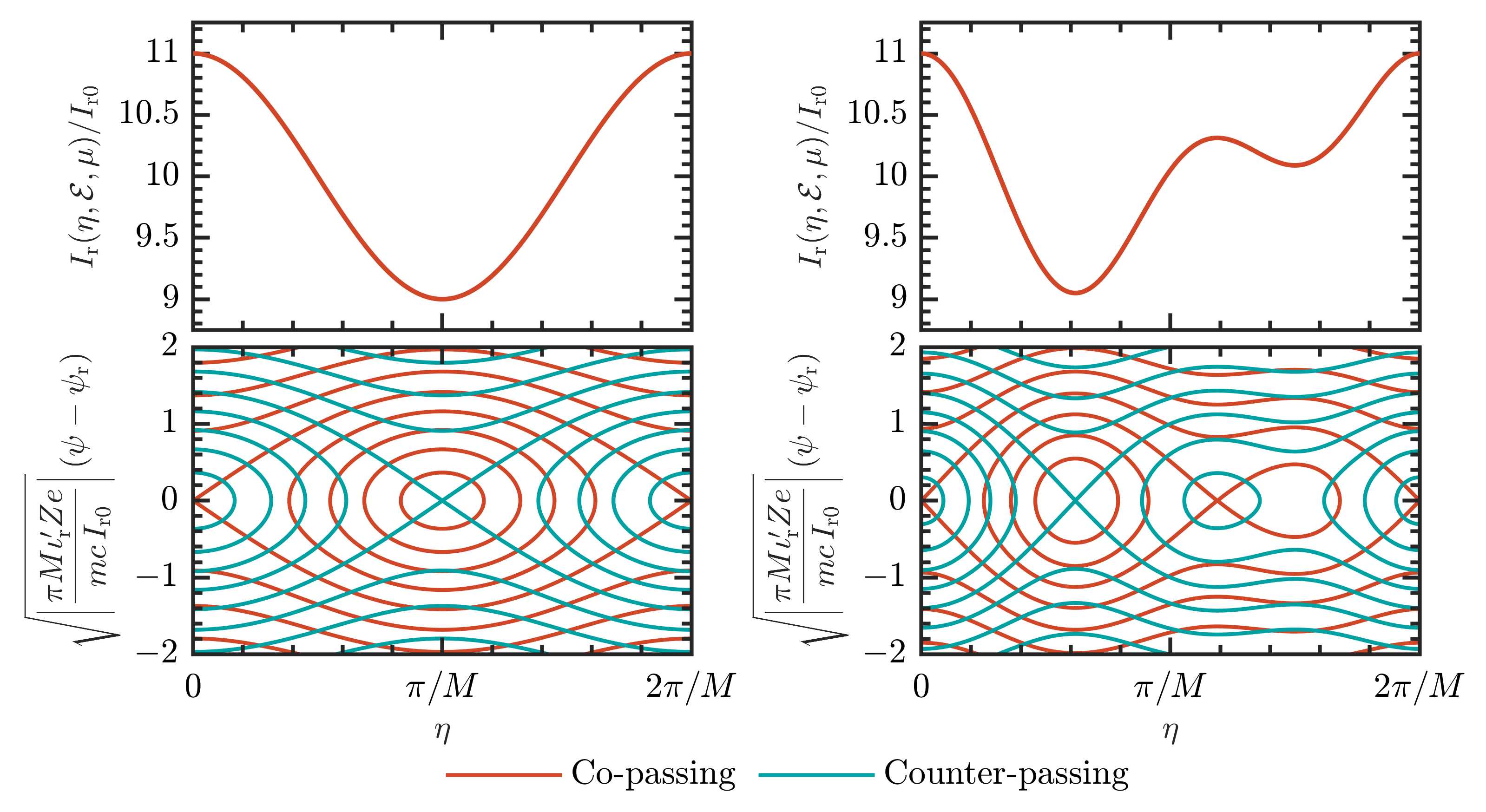}
    \caption{(top row) Examples of possible $I_{\rm{r}}(\eta,\mathcal{E},\mu)$, normalized by a typical value $I_{\rm{r}0}\sim vL$, plotted against $\eta$.
    (bottom row) Particle orbits near the rational surface, visualized by plotting level sets of the invariant $\mathcal{I}_{\rm r}(\psi,\eta,\mathcal{E},\mu,\sigma)$ (defined in \eqref{eq:simpleinvariant}) corresponding to the $I_{\rm{r}}$ functions above. These level sets are plotted against $\psi-\psi_{\rm r}$ (normalized to be dimensionless) on the vertical axis and $\eta$ on the horizontal axis. Co-passing ($\sigma=1$) and counter-passing ($\sigma=-1$) orbits are plotted in red and blue, respectively. For these plots, $\iota_{\rm r}'$ was assumed to be negative.}
    \label{fig:theory_island_plots}
\end{figure}

Before making such a plot, we can further simplify formula \eqref{eq:invariant} for $\mathcal{I}$. So far, the only assumption we have made regarding proximity to the rational surface is ${|\iota - N/M|\ll 1}$, which, if the shear is low, may still be satisfied at relatively large values of ${|\psi-\psi_{\rm{r}}|}$. We now make the stronger assumption that ${|\psi-\psi_{\rm{r}}|\ll \Psi_{\rm{t}}}$, since the drift islands usually fall inside a small region around $\psi_{\rm{r}}$. We will see, in \cref{subsec:lowshear}, that this is equivalent to assuming that the magnetic shear satisfies $s \gg\rho_\star$. The condition ${|\psi-\psi_{\rm{r}}|\ll \Psi_{\rm{t}}}$ allows us, in \eqref{eq:invariant}, to replace $I(\psi,\eta,\mathcal{E},\mu)$ with its value at the rational surface, for which we write
\begin{equation}\label{eq:Jr}
    I_{\rm{r}}(\eta, \mathcal{E},\mu)\coloneq I(\psi_{\rm{r}}, \eta, \mathcal{E},\mu) = \oint_{\rm{r}} \,\frac{|v_\parallel|}{\bh\0\grad{\zeta}} \, \rmd \zeta\,.
\end{equation}
Throughout the paper, the subscript `$\rm{r}$' denotes an integral taken along the rational-surface field line labelled by $(\psi_{\rm{r}}, \eta)$. With this replacement, together with the approximation $\iota(\psi)-N/M\simeq \iota'_{\rm r}(\psi - \psi_{\rm r})$, the transit invariant becomes\footnote{The approximation \eqref{eq:simpleinvariant} for the transit invariant is equivalent to the standard `pendulum approximation' for the Hamiltonian close to a resonance in a quasi-integrable Hamiltonian system \mycitep[\eg]{Lichtenberg2013, Hamilton2023}.}
\begin{equation}\label{eq:simpleinvariant}
    \mathcal{I}(\psi,\eta,\mathcal{E},\mu,\sigma)\simeq \mathcal{I}_{\rm r}(\psi,\eta,\mathcal{E},\mu,\sigma) \coloneq \sigma I_{\rm{r}}(\eta, \mathcal{E},\mu) - \frac{\upi M Z e}{mc}\?\iota_{\rm r}'(\psi-\psi_{\rm{r}})^2\,.
\end{equation}
The level sets of this function are drawn in the ($\eta, \psi$) plane in \cref{fig:theory_island_plots} for two possible forms of $I_{\rm{r}}(\eta,\mathcal{E},\mu)$. In the top row of \cref{fig:theory_island_plots}, we show the two example functional forms of $I_{\rm{r}}(\eta,\mathcal{E},\mu)$. In the bottom row, we plot the levels sets of $\mathcal{I}_{\rm r}(\psi,\eta,\mathcal{E},\mu,\sigma)$ corresponding to the $I_{\rm{r}}$ functions above, for both possible signs of $\sigma$: co-passing ($\sigma=1$) and counter-passing ($\sigma=-1$) orbits are plotted in red and blue, respectively. The two functional forms of $I_{\rm{r}}(\eta,\mathcal{E},\mu)$ in the top row were chosen because they give rise to qualitatively different drift-island shapes in the bottom row; the case on the left is more representative of what is observed in real stellarators.

In each case, the level sets of $\mathcal{I}_{\rm r}(\psi,\eta,\mathcal{E},\mu,\sigma)$ display a clear island structure. We also see that co- and counter-passing particles trace out different drift islands. The reason is that, in \eqref{eq:simpleinvariant}, the sign of $I_{\rm{r}}$ is different for co- and counter-passing particles, while the sign of the geometric term is not. This is a consequence of the fact that, in the canonical momentum $\bb{p} = m\bb{v} + (Ze/c)\bb{A}$, the first term depends on the direction of circulation around the torus, while the second term does not.

%
%
\begin{figure}
\centering
\begin{tikzpicture}[x=\textwidth]
\node[anchor=south west,inner sep=0] (image) at (0.0,0) {\includegraphics[width=\textwidth]{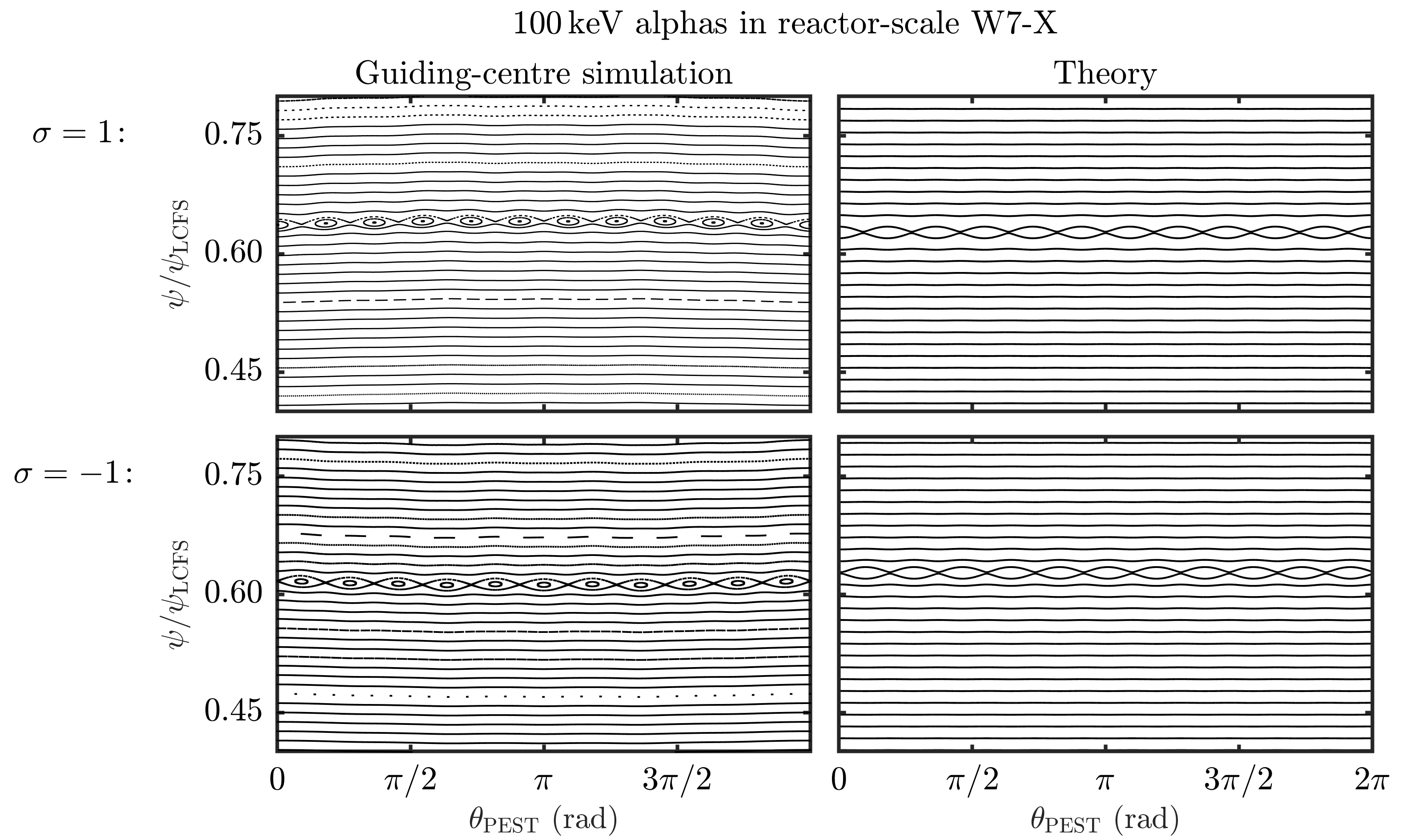}};
    \begin{scope}[x={(image.south east)},y={(image.north west)}]
    \end{scope}
\end{tikzpicture}
    \caption{(left panels) Poincar\'e plot of the guiding-centre orbits of passing alpha particles in a reactor-scale W7-X equilibrium, simulated using ASCOT5. These Poincar\'e plots are taken at cylindrical angle $\phi = 0$. The radial coordinate is $\psi$ normalized by its value at the last closed flux surface, $\psi_{\rm LCFS}$. The coordinate on the horizontal axis is the poloidal straight-field-line angle $\theta_{\rm PEST}$ that corresponds to the cylindrical toroidal angle. The alphas are freely passing, with energy $\SI{100}{\kilo\electronvolt}$ and $\mu = 0$. They are co-passing ($\sigma = 1$) in the upper panel and counter-passing ($\sigma=-1$) in the lower panel. (right panels) Level sets of $\mathcal{I}_{\rm r}(\psi, \eta, \mathcal{E}, \mu, \sigma)$, defined in \eqref{eq:simpleinvariant}, for particles near the $\iota = 10/11$ rational surface. In each panel on the right, the sign of $\sigma$ is the same as in the corresponding panel on the left.}
    \label{fig:W7X_lowenergy_lowestorder}
\end{figure}

%
%
\begin{figure}
\centering
\begin{tikzpicture}[x=\textwidth]
\node[anchor=south west,inner sep=0] (image) at (0.0,0) {\includegraphics[width=\textwidth]{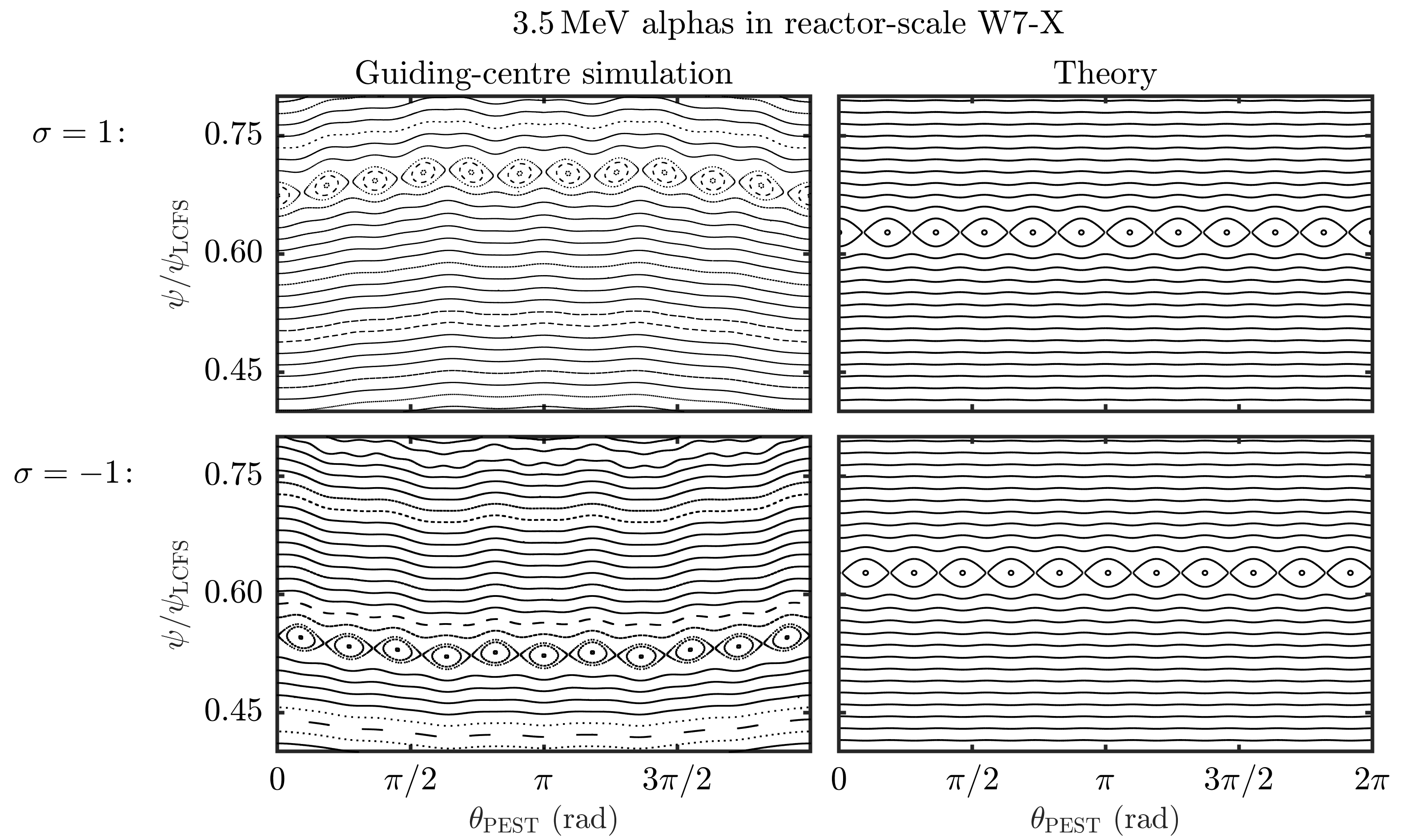}};
    \begin{scope}[x={(image.south east)},y={(image.north west)}]
    \end{scope}
\end{tikzpicture}
    \caption{Poincar\'e plot and theory comparison similar to \cref{fig:W7X_lowenergy_lowestorder} but for alphas with energy $\SI{3.5}{\mega\electronvolt}$. We see that the theory predicts the drift-island width accurately but fails to capture the $\theta$ dependence of the shape of the island chain or its radial shift away from the rational surface.}
    \label{fig:W7X_lowestorder}
\end{figure}

%
%
\begin{figure}
\vspace{7mm}
\centering
\includegraphics[width=\textwidth]{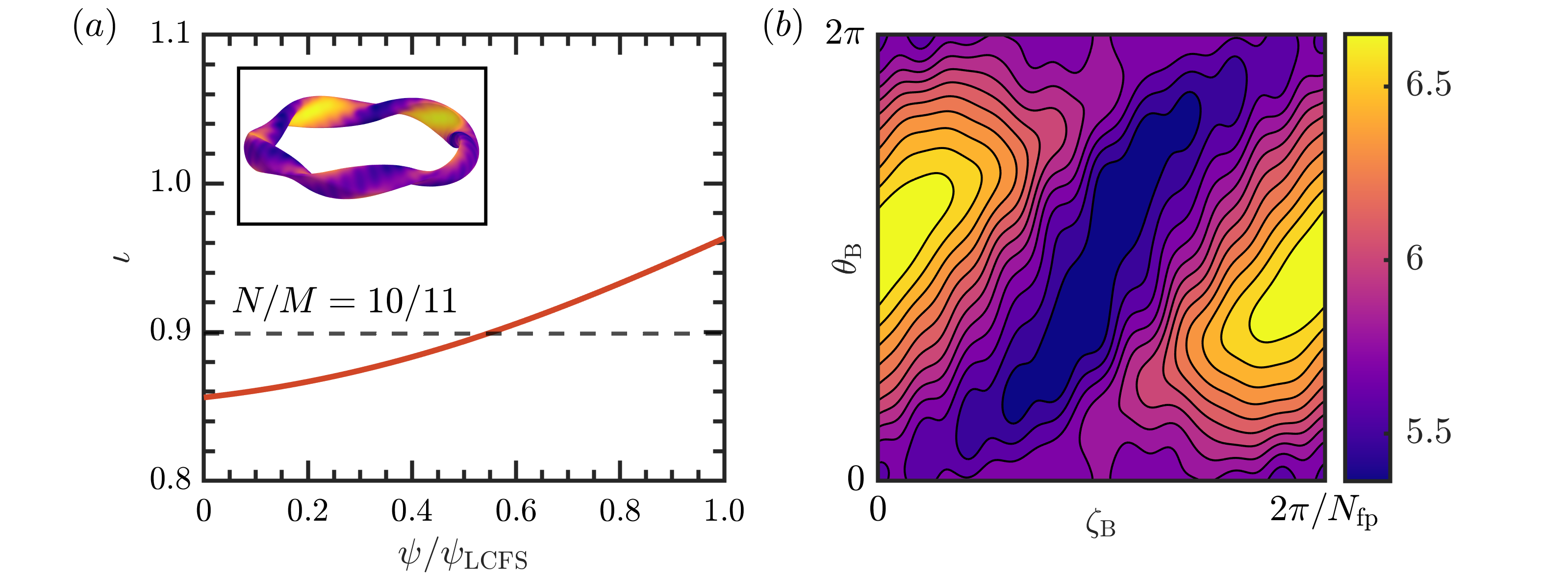}
    \caption{$(a)$ Rotational transform of the W7-X equilibrium we use to trace alpha orbits in \crefand{fig:W7X_lowenergy_lowestorder}{fig:W7X_lowestorder}. $(b)$ Plot of the magnetic field strength in Boozer coordinates on the rational surface with $\iota = N/M = 10/11$.}
    \label{fig:W7Xeqm}
\vspace{4mm}
\end{figure}

The discussion in this subsection shows that the transit-invariant equation, \eqref{eq:invariant} or \eqref{eq:simpleinvariant}, can be thought of as an equation for the drift-island shape. In \crefand{fig:W7X_lowenergy_lowestorder}{fig:W7X_lowestorder}, we compare this theory with Poincar\'e plots obtained from collisionless, guiding-centre simulations conducted using the ASCOT5 orbit-following code \mycitep{Varje2019}. The guiding-centre equations simulated by ASCOT5 are not exactly the same as the guiding-centre equations we used for our calculations because ASCOT5 includes relativistic effects; this makes very little difference to the alpha-particle trajectories at the relevant energies..

The equilibrium used for these simulations was a $\beta = 2\%$ W7-X \mycitep{Beidler1990} standard-configuration equilibrium scaled up to reactor size (throughout this article, all equilibria were scaled to the same volume, $\SI{444}{\meter\cubed}$, and volume-averaged magnetic field strength, $\SI{5.86}{\tesla}$, as the ARIES-CS reactor design, as in \citet{Bader2021}). We provide more information about this equilibrium in \cref{fig:W7Xeqm}. The rotational-transform profile shows that there is a rational surface with $\iota = N/M = 10/11$ in the plasma; drift islands form around this flux surface. This W7-X equilibrium is from the examples directory for the DESC stellarator optimization package \mycitep{Dudt2020}. We used a scaled-up DESC equilibrium to evaluate our theoretical expressions. In order to load the magnetic-field data into ASCOT5, we created a VMEC file \mycitep{Hirshman1986} for the equilibrium and evaluated the magnetic field on a cylindrical grid using the EXTENDER code \mycitep{Drevlak2005}.

In \crefand{fig:W7X_lowenergy_lowestorder}{fig:W7X_lowestorder}, we see that the level sets of $\mathcal{I}_{\rm r}$ accurately match the Poincar\'e plots from the simulations at $\SI{100}{\kilo\electronvolt}$ but differ at $\SI{3.5}{\mega\electronvolt}$. Typically, the theory begins to deviate from the simulations at higher energies. Nevertheless, while \eqref{eq:invariant} may not satisfactorily capture all of the features of the drift islands at high energies, it does usually give an accurate prediction of the drift-island \emph{width}. For this reason, in the next two sections, we use \eqref{eq:invariant} to investigate how the drift-island width depends on certain properties of the stellarator equilibrium. Later, in \cref{sec:higherorder}, we will carry out a more accurate calculation of the drift-island shape that precisely matches the simulations even at very high energies.

\section{Drift-island width in different stellarators}\label{sec:optimisation}

Although the lowest-order transit adiabatic invariant fails to capture all features of the drift islands at high energies, it does provide an accurate prediction of the drift-island width. We can, therefore, use \eqref{eq:invariant} or \eqref{eq:simpleinvariant} to find out how the stellarator magnetic field should be designed to ensure that any drift islands are small. In this section, we investigate how the drift-island width depends on properties of the equilibrium magnetic field, such as the number of field periods (\cref{subsec:resonance}), aspect ratio (\cref{subsec:lar}), and shear (\cref{subsec:shear}).

\subsection{Number of field periods}\label{subsec:resonance}

First, we discuss how the drift-island width depends on the number of field periods, $N_{\rm{fp}}$. In simulations, \citet{White2022} and \citet{White2022b} observed drift islands only around rational surfaces whose values of $N$ share common factors with $N_{\rm{fp}}$; they describe this as a `resonance with the modulation of the background field'. Below, we explain this observation by showing that the drift-island width predicted by \eqref{eq:simpleinvariant} is significantly larger if $N$ and $N_{\rm{fp}}$ share common factors.

For particles that cross the rational surface at $\eta = \eta_{\rm{r}}$, \eqref{eq:simpleinvariant} is equivalent to
\begin{equation}\label{eq:psisolution}
	\psi - \psi_{\rm{r}} = \pm \sqrt{\frac{mc\sigma\?[I_{\rm{r}}(\eta, \mathcal{E},\mu) - I_{\rm{r}}(\eta_{\rm{r}}, \mathcal{E},\mu)]}{\upi M Z e \?\iota_{\rm r}'}}\,.
\end{equation}
The drift-island half-width is the maximum value of this square root, so we must determine the maximum value of $\smash{|I_{\rm{r}}(\eta, \mathcal{E},\mu) - I_{\rm{r}}(\eta_{\rm{r}}, \mathcal{E},\mu)|}$ as $\eta$ varies. Recall that $I_{\rm{r}}$ is defined by
\begin{equation}\label{eq:Jrepeated}
    I_{\rm{r}} = \frac{Ze\sigma}{mc}\oint_{\rm{r}} \, V\,\rmd\zeta\,,
\end{equation}
where we introduce $\smash{V \coloneq (mc/Ze)(v_\parallel/\bh \bb{\cdot}\bb{\nabla}\zeta)}$ for later convenience. We may expand $V$ in \eqref{eq:Jrepeated} as a Fourier series in the angular coordinates,
\begin{equation}
    V = \sum_{(p,q) \in \mathbb{Z}^2} V_{pq}\,\rme^{\rmi (p\theta - q \zeta)}\,.
\end{equation}
We assume that the straight-field-line coordinates are chosen to have the same periodicity as the magnetic field strength; that is, the straight-field-line coordinates are chosen so that $\smash{\bh \bb{\cdot}\bb{\nabla}\zeta}$ is periodic under $\zeta$-increments of $2\upi/N_{\rm{fp}}$. An arbitrary straight-field-line coordinate system need not have this property \Dash Boozer coordinates are one example of a coordinate system that does. With this choice, many of the coefficients $V_{pq}$ are zero; the fact that there are $N_{\rm{fp}}$ field periods means the only non-zero harmonics have $q=N_{\rm{fp}}n$ for some $n\in\mathbb{Z}$. Therefore, we have
\begin{equation}\label{eq:Jint}
I_{\rm{r}} = \frac{Ze\sigma}{mc}\int_0^{2\upi M}\mspace{-10mu}\sum_{(p,n)\in\mathbb{Z}^2}\mspace{-8mu} V_{p,\?  N_{\rm{fp}}l}\,\rme^{\rmi p\eta}\? \rme^{\rmi ((N/M) p - N_{\rm{fp}}n)\zeta} \,\rmd\zeta
\end{equation}
since the integral in \eqref{eq:Jrepeated} is taken over a curve with $\eta = \theta - (N/M)\?  \zeta$ fixed. In \eqref{eq:Jint}, the only terms in the sum which do not integrate to zero are those with
\begin{equation}\label{eq:diophantine}
Np - M N_{\rm{fp}} n = 0\,.
\end{equation}
The solutions to this linear Diophantine equation take the form $(p,n) = \smash{(k\widetilde{M}, k\widetilde{N})}$ for some $k\in\mathbb{Z}$, where $\smash{\widetilde{M}} \coloneq MN_{\rm{fp}} / \operatorname{gcd}(N,N_{\rm{fp}})$, $\smash{\widetilde{N}}\coloneq NN_{\rm{fp}} / \operatorname{gcd}(N,N_{\rm{fp}})$, and we have used the fact that $N$ and $M$ are coprime. Here, $\operatorname{gcd}(a,b)$ stands for the greatest common divisor of $a$ and $b$; thus, $\smash{|\widetilde{N}|}$ is the smallest positive integer divisible by both $N$ and $N_{\rm fp}$, while $\smash{\widetilde{M}}$ is the integer that makes $\smash{\widetilde{N}/\widetilde{M}} = N/M$. Consequently, \eqref{eq:Jint} is equivalent to
\begin{equation}
I_{\rm{r}} = \frac{2\upi M Ze\sigma}{mc}\sum_{k\in\mathbb{Z}}V_{k\widetilde{M}\!,\? k\widetilde{N}}\, \rme^{\rmi k\widetilde{M}\eta}\,.
\end{equation}
Typically, the dominant harmonics are those with the smallest values of $|p|$ and $|q|$, so we can approximate $I_{\rm{r}}$ by keeping the $k=0$ and $k = \pm 1$ terms only. Then,
\begin{equation}\label{eq:oneharmonic}
    I_{\rm{r}}\approx  \frac{2\upi M Ze\sigma}{mc} \Bigl( V_{00} + V_{\widetilde{M}\widetilde{N}}\,\rme^{\rmi \widetilde{M}\eta} + V_{\widetilde{M}\widetilde{N}}^{\raisemath{1.5pt}{*}}\,\rme^{-\rmi \widetilde{M}\eta} \Bigr)\,,
\end{equation}
where the asterisk denotes complex conjugation. Substituting this approximation for $I_{\rm{r}}$ into \eqref{eq:psisolution}, we find that the drift-island half-width is
\begin{equation}\label{eq:widthNM}
\Delta \psi_{\rm{hw}} \approx \sqrt{\biggl\lvert\frac{8\?  V_{\widetilde{M}\widetilde{N}}}{\iota_{\rm r}'}\biggr\rvert}\,.
\end{equation}
We learn that the drift-island width in a stellarator is determined by resonant harmonics of $V$. If the magnetic field is analytic, then the amplitudes of these harmonics decay exponentially with $|N|$ and $|M|$, which means the drift-island width also decays exponentially with $|N|$ and $|M|$. 

%
%
\begin{figure}
\vspace{2mm}
\centering
\includegraphics[width=\textwidth]{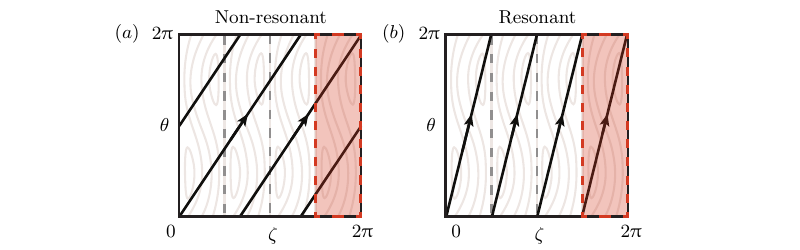}
    \caption{Illustration of the difference between $(a)$ a generic rational surface (in this case, $M=2$, $N=3$, and  $N_{\rm{fp}}=4$) and $(b)$ a rational surface for which $N$ shares common factors with $N_{\rm{fp}}$ (in this case, $M=1$, $N=4$, and $N_{\rm{fp}}=4$). When $N$ and $N_{\rm{fp}}$ share common factors, field lines visit equivalent points in different field periods. In each case, one field period is highlighted in red.}
    \label{fig:fieldperiods}
\end{figure}

Using \eqref{eq:widthNM}, we can now explain why the drift islands are widest when $N$ and $N_{\rm{fp}}$ share common factors: increasing $\operatorname{gcd}(N, N_{\rm{fp}})$ reduces $\smash{|\widetilde{N}|}$ and $\smash{|\widetilde{M}|}$, which typically makes $\smash{|V_{\widetilde{M}\widetilde{N}}|}$ larger. There is a simple physical explanation for this effect (see \cref{fig:fieldperiods}). We remind the reader that, on irrational surfaces, the radial drift experienced by passing particles rapidly averages out because they sample the entire flux surface. Meanwhile, near rational surfaces, passing particles experience the same radial drift again and again, so they accumulate much larger radial displacements. If there are multiple field periods, then particles can experience the same radial drift before returning to their starting point; their orbit simply has to pass through a point \emph{equivalent to} their initial position but in a different field period. For such particles, the extent to which their radial drifts reinforce, instead of averaging out, is amplified, leading to wider drift islands. For a rational-surface field line to pass through a point equivalent to the initial point but in a different field period, $N_{\rm{fp}}$ and $N$ must share common factors; for example, see \crefsub{fig:fieldperiods}{b}. The more factors $N$ and $N_{\rm{fp}}$ share, the more times particles will experience the same drifts before returning to their starting points, and the larger the drift islands will be. As an example of this effect, \crefand{fig:W7X_lowenergy_lowestorder}{fig:W7X_lowestorder} show that drift islands arise in W7-X at the rational surface with $\iota = 10/11$, for which the numerator $N=10$ is divisible by $N_{\rm fp}=5$.

In summary, to avoid wide drift islands, the most important low-order rationals for the rotational-transform profile to avoid are those whose numerators $N$ share common factors with $N_{\rm{fp}}$.

\subsection{Aspect ratio}\label{subsec:lar}

Usually, the major radius $R$ of a stellarator is much larger than the minor radius $a$; most stellarators constructed to date have values of $\epsilon \coloneq a/R$ around $0.1$--$\,0.3$ \mycitep{Beidler2011}. However, so far, in our estimates of the drift-island width, we have ordered gradients of the field using a single length scale $L$. In this section, we investigate how the drift-island width scales in a \emph{large-aspect-ratio} stellarator with $\epsilon \ll 1$. For this discussion, we define $\rho_\star \coloneq \rho/a$; this is the only subsection in which we do not take $a \sim R \sim L$.

It will be convenient to use Boozer coordinates ${(\psi, \theta_{\rm{B}}, \zeta_{\rm{B}})}$ \mycitep{Boozer1981}, since ${\bh \? \0\grad{\zeta_{\rm{B}}} = B/(G_{\rm pol} + \iota\? G_{\rm tor})}$ depends on the angular coordinates via the magnetic field strength only. Here, $c\?G_{\rm tor}(\psi)/2$ is the toroidal current inside flux surface $\psi$ and $c\?G_{\rm pol}(\psi)/2$ is the poloidal current linked outside the flux surface. If we define $G(\psi)\coloneq G_{\rm pol}(\psi) + (N/M)\? G_{\rm tor}(\psi)$, then in Boozer coordinates \eqref{eq:Jrepeated} reads
\begin{equation}\label{eq:JBoozer}
    I_{\rm{r}} = G_{\rm r}\oint_{\rm{r}} \frac{\sqrt{2(\mathcal{E}-\mu B )}}{B}\,\rmd\zeta_{\rm{B}}\,,
\end{equation}
where $G_{\rm r}\coloneq G(\psi_{\rm r})$. In a large-aspect-ratio stellarator with ${\beta \coloneq 8\upi P/ B^2 \lesssim \epsilon}$ ($P$ is pressure), the magnetic field strength satisfies
\begin{equation}\label{eq:lardef2}
    B(\psi,\theta_{\rm{B}},\zeta_{\rm{B}}) = B_0(\zeta_{\rm{B}}) + B_1(\psi,\theta_{\rm{B}},\zeta_{\rm{B}})\,,
\end{equation}
where $B_1\sim \epsilon B_0$ \mycitep{dHerbement2022}. The currents in such a stellarator satisfy $G_{\rm tor}\sim RB$ and $G_{\rm pol}\sim aB$. We expand $I_{\rm{r}}$ in $\epsilon\ll 1$ by substituting \eqref{eq:lardef2} into \eqref{eq:JBoozer} and ordering $v_\parallel \sim v$ for passing particles. The result is
\begin{equation}\label{eq:Jlar}
I_{\rm{r}} = G_{\rm r}\oint_{\rm{r}}\frac{\sqrt{2(\mathcal{E}-\mu B_0 )}}{B_0}\biggl( 1 - \frac{2\mathcal{E}-\mu B_0}{2(\mathcal{E}-\mu B_0)}\frac{B_1}{B_0} + O(\epsilon^2) \biggr)\,\rmd\zeta_{\rm{B}}\,.
\end{equation}
The first term in \eqref{eq:Jlar} is independent of $\eta$, since $B_0$ depends on $\zeta_{\rm{B}}$ only, so ${I_{\rm{r}}(\eta,\mathcal{E},\mu) - I_{\rm{r}}(\eta_{\rm{r}},\mathcal{E},\mu)\sim \epsilon v R \sim v a}$. Using this result in \eqref{eq:psisolution}, with $\Psi_{\rm t} \sim a^2 B$ and $\iota_{\rm r}' \sim s/a^2B$, we find that the drift-island width in a large-aspect-ratio stellarator is 
\begin{equation}\label{eq:larwidth}
    \frac{\Delta\psi}{\Psi_{\rm{t}}}\sim \Bigl(\frac{\rho_\star}{s}\Bigr)^{\!1/2}\,.
\end{equation}
Thus, increasing the aspect ratio by making the major radius larger, at fixed ${\rho_\star = \rho/a}$, does not reduce the drift-island width as a fraction of the minor radius.

Interestingly, there is one case where increasing the aspect ratio in this way does lead to smaller drift islands. This is the case of stellarators with `mirror ratios close to unity', which means the variation of the magnetic field strength along each field line is small in $\epsilon$, as in a large-aspect-ratio tokamak. The magnetic field strength in such a device takes the form \mycitep{dHerbement2022}
\begin{equation}\label{eq:mirrorratiounity}
    B(\psi,\theta_{\rm{B}},\zeta_{\rm{B}}) = B_0 + B_1(\psi,\theta_{\rm{B}},\zeta_{\rm{B}})\,,
\end{equation}
where $B_1\sim\epsilon B_0$ and $B_0$ is now a constant rather than a function of $\zeta_{\rm{B}}$. With \eqref{eq:mirrorratiounity}, and assuming that $\mathcal{E} - \mu B_0 \gg \mu B_1$ for passing particles sufficiently far from the trapped--passing boundary, the expansion \eqref{eq:Jlar} for $I_{\rm{r}}$ becomes%
\begin{equation}\label{eq:jclosetounity}
I_{\rm{r}} = \frac{G_{\rm r}\sqrt{2(\mathcal{E}-\mu B_0 )}}{B_0}\biggl(2\upi M - \frac{2\mathcal{E} - \mu B_0}{2(\mathcal{E}-\mu B_0)}\frac{1}{B_0}\oint_{\rm{r}}\?B_1\?\rmd\zeta_{\rm{B}} + O(\epsilon^2)\biggr)\,.
\end{equation}
Important information about the integral appearing in \eqref{eq:jclosetounity} can be derived from MHD force balance. In MHD equilibrium, it is well known that, for the current density on $\psi_{\rm{r}}$ to be divergence-free, the `Hamada condition' \mycitep{Hamada1962} must be satisfied:
\begin{equation}\label{eq:Hamada}
    P_{\rm r}' \,\partial_\eta\biggl(\?\oint_{\rm{r}}\,\frac{\rmd l}{B}\biggr) = 0\,,
\end{equation}
where ${P_{\rm r}' \coloneq (\rmd P/\rmd \psi)(\psi = \psi_{\rm{r}})}$ is the pressure gradient at the rational surface. For a large-aspect-ratio magnetic field with mirror ratio close to unity and a non-zero pressure gradient, expanding the Hamada condition \eqref{eq:Hamada} to order $\epsilon$ gives
\begin{equation}
    \partial_\eta \biggl(\?\oint_{\rm{r}}\? B_1\? \rmd \zeta_{\rm{B}}\!\? \biggr) = O( \epsilon^2 B )\,.
\end{equation}
This tells us that the integral in \eqref{eq:jclosetounity} does not vary with $\eta$ to leading order in $\epsilon\ll 1$, which implies $I_{\rm{r}}(\eta,\mathcal{E},\mu) - I_{\rm{r}}(\eta_{\rm{r}},\mathcal{E},\mu)\sim \epsilon^2 v R$.\footnote{This result can also be obtained using the equations for the radial drift velocity and MHD equilibrium in a large-aspect-ratio stellarator derived by \citet{dHerbement2022}. Their equation (3.19) can be combined with their (3.14) and (B4), provided $P'\neq 0$, to conclude that $\bb{v}_{\rm{d}}\bb{\cdot}\bb{\nabla}\psi$ is the partial derivative with respect to $l$ of a complicated function, up to corrections of order $\epsilon^2 \rho_\star v\,\Psi_{\rm{t}}/L$. This partial derivative vanishes under a transit average for passing particles that are sufficiently far from the trapped--passing boundary. So, the transit-averaged radial drift rate of these particles is order $\epsilon^2 \rho_\star v\,\Psi_{\rm{t}}/L$, a factor of $\epsilon$ smaller than it would be in a generic large-aspect-ratio stellarator.} Therefore, the drift-island width becomes
\begin{equation}
    \frac{\Delta\psi}{\Psi_{\rm{t}}}\sim\Bigl( \frac{\epsilon\rho_\star}{s} \Bigr)^{\!1/2}\,,
\end{equation}
which is smaller than the result \eqref{eq:larwidth} for a generic large-aspect-ratio stellarator. Thus, if the mirror ratio is close to unity and the Hamada condition is satisfied with a non-zero pressure gradient, then, for passing particles sufficiently far from the trapped--passing boundary, a large aspect ratio helps to reduce the drift-island width as a fraction of the minor radius.

\subsection{Magnetic shear}\label{subsec:shear}

We conclude this section with a few observations about the relationship between the drift-island width and the equilibrium magnetic shear. Scaling \eqref{eq:scaling} shows that the drift-island width increases as the shear decreases. It might, therefore, seem that drift-island overlap is more likely to occur when the shear is low. However, the distance between two given rational surfaces scales with $1/s$, while the drift-island width scales with $1/\sqrt{s}$. Thus, as $s$ decreases, the rational surfaces move apart faster than the drift islands get wider \mycitep[similar reasoning is valid for islands due to other types of resonance, \eg]{Paul2023}. Therefore, drift-island overlap is, in fact, less likely to occur at low shear, for a given $\rho_\star$.

In principle, drift-island overlap will eventually occur if the particle energy is increased (\ie making $\rho_\star$ larger) at a fixed $s$. In our simulations, we did not observe drift-island overlap for passing alpha particles away from the trapped--passing boundary\footnote{From a Poincar\'e plot, it can be difficult to say whether drift-island overlap is occurring close to the trapped--passing boundary and causing orbit stochasticity there. This is because, usually, the trapped--passing boundary is already surrounded by a region of stochastic orbits due to particles undergoing transitions between different trapping states.} in any of the stellarator equilibria that we studied, even at energies of $\SI{3.5}{\mega\electronvolt}$ or more. As a result, we expect stochastic transport due to drift-island overlap not to be a relevant transport mechanism for passing alpha particles in the equilibrium magnetic field of a reactor-scale stellarator.

Low magnetic shear has one other important consequence for the drift islands: it allows them to shift, radially, away from their associated rational surface. This phenomenon will be discussed in \cref{subsec:lowshear} and \cref{sec:higherorder}.

\section{Cyclometric stellarators}\label{sec:cyclometry}

We have shown that scaling certain parameters of the stellarator magnetic field \Dash for example, increasing its aspect ratio or magnetic shear \Dash can reduce the drift-island width. In this section, we investigate whether there are stellarator magnetic fields for which the drift-island width vanishes completely. In \cref{subsec:cyclometry}, we show that the there are no drift islands in perfectly omnigeneous stellarators, which is equivalent to a result in appendix C of \citet{Burby2023}. This is fortunate because modern stellarator designs are invariably optimized for omnigeneity \mycitep[an omnigeneous stellarator is one in which the bounce-averaged radial drift velocity of every trapped particle vanishes; see, \eg]{Hall1975, Cary1997, Cary1997b, Parra2015}. Surprisingly, we also find that there are no drift islands in some highly non-omnigeneous configurations: we show that the drift-island width is zero for all passing particles if and only if the magnetic field has a property we call \emph{cyclometry}. We discuss nearly cyclometric stellarators in \cref{subsec:omnigeneity} and determine how the drift-island width scales with the deviation from cyclometry. The scaling changes when the magnetic shear is extremely low (\viz $s\lesssim \rho_\star$); we discuss nearly cyclometric stellarators with such low values of the magnetic shear in \cref{subsec:lowshear}.

\subsection{Stellarators without drift islands}\label{subsec:cyclometry}

According to \eqref{eq:psisolution}, the width of any drift islands around rational surface $\psi_\mathrm{r}$ will be zero if and only if
\begin{equation}\label{eq:zeroderiv}
\partial_\eta I_{\rm r} = \partial_\eta\biggl(\?\oint_\mathrm{r} \sqrt{2(\mathcal{E} - \mu B)}\,\rmd l\biggr) = 0
\end{equation}
for all passing particles \mycitep{Grad1967, Burby2023}. Physically, \eqref{eq:zeroderiv} ensures that the net radial drift vanishes for particles that stream all the way around a closed field line on the rational surface.

In this subsection, we assume that \eqref{eq:zeroderiv} is satisfied for all passing particles and investigate what this implies for the magnetic field. In fact, our arguments only require that there exists some open neighbourhood $\mathfrak{N}$ of zero such that \eqref{eq:zeroderiv} is satisfied for all passing particles with $\mu/\mathcal{E}\in \mathfrak{N}$. The arguments in this subsection only apply to energetic particles, for which the electrostatic potential is negligible, but could be generalized to thermal particles if the electrostatic potential is close enough to a flux function that the dominant radial drift is the magnetic drift. For low-collisionality main ions in nearly omnigeneous stellarators \mycitep{Calvo2017} and large-aspect-ratio stellarators \mycitep{dHerbement2022}, the electrostatic potential satisfies this condition in some, but not all, collisionality regimes, as discussed in \citet{Calvo2018}. Initially, we assume the magnetic field is a smooth, well-behaved function; we consider more general stellarators with discontinuous magnetic field strengths in \cref{subsec:discussion}.

\subsubsection{Constraint on the minimum and maximum field strength}\label{subsubsec:freelypassing}

Suppose \eqref{eq:zeroderiv} is satisfied for passing particles with $\mu/\mathcal{E}$ in some open neighbourhood $\mathfrak{N}$ of zero. Then, we can Taylor expand the integrand in \eqref{eq:zeroderiv} to find
\begin{equation}\label{eq:Iseries}
\sqrt{2\mathcal{E}}\sum_{n=1}^\infty c_n \Bigl(\frac{\mu}{\mathcal{E}}\Bigr)^{\!n}\?\partial_\eta\biggl(\?\oint_\mathrm{r} B^n\,\rmd l\biggr) = 0\,, \qquad c_n = \begin{cases}
        \displaystyle \frac{1}{2^{2n}}\frac{(2n)!}{(1-2n)(n!)^2} &\text{if $n>0$,}\\
        \,\,\,1 &\text{if $n = 0\,$,}
    \end{cases}
\end{equation}
for values of $\mu/\mathcal{E}\in \mathfrak{N}$ that are within the radius of convergence. For these values of $\mu/\mathcal{E}$, the series in \eqref{eq:Iseries} is identically zero, so its coefficients must all vanish. This means
%
\begin{equation}\label{eq:powersofB}
		\partial_\eta\biggl(\?\oint_\mathrm{r} B^n\,\rmd l\biggr) = 0\qquad\text{for }n = 0, 1, 2, \ldots
\end{equation}
For example, when $n=0$, \eqref{eq:powersofB} says that every field line on the rational surface is a closed curve with equal length. More generally, if $Y(B,\mathcal{E}, \mu, \sigma)$ is any analytic function of $B$, then \eqref{eq:powersofB} implies
\begin{equation}\label{eq:Lambdas}
		\partial_\eta\biggl(\?\oint_\mathrm{r}\,Y(B(\psi_\mathrm{r},\eta,l),\mathcal{E}, \mu, \sigma)\,\rmd l\biggr) = 0\,.
\end{equation}
We can use \eqref{eq:Lambdas}, with a specific choice of $Y$, to derive a constraint on the minimum and maximum values of $B$ along any field line on the rational surface. Let these values be $B_m(\eta)$ and $B_M(\eta)$, respectively, for a field line $\eta$. We will use \eqref{eq:Lambdas} to prove, by contradiction, that $B_m(\eta)$ and $B_M(\eta)$ must be independent of $\eta$. 

Starting with $B_m$, suppose ${B_m(\eta_1) = B_1}$ and ${B_m(\eta_2) = B_2 \neq B_1}$ for two field lines $\eta_1$ and $\eta_2$. Without loss of generality, assume $B_1 < B_2$. Now, choose $Y$ in \eqref{eq:Lambdas} to be highly concentrated around $B_1$ and extremely small for $B \geq B_2$ (we need $Y$ to be `extremely small', rather than zero, because an analytic $Y$ that vanishes for $B \geq B_2$ would have to be zero everywhere). This choice would make the integral in \eqref{eq:Lambdas} much larger on field line $\eta_1$ than it is on field line $\eta_2$, but \eqref{eq:Lambdas} says these integrals must be equal. This contradiction proves that, in fact,  ${B_m(\eta_1) = B_m(\eta_2)}$. Since $\eta_1$ and $\eta_2$ were arbitrary, we conclude that $B_m$ must be independent of $\eta$. Identical reasoning can be used to show that $B_M$ must also be independent of $\eta$.

It is well known that a similar constraint must be satisfied by the value of $B$ at any local minimum or maximum of the field strength on an omnigeneous flux surface \citep{Cary1997, Cary1997b, Parra2015}. The traditional proof of this result is quite different from the argument we gave above; it is worth explaining the reason for this difference. The precise statement of the constraint for omnigeneous flux surfaces is as follows. If we use $\alpha$ as a field-line label and write $B_{lm}$ ($B_{lM}$) for the value of $B$ at a local minimum (maximum), then, on an omnigeneous flux surface, we must have $\partial_\alpha B_{lm} = 0$ ($\partial_\alpha B_{lM} = 0$), for all $\alpha$. The usual proof of $\partial_\alpha B_{lm} = 0$ considers the orbits of deeply trapped particles that only sample field-strength values close to $B_{lm}$ in a single well \citep{Cary1997, Cary1997b, Parra2015}. We cannot use this approach to prove that $B_m(\eta)$ is independent of $\eta$ because passing particles stream over multiple wells and never sample just the values of the field strength near $B_m$. 

\subsubsection{Constraint on the well widths}\label{subsec:bulk}

\begin{figure}
\centering
\includegraphics[width=\textwidth]{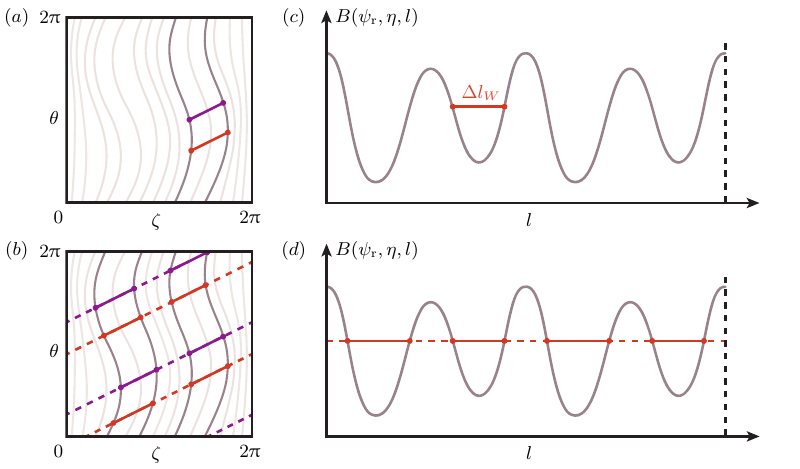}
    \caption{Difference between omnigeneous (isometric) and cyclometric magnetic fields on a rational surface (with $N_{\rm fp} = 1$). $(a)$~An omnigeneous magnetic field. The red and purple lines show two wells with the same bounce-point magnetic field strength, which must have equal widths. $(b)$~Sketch of the field strength against distance along a field line in this omnigeneous field, with a single well indicated. $(c)$~A cyclometric magnetic field. The red and purple dashed lines represent two closed field lines. The solid line segments show the different wells along these field lines, for a given bounce-point field strength. The sum of the widths of these wells must be the same for each field line. $(d)$~Sketch of the field strength as a function of distance along one of these field lines, with the well widths that must be summed to obtain $L(\eta, B)$ indicated.}
    \label{fig:cyclovsomni}
\end{figure}

Next, we use \eqref{eq:zeroderiv} to obtain a constraint on the magnetic field involving a quantity we call the well width, $\Delta l_W(\eta,B)$. To define $\Delta l_W(\eta,B)$, consider a field line $\eta$ and let $B$ satisfy $B_m \leq B \leq B_M$. In general, as $l$ varies, there will be multiple intervals within which $B(\psi_\mathrm{r}, \eta, l) \leq B$. We refer to these intervals as different `wells'; they are the regions accessible to trapped particles with $\mu/\mathcal{E} = 1/B$. Then, $\Delta l_W(\eta,B)$ is the width of well $W$, treated as an interval in $l$ (examples are shown in \crefsub{fig:cyclovsomni}{b}).

The sum of the well widths of every well going around a rational-surface field line $\eta$ will be denoted by $L(\eta, B)$:
\begin{equation}\label{eq:LWdef}
L(\eta, B) \coloneq \mspace{-9mu} \sum_{W\in\mathcal{W}(\eta,B)} \mspace{-9mu} \Delta l_{W}(\eta, B)\,,
\end{equation}
where $\mathcal{W}(\eta,B)$ is the set of wells, for a given $B$, along this field line. We will show that, if \eqref{eq:zeroderiv} is satisfied for all passing particles with $\mu/\mathcal{E}\in \mathfrak{N}$, then $L(\eta, B)$ must be independent of $\eta$. To prove this statement, we begin by changing integration variables in
\begin{equation}\label{eq:integraleq}
I_{\rm r} = \oint_\mathrm{r} \sqrt{2(\mathcal{E} - \mu B(\psi_\mathrm{r},\eta,l))}\,\rmd l
\end{equation}
from $l$ to $B = B(\psi_\mathrm{r},\eta,l)$. Taking into account the multi-valued nature of this substitution, we find
\begin{equation}
I_{\rm r} = \int_{B_m}^{B_M}\!\!\sqrt{2(\mathcal{E} - \mu B)}\,\sum_i\biggl|\frac{\partial l_i}{\partial B}\biggr|\,\rmd B\,,
\end{equation}
where $l_i(\eta, B)$ is the $l$ coordinate of the $i$th point at which the field strength equals $B$ along the field line $(\psi_{\rm{r}}, \eta)$. The sign of $\partial l_i/\partial B$ alternates on opposite sides of each well, so $\sum_i |\partial l_i / \partial B| = \partial L / \partial B$. Thus, integrating by parts gives
\begin{equation}\label{eq:djdalphapassing}
I_{\rm r} = \sqrt{2(\mathcal{E} - \mu B_M)}\,L_M +  \int_{B_m}^{B_M}\!\!\frac{\mu\? L(\eta, B)}{\sqrt{2(\mathcal{E} - \mu B)}}\,\rmd B\,.
\end{equation}
The quantity $L_M \coloneq L(\eta,B_M)$ is the total length of the closed field line $\eta$. In \cref{subsubsec:freelypassing}, we showed that $L_M$, $B_m$, and $B_M$ must all be independent of $\eta$. Therefore, the only $\eta$ dependence in \eqref{eq:djdalphapassing} is contained in $L(\eta, B)$, inside the integral. Since the integration limits are constants, we cannot apply the Abel-inversion technique of \citet{Cary1997b} to prove that $\partial_\eta I_{\rm r} = 0$ implies $\partial_\eta L(\eta,B) = 0$. Instead, as before, we use the fact that ${\mu/\mathcal{E} < 1/B_M \leq 1/B}$ to Taylor expand the integrand, finding
\begin{equation}\label{eq:expandedintegraleq}
I_{\rm r} = \sqrt{2(\mathcal{E} - \mu B_M)}\,L_M + \frac{\mu}{\sqrt{2\mathcal{E}}}\sum_{n=0}^\infty \frac{1}{2^{2n}}\frac{(2n)!}{(n!)^2}\Bigl(\frac{\mu}{\mathcal{E}}\Bigr)^{\!n}\!\int_{B_m}^{B_M}\!B^n L(\eta, B)\,\rmd B\,.
\end{equation}
For $I_{\rm r}$ to be independent of $\eta$ for passing particles with $\mu/\mathcal{E}\in \mathfrak{N}$ and within the radius of convergence of the sum in \eqref{eq:expandedintegraleq}, every coefficient of $(\mu/\mathcal{E})^n$ in the sum must be independent of $\eta$; that is,
\begin{equation}
\partial_\eta\biggl(\?\int_{B_m}^{B_M}\!B^n L(\eta, B)\,\rmd B\biggr) = 0\qquad\text{for }n = 0, 1, 2, \ldots
\end{equation}
A standard result in real analysis states that two continuous real functions on a closed interval with equal moments must be equal \mycitep[\eg][, p.169]{Rudin1976}. Applying this result to $L(\eta_1, B)$ and $L(\eta_2, B)$, for any two field lines $\eta_1$ and $\eta_2$, we conclude that $L(\eta, B)$ must be independent of $\eta$.\footnote{Note that this argument uses the fact that $B_m$ and $B_M$ are independent of $\eta$. Also note that we are careful to apply this argument to $L(\eta, B)$, which is a continuous function of $B$, rather than to $\partial_\eta L(\eta, B)$, which need not be a continuous function of $B$, as can be seen from \eqref{eq:changeindeltaL}.} We refer to magnetic fields with this property as `cyclometric'.

Cyclometry is a strong constraint, from which we can deduce two further statements about the magnetic field. First of all, \eqref{eq:Lambdas} must hold, regardless of whether the integrand is analytic; that is,
\begin{equation}\label{eq:Lambda2}
		\partial_\eta\biggl(\?\oint_\mathrm{r}\,Y( B(\psi_\mathrm{r},\eta,l),\mathcal{E}, \mu, \sigma)\,\rmd l\biggr) = 0\,,
\end{equation}
for any $Y(B, \mathcal{E}, \mu, \sigma)$. To show this, we change variables in \eqref{eq:Lambda2} from $l$ to $B(\psi_\mathrm{r}, \eta, l)$, integrate by parts, and use $\partial_\eta L(\eta, B) = 0$. Second, on the rational surface, there cannot be any tangential intersections between field lines and contours of constant $B$. Therefore, these contours must close toroidally, helically, or poloidally; the magnetic field must be `pseudosymmetric' \citep{Mikhailov2002, Skovoroda2005}. We provide a proof of this second statement in \cref{app:topology}. As in \cref{subsubsec:freelypassing}, the proof is different from the proof of the analogous result for omnigeneous stellarators. In omnigeneous stellarators, the possibility of tangential intersections can be ruled out by considering particles that are trapped in wells near a hypothetical tangential intersection. This kind of argument cannot be used in the cyclometric case because passing particles stream over multiple wells and never sample just the values of the field strength near a tangential intersection. Note that the proof we give in \cref{app:topology} uses the fact that the magnetic field is smooth. Cyclometric fields that are not pseudosymmetric are possible if the field strength is allowed to have discontinuities; we discuss this possibility further in \cref{subsec:discussion}. 

\subsubsection{Boozer-coordinate formulation}\label{subsec:Boozing}

Using Boozer coordinates, it is possible to define cyclometry without making reference to arc lengths along field lines. This alternative formulation will be necessary when we discuss nearly cyclometric magnetic fields in \crefrange{subsec:omnigeneity}{subsec:lowshear}. Furthermore, it provides a convenient way of extending the definition of cyclometry to flux surfaces other than the rational surface $\psi_{\rm r}$, which will be important in \cref{subsec:lowshear}.

The condition that cyclometric fields must satisfy can be written in Boozer coordinates as
\begin{equation}\label{eq:target}
\partial_\eta I =  \partial_\eta\biggl(G\oint \frac{\sqrt{2(\mathcal{E} - \mu B)}}{B}\,\rmd\zeta_{\rm B}\biggr) = 0 \quad \text{for passing particles with $\mu/\mathcal{E}\in\mathfrak{N}$.}
\end{equation}
Here, we have used $I(\psi, \eta, \mathcal{E}, \mu)$ instead of $I_{\rm r}(\eta, \mathcal{E}, \mu)$; this will allow us to define cyclometry on a general flux surface. In writing \eqref{eq:target}, we also used the fact that $\bh\?\0\?\con{\zeta}$ becomes $[G_{\rm pol} + (N/M)\?G_{\rm tor}]/B = G/B$ in Boozer coordinates. Note that, while $I(\psi, \eta, \mathcal{E}, \mu)$ is defined on any flux surface, its definition involves an integration curve with rotational transform equal to a specified rational number, $N/M$. Cyclometry on a general flux surface $\psi$ is only useful if $N/M$ is approximately equal to the magnetic rotational transform, $\iota(\psi)$, on that surface.

Let $\Delta\zeta_{\mathrm{B}, W}(\psi, \eta, B)$ be the cumulative change in the Boozer toroidal angle when we move, along the curve of constant $\eta = \theta_{\rm B} - (N/M)\?\zeta_{\rm B}$, between the endpoints of well $W$. We define wells on a general flux surface in a similar manner to wells on the rational surface, except as intervals along curves of constant $\eta$ instead of along field lines. Let
\begin{equation}\label{eq:ZBdef}
    Z_{\rm B}(\psi, \eta, B) \coloneq \mspace{-9mu} \sum_{W\in\mathcal{W}(\eta,B)} \mspace{-9mu} \Delta\zeta_{\mathrm{B}, W}(\psi, \eta, B)\,,
\end{equation}
which is the Boozer-coordinate analogue of \eqref{eq:LWdef}. Following exactly the same steps as in \cref{subsec:cyclometry}, we can show that \eqref{eq:target} is satisfied if and only if $Z_{\rm B}(\psi, \eta, B)$ is independent of $\eta$, for all $B$. Thus, cyclometry is equivalent to the condition that, in Boozer coordinates, the sum of the toroidal-angle well widths is the same along each curve of constant $\eta$, for all $B$.

\subsubsection{Discussion}\label{subsec:discussion}

We have shown that no drift islands exist at a given rational surface if and only if the magnetic field is cyclometric on that surface. Cyclometry is a weaker condition than omnigeneity; in an omnigeneous field, the width of each well individually must be independent of the field-line label.\footnote{This property is sometimes called `isometry' \mycitep{Cary1997b, Skovoroda2005}, which is the reason we chose the name `cyclometry'.} Thus, perfectly omnigeneous stellarators are automatically cyclometric on every rational surface and do not contain drift islands \mycitep{Burby2023}. The difference between omnigeneous and cyclometric magnetic fields is illustrated in \cref{fig:cyclovsomni}.

\begin{figure}
\vspace{2mm}
\centering
\includegraphics[width=\textwidth]{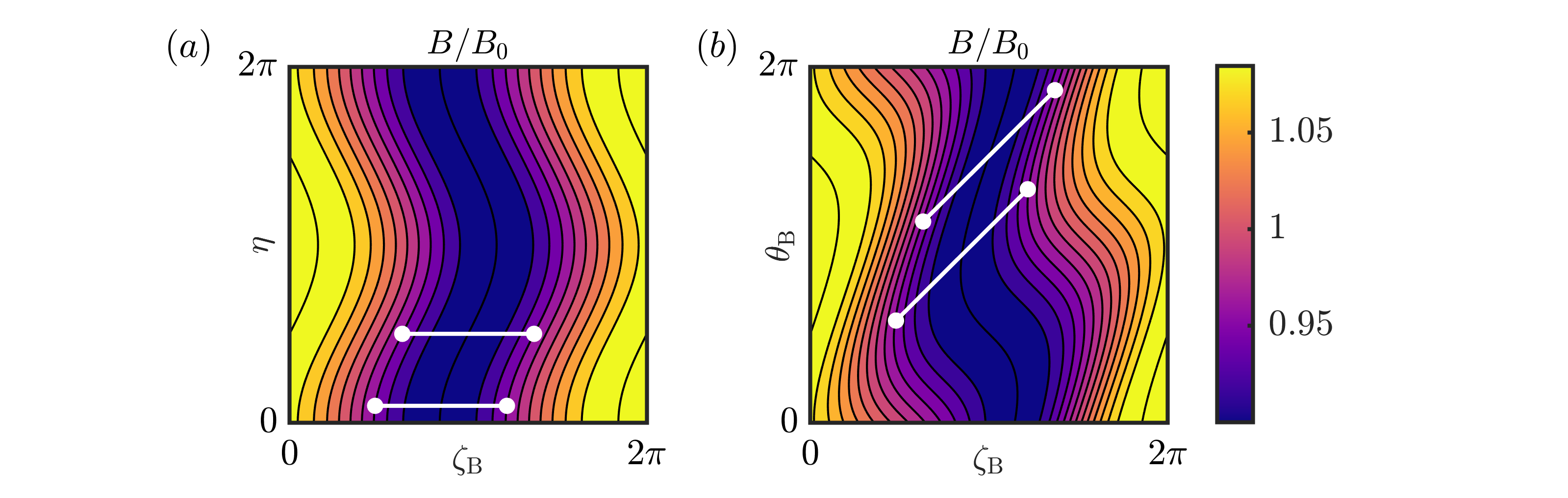}
    \caption{Example of how an analytic, omnigeneous magnetic field strength may be constructed on a rational surface. $(a)$ First, it is simplest to write down such a field as a function of $\zeta_{\rm B}$ and $\eta = \theta_{\rm B} - (N/M)\?\zeta_{\rm B}$. In a contour plot of $B$ in the $(\zeta_{\rm B}, \eta)$ plane, magnetic field lines are horizontal lines. Thus, for the field strength to be omnigeneous, the horizontal distance between opposing sides of each well much be independent of $\eta$. This can be achieved, for example, if the field strength is simply shifted horizontally by different amounts at each $\eta$. In the case shown, the field strength is $B(\eta,\zeta_{\rm B}) = B_0[1 + 0.1 \cos(\zeta_{\rm B} + 0.5 \cos\eta)]$, where $B_0$ is a normalization constant, and $M = 1$. $(b)$ Contour plot of the same field strength in the conventional $(\zeta_{\rm B}, \theta_{\rm B})$ plane, obtained by changing the vertical-axis coordinate from $\eta$ to $\theta_{\rm B} = \eta + (N/M)\?\zeta_{\rm B}$, which amounts to shearing the plot vertically. In this case, we took the rational surface to have $N = M = 1$.}
    \label{fig:analytic}
\end{figure}

The fact that omnigeneous fields are cyclometric on rational surfaces raises the question of whether cyclometric fields might possess other properties of omnigeneous fields. For example, it is well known that omnigeneous fields have a contour of maximum $B$ that is a straight line in Boozer coordinates, and also that analytic, non-quasisymmetric omnigeneous fields do not exist \mycitep{Cary1997b}. In fact, neither of these properties carry over to cyclometric fields because the proofs in the omnigeneous case require the rotational transform to be irrational. To demonstrate this, \cref{fig:analytic} shows how an analytic, omnigeneous magnetic field strength, with a contour of maximum $B$ that is not straight in Boozer coordinates, may be constructed when the rotational transform is rational.

\begin{figure}
\vspace{3mm}
\centering
\includegraphics[width=\textwidth]{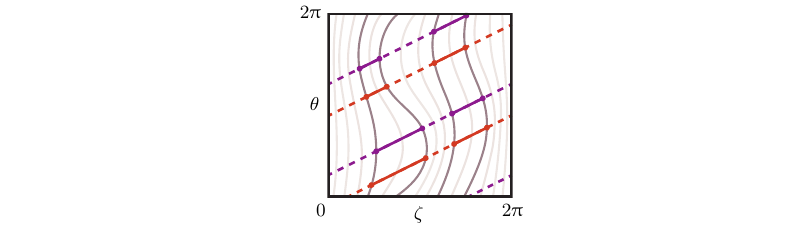}
    \caption{Cartoon depiction of a magnetic field strength (with $N_{\rm fp} = 1$) that is cyclometric but not isometric. The sum of the well widths along the red and purple field lines is the same, even though individual wells change size.}
    \label{fig:cyclonotiso}
\end{figure}

Omnigeneous magnetic fields are automatically cyclometric on rational surfaces but the converse is not true. That is, it is possible for there to be no drift islands around a highly non-omnigeneous rational surface. The reason is that the field may be cyclometric despite the fact that individual wells widen or shrink as $\eta$ varies; an example is shown in \cref{fig:cyclonotiso}.

\begin{figure}
\vspace{3mm}
\centering
\includegraphics[width=\textwidth]{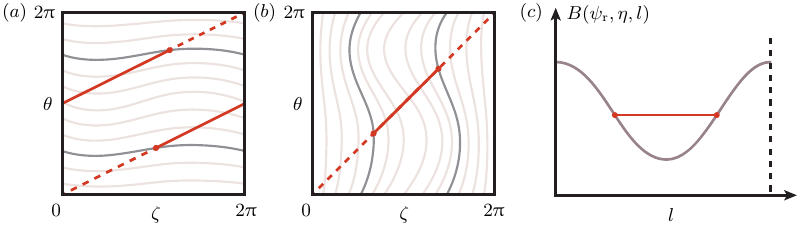}
    \caption{Cartoon showing two magnetic field strengths (with $N_{\rm fp} = 1$) in straight-field-line coordinates for which cyclometry implies omnigeneity. $(a)$ A case with $\iota = 1/2$. $(b)$ A case with $\iota = 1$. $(c)$ Sketch of how the field strength varies, in both cases, as a function of distance along a closed field line.}
    \label{fig:onewell}
\end{figure}

Interestingly, there is a class of magnetic fields for which cyclometry and omnigeneity \emph{are} equivalent. This is the case if, for every $B$, there is only one (non-identical) well along each rational-surface field line; then, $L(\eta, B)$ is a multiple of the width $\Delta l(\eta,B)$ of this single well. This can occur, for example, if the field-strength contours wrap around the torus toroidally and the rotational transform is $\iota = 1/M$, for some $M\in\mathbb{Z}^{>0}$ (see \crefsub{fig:onewell}{a}). It can also occur if the field-strength contours wrap around the torus poloidally, there is a single well in every field period, and $\iota$ is a multiple of $N_\mathrm{fp}$ (see \crefsub{fig:onewell}{b}). This second scenario is probably of academic interest only because the required rotational transform is very high.  For magnetic fields such as these, cyclometry forces $\Delta l(\eta,B)$ to be independent of $\eta$ for all $B$, and hence cyclometry implies omnigeneity on $\psi_\mathrm{r}$.

%
%

\begin{figure}
\vspace{3mm}
\centering
\includegraphics[width=\textwidth]{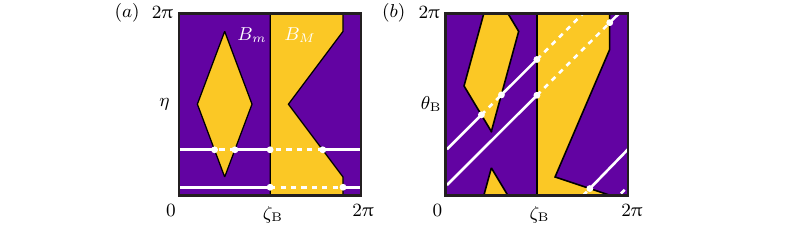}
    \caption{Plots of the magnetic field strength on a flux surface, in Boozer coordinates, for a piecewise-cyclometric field. The field strength is $B_m$ in the purple region and $B_M$ in the orange region. $(a)$ Plot in $(\eta,\zeta_\mathrm{B})$ coordinates, where $\eta = \theta_\mathrm{B} - (N/M)\?\zeta_\mathrm{B}$. The white lines are two curves of constant $\eta$ on the flux surface. $(b)$ Plot of the same field strength in conventional $(\theta_\mathrm{B}, \zeta_\mathrm{B})$ coordinates, assuming $N=1$ and $M=2$. The same constant-$\eta$ curves are shown in white.}
    \label{fig:piecewise}
\end{figure}

To complete our discussion of cyclometric magnetic fields, it is worth pointing out that some of the constraints derived in this subsection may be relaxed if we permit the magnetic field strength to be a discontinuous function. Recently, by allowing the field strength to be discontinuous, more general omnigeneous magnetic field strengths were discovered \citep{Velasco2024, Calvo2025, Velasco2025}. These \emph{piecewise omnigeneous} field strengths are automatically cyclometric on rational surfaces because their values of $Z_{\rm B}$ (see \eqref{eq:ZBdef}) are independent of $\eta$ on $\psi_{\rm r}$. Similarly, it is possible to construct more general cyclometric field strengths that are discontinuous; we refer to these fields as \emph{piecewise cyclometric}. The idea behind this construction is easiest to understand with a picture such as \cref{fig:piecewise}. This figure shows a piecewise-cyclometric magnetic field strength that assumes only two values, $B_m$ and $B_M$. In \crefsub{fig:piecewise}{a}, the field strength is plotted against $\zeta_\mathrm{B}$ and $\eta$. In these coordinates, the constant-$\eta$ curves are straight horizontal lines; measured along these lines, the widths $\Delta\zeta_{\mathrm{B}, W}$ of the purple regions change as $\eta$ varies. Thus, on $\psi_{\rm r}$, a particle trapped in a single well (within which the field strength is $B_m$) would experience a changing well width as $\eta$ varies, so this field strength is not omnigeneous. Crucially, when the width of one purple region increases, the corresponding width of the other purple region along the same horizontal line decreases, so that sum of these widths, $Z_{\rm B}$, is constant. Therefore, this field strength is piecewise cyclometric. In \crefsub{fig:piecewise}{b}, the same field strength is replotted against $\zeta_{\rm B}$ and $\theta_{\rm B}$, which amounts to shearing the plot in \crefsub{fig:piecewise}{a} vertically. Since there are closed constant-$B$ contours that do not encircle the torus toroidally, helically, or poloidally, this figure demonstrates that piecewise-cyclometric fields do not need to be pseudosymmetric.

The results of this subsection may be useful for stellarator optimization. In a stellarator containing a low-order rational surface, the size of any drift islands could be reduced by optimizing the magnetic field for cyclometry on that surface. Since cyclometry is a property of the magnetic field only, this may be more convenient than optimizing for $\partial_\eta I_{\rm r} = 0$ at a number of different pitch angles. Of course, the drift islands could also be removed by optimizing for better omnigeneity around the rational surface; however, the fact that cyclometry can be achieved even when there are relatively large deviations from omnigeneity means cyclometry is a less stringent optimization target that might be more compatible with other objective functions.

\subsection{Nearly cyclometric stellarators}\label{subsec:omnigeneity}

In real stellarators, small deviations from perfect cyclometry are inevitable. In this subsection, we investigate how the the size of these deviations affects the drift-island width. For this discussion, we assume an order-unity aspect ratio (\ie $\epsilon\sim 1$).

For any flux surface $\psi$ close to $\psi_{\rm r}$, we will refer to a magnetic field as `nearly cyclometric on $\psi$' if it is possible to split
\begin{equation}\label{eq:Bnearlyom}
B(\psi,\theta_\mathrm{B},\zeta_\mathrm{B}) = B_0(\psi, \theta_\mathrm{B},\zeta_\mathrm{B}) + B_1(\psi, \theta_\mathrm{B},\zeta_\mathrm{B})\,,
\end{equation}
where $B_0$ is a cyclometric field strength and $B_1 \sim \delta B_0$, with $\delta\ll 1$, is a small perturbation. That is, $B_0$ is a function of the Boozer angles for which $Z_{\rm B}$ in \eqref{eq:ZBdef} is independent of $\eta$, for all $B$.

In this subsection, we only consider magnetic fields that are nearly cyclometric on the rational surface $\psi_\mathrm{r}$. Expanding \eqref{eq:JBoozer} for such a field, using \eqref{eq:Bnearlyom}, we find
\begin{equation}\label{eq:Jexpanded}
I_{\rm r} = G_{\rm r}\oint_{\rm r}\frac{\sqrt{2(\mathcal{E}-\mu B_0 )}}{B_0}\biggl( 1 - \frac{2\mathcal{E}-\mu B_0}{2(\mathcal{E}-\mu B_0)}\frac{B_1}{B_0} + O(\delta^2) \biggr)\,\rmd\zeta_\mathrm{B}\,.
\end{equation}
Since $B_0$ is cyclometric, \eqref{eq:target} is satisfied for this magnetic field strength. This tells us that the leading term in \eqref{eq:Jexpanded} is independent of $\eta$, so $I_{\rm r}(\eta,\mathcal{E},\mu) - I_{\rm r}(\eta_\mathrm{r},\mathcal{E},\mu)\sim \delta v L$. Therefore, the drift-island width in a nearly cyclometric stellarator scales as\footnote{One might think that \eqref{eq:deltawidth} requires the radial drift due to $B_1$ to be smaller, pointwise, than the radial drift due to $B_0$, by a factor of $\delta$. In fact, the argument leading to \eqref{eq:deltawidth} is valid even when the radial drift due to $B_1$ is of the same order as the drift due to $B_0$; this occurs when $B_1$ varies on short lengthscales $\sim\!\delta L$, so that its gradient is $\boldsymbol{\nabla}B_1\sim \boldsymbol{\nabla}B_0$. The fact that \eqref{eq:deltawidth} is valid even when the perturbation varies on lengthscales $\sim\!\delta L$ means it is possible to use \eqref{eq:deltawidth} to derive the drift-island width \eqref{eq:larwidth} for large-aspect-ratio stellarators. To do this, we set $L \sim R$ and consider $B_1$ in \eqref{eq:lardef2} as a deviation from omnigeneity of order $\epsilon$ varying on lengthscales $a\sim\epsilon L$ (we must also remember that $\rho_\star$ in the large-aspect-ratio case is defined by $\rho_\star = \rho/a$).}
\begin{equation}\label{eq:deltawidth}
    \frac{\Delta\psi}{\Psi_\mathrm{t}}\sim \biggl(\frac{\rho_\star\delta}{s}\biggr)^{\!1/2}\,.
\end{equation}
This result shows that optimizing a stellarator for omnigeneity should reduce the width of any drift islands, unless this optimization also forces the magnetic shear to be very low. In the next subsection, we show that, when the shear is low, scaling \eqref{eq:deltawidth} must be modified.

Scaling \eqref{eq:deltawidth} agrees with the estimate obtained in \cref{subsec:orbitwidth}. In \cref{subsec:magneticislands}, we used an analogy with magnetic islands to argue that the drift-island width scales with $(\rho_\star/s)^{1/2}$, but we did not obtain the factor of $\sqrt{\delta}$ which reduces the drift-island width in nearly cyclometric stellarators. Now we are in a position to explain how the argument of \cref{subsec:magneticislands} could be extended to obtain this factor. The key point is that, in a perfectly cyclometric field, the field $\bb{B}^\star$ that the guiding centres follow is integrable (up to order $\rho_\star^2$ terms, which are beyond the level of precision of our calculations so far). Therefore, the size of the perturbation that should be used in the formula for magnetic-island width is not the $\sim\!\rho_\star$ drift term in $\bb{B}^\star$ (see \eqref{eq:B*}), but rather the $\sim\!\rho_\star\delta$ drift due to the non-cyclometric piece of the magnetic field.\footnote{For this order-$\rho_\star\delta$ term to be more important than order-$\rho_\star^2$ corrections that we neglect, we need $\delta\gg\rho_\star$. In fact, all our estimates and calculations up to this point require $\delta\gg\rho_\star$; this condition ensures the particle drifts a radial distance of $\Delta\psi\sim(\rho_\star\delta)\?\Psi_{\rm t}$ in a single transit, as we assumed for our estimates in \cref{subsec:orbitwidth}. Once $\delta\lesssim\rho_\star$, the stellarator is close enough to perfect cyclometry that the radial distance drifted in a single transit is affected by higher-order guiding-centre drifts, which we do not consider in this article. Thus, for particles with normalized gyroradius $\rho_\star$, optimization for cyclometry beyond $\delta\sim\rho_\star$ does not reduce the orbit width any further.} With this modification, the magnetic-island width formula also predicts scaling \eqref{eq:deltawidth}.

\subsection{Nearly cyclometric stellarators with low shear}\label{subsec:lowshear}

Stellarators that have been optimized to be highly quasisymmetric or quasi-isodynamic \Dash and, therefore, highly cyclometric \Dash often have very low magnetic shear \mycitep[\eg]{Landreman2022, Goodman2023, Nies2024}. This may be concerning in light of scaling \eqref{eq:deltawidth}, which suggests that, even if $\delta$ is small, a low enough shear will produce large drift islands. Fortunately, it turns out that \eqref{eq:deltawidth} does not apply when the shear is sufficiently low. In this subsection, we show that, as the shear is reduced, the drift-island width saturates at $\Delta\psi \sim \sqrt{\delta}\,\Psi_\mathrm{t}$ once $s\lesssim\rho_\star$. We demonstrate this by deriving an equation for the drift-island shape in a nearly cyclometric stellarator with $s \sim \rho_\star$.

\begin{figure}
\vspace{3mm}
\centering
\includegraphics[width=\textwidth]{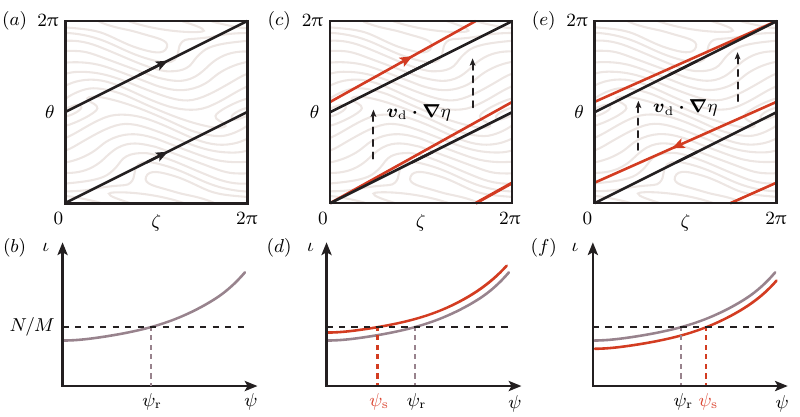}
    \caption{Cartoon illustrating how the tangential magnetic drift causes the drift islands to shift radially. $(a)$ A magnetic field line on a rational surface is plotted in straight-field-line coordinates. Grey contours represent the magnetic field strength. $(b)$ The rotational-transform profile for this magnetic field, with the rational surface $\psi_{\rm r}$ indicated. $(c)$ Now consider a passing particle that streams along this field line, while experiencing a tangential drift in the poloidal direction, as shown. The resulting orbit is depicted in red. The slope of this line is the kinetic rotational transform, which, in this case, is larger than $\iota$. $(d)$ The red curve shows the profile of the kinetic rotational transform on each flux surface for this particle. The drift islands will appear on the surface $\psi_{\rm s}$, where the kinetic rotational transform is rational. Thus, the islands shift away from $\psi_{\rm r}$ by an amount that scales with the size of the tangential drift and scales inversely with the shear. $(e)$ A passing particle streaming in the opposite direction to the particle in $(c)$ and $(d)$ still experiences the same tangential drift as that particle, so the kinetic rotational transform is now lower than $\iota$. $(f)$ This will cause the drift islands to shift radially in the opposite direction. }
    \label{fig:radialshift}
\end{figure}

Importantly, when $s \sim \rho_\star$, the drift islands shift radially away from their associated rational surface. This radial shift has been observed in simulations of alpha particles in stellarators \mycitep[\eg]{White2022, White2022b, Chambliss2024} and, in some cases, can cause the drift islands to move all the way into or out of the plasma volume \mycitep{White2022, White2022b}. The same radial shift is known to affect runaway-electron and energetic ion orbits in perturbed tokamaks \mycitep[\eg]{deRover1996, Motojima1999, Heyn2012, Matsuyama2014, Shinohara2020, He2020, Liu2021}, as well as thermal-ion orbits in the presence of small-scale neoclassical tearing modes \mycitep{Imada2018, Dudkovskaia2023}. Physically, the radial shift is caused by the tangential magnetic drift, for the following reason. Drift islands form around closed orbits, which occur when the rotational transform of the passing orbits \Dash the \emph{kinetic rotational transform} \Dash is rational. Tangential drifts cause the kinetic rotational transform $\mybarup{\iota}$ to deviate from the magnetic rotational transform by an amount $\mybarup{\iota} - \iota \sim \rho_\star$ (the precise definition that we use for $\mybarup{\iota}$ is not needed for the present discussion, but it can be found in \eqref{eq:kiota}). As a result, the flux surface $\psi_\mathrm{s}$ on which the kinetic rotational transform assumes a given rational value is shifted, with respect to the corresponding rational surface $\psi_\mathrm{r}$, by ${\psi_\mathrm{s}-\psi_\mathrm{r}\sim (\rho_\star/s)\? \Psi_\mathrm{t}}$, as shown in \cref{fig:radialshift}. This estimate for the radial shift will be confirmed by the calculations below. Note that the shift becomes order unity when $s\sim\rho_\star$.

To study particle orbits in a device with $s\sim\rho_\star$, we can still use the transit-invariant formula \eqref{eq:invariant}. The derivation of \eqref{eq:invariant} in \cref{sec:adiabaticinvariant} remains valid because $\eta$ and $\psi$ still vary slowly compared with the fast angle $\zeta$. However, the simplified island-shape equation \eqref{eq:simpleinvariant} is no longer correct since it assumes that the drift islands are close to $\psi_\mathrm{r}$.

We first determine the shifted location of the drift islands, $\psi_\mathrm{s}$. The closed particle orbits at the centre (O-point) of the drift islands must have ${\langle \rmd \eta / \rmd t \rangle_\mathrm{t} = 0}$. In nearly cyclometric stellarators, equation \eqref{eq:avetachange} for $\langle \rmd \eta / \rmd t \rangle_\mathrm{t}$, which includes tangential drifts, reads
\begin{equation}\label{eq:detadtoptimised}
    \biggl\langle\frac{\rmd\eta}{\rmd t}\biggr\rangle_{\!\!\mathrm{t}} = \frac{2\upi M\sigma}{\tau_\mathrm{t}}\biggl( \iota(\psi) - \frac{N}{M} \biggr) - \frac{mc}{Ze\tau_\mathrm{t}}\?\partial_\psi I_0 + O\biggl(\frac{\rho_\star\delta\?  v}{L}\biggr)\,.
\end{equation}
Here, $I_0(\psi,\mathcal{E},\mu)$ is the leading-order term when $I(\psi,\eta,\mathcal{E},\mu)$, defined in \eqref{eq:Jdef}, is expanded for a nearly cyclometric magnetic field: $I = I_0 + I_1 + O(\delta^2vL)$, where
\begin{subequations}
\begin{align}
\label{eq:I0def}I_0 &\coloneq G \oint\frac{\sqrt{2(\mathcal{E}-\mu B_0 )}}{B_0}\,\rmd\zeta_\mathrm{B}\,,\\
I_1 &\coloneq - G \oint  \frac{2\mathcal{E}-\mu B_0}{B_0\sqrt{2(\mathcal{E}-\mu B_0)}}\frac{B_1}{B_0}\,\rmd\zeta_\mathrm{B}\,.
\end{align}
\end{subequations}
The stellarator needs to be nearly cyclometric to ensure that, despite the low shear, the drift-island width is smaller than the minor radius. In \cref{subsec:omnigeneity}, we considered magnetic fields that are nearly cyclometric on the rational surface $\psi_{\rm r}$. Now that the drift islands can shift an order-unity distance away from $\psi_{\rm r}$, the magnetic field strength on $\psi_{\rm r}$ is no longer relevant; instead, we need to assume that the magnetic field is nearly cyclometric in a region containing $\psi_{\rm s}$. The reader might wonder whether, in a device with $s \sim \rho_\star$, cyclometry is actually a stronger condition than necessary to avoid drift islands at $\psi_{\rm s}$. After all, we derived cyclometry in \cref{subsec:cyclometry} by requiring that the transit-averaged radial drift vanishes on $\psi_{\rm r}$ for \emph{all} passing particles; but, when $s \sim \rho_\star$, the transit-averaged radial drift only needs to vanish on $\psi_{\rm s}$ for those passing particles \emph{whose drift islands arise on} $\psi_{\rm s}$. Nevertheless, it turns out that cyclometry is still necessary; in fact, to remove drift islands from a stellarator with $s \sim \rho_\star$, every flux surface that the islands can shift to needs to be cyclometric. We prove this claim in \cref{app:lowshearcyclometry}.

Resuming our calculation of the shifted island position, we find that $\psi_\mathrm{s}$ satisfies
\begin{equation}\label{eq:psis}
    2\upi M\sigma\biggl( \iota(\psi_\mathrm{s}) - \frac{N}{M} \biggr) - \frac{mc}{Ze}\?(\partial_\psi I_0)_\mathrm{s} = 0\,,
\end{equation}
since \eqref{eq:detadtoptimised} must vanish at $\psi_\mathrm{s}$. Here, the subscript `$\mathrm{s}$' indicates quantities evaluated at $\psi_\mathrm{s}$. Note that \eqref{eq:psis} implies ${\psi_\mathrm{s} - \psi_\mathrm{r}\sim(\rho_\star/s)\? \Psi_\mathrm{t}}$, confirming our estimate above. Also, note that $\psi_\mathrm{s}$ depends on $\mathcal{E}$ and $\mu$, which means the drift islands appear at different radial positions for particles with different speeds and pitch angles. Furthermore, the drift islands shift radially in opposite directions for co- and counter-passing particles; this can be understood from \cref{fig:radialshift}. Therefore, in a stellarator whose rotational transform profile has been designed to pass through a low-order rational at the plasma edge to allow an island divertor, drift islands will shift into the plasma for either co- or counter-passing alpha particles. These drift islands may be wide because the low-order rational surface at the edge typically has $N = N_{\rm fp}$, which, as explained in \cref{subsec:resonance}, promotes large drift islands.

To obtain an equation for the drift-island shape, we expand $\mathcal{I}$ for particles that satisfy ${\psi-\psi_\mathrm{s}\sim \sqrt{\delta}\? \Psi_\mathrm{t}}$. We find
\begin{align}\label{eq:lowshearI}
    \mathcal{I}(\psi,\eta) &= \sigma \biggl( I_{0\mathrm{s}} + I_{1\mathrm{s}}(\eta) + \frac{1}{2} (\partial_\psi^2 I_0)_\mathrm{s}(\psi-\psi_\mathrm{s})^2\biggr) \nn
    \al - \frac{2\upi M Ze}{mc}\!\int_{\psi_\mathrm{r}}^{\psi_\mathrm{s}}\biggl(\iota(\psi') - \frac{N}{M} \biggr)\,\rmd\psi'
     - \frac{\upi M Ze}{mc}\?\iota'_\mathrm{s}(\psi-\psi_\mathrm{s})^2 + O(\delta^{3/2} vL)\,,
\end{align}
where we have suppressed any velocity-space arguments. The level sets of $\mathcal{I}(\psi, \eta)$ describe the drift-island shape in a low-shear device; instead of \eqref{eq:psisolution}, which was valid when $s\gg \rho_\star$, we now have
\begin{equation}\label{eq:psisolnlowshear}
	\psi - \psi_\mathrm{s} = \pm \sqrt{\frac{mc\sigma\?[I_{1\mathrm{s}}(\eta, \mathcal{E},\mu) - I_{1\mathrm{s}}(\eta_\mathrm{s}, \mathcal{E},\mu)]}{\upi M Z e \?  [\iota'_\mathrm{s}- (mc\sigma/2\upi M Ze)(\partial_\psi^2 I_0)_\mathrm{s}]}}\,,
\end{equation}
for particles that cross $\psi_\mathrm{s}$ at $\eta = \eta_\mathrm{s}$. This equation confirms that our initial estimate ${\psi-\psi_\mathrm{s}\sim\sqrt{\delta}\? \Psi_\mathrm{t}}$ was correct.

From \eqref{eq:psisolnlowshear}, we learn that there are two important modifications to the drift-island shape when the shear is small. First, the radial shift means that all functions of position are evaluated at $\psi_\mathrm{s}$ instead of $\psi_\mathrm{r}$. Second, the magnetic shear $\iota_{\rm r}'$ has been replaced with $\iota_{\rm s}' - (mc\sigma/2\upi M Ze)(\partial_\psi^2 I_0)$, which is the shear in the kinetic rotational transform of the particle orbits. The new term $(mc/2\upi M Ze)(\partial_\psi^2 I_0)$ is the radial derivative of the (transit-averaged) tangential drift rate, which is an order-$\rho_\star$ correction. Even when the magnetic shear becomes vanishingly small, this term will allow particles to move tangentially across $\psi_{\rm s}$ and will set a finite width for the drift islands. Indeed, taking the subsidiary limit $s\ll \rho_\star$ in \eqref{eq:psisolnlowshear}, we see that the island width $\Delta\psi$ does not increase beyond $\sim\!\sqrt{\delta}\?  \Psi_\mathrm{t}$, as claimed at the start of this subsection. Once $s\ll \rho_\star$, the orbits essentially follow level sets of $I(\psi, \eta, \mathcal{E}, \mu)$, which is the adiabatic invariant discussed by \citet{Hastie1967} for passing particles in magnetic fields with closed field lines \mycitep[see also ]{Elbarmi2020}.\footnote{A similar result is known to hold for one-and-a-half dimensional Hamiltonian systems whose unperturbed Hamiltonian $\mathcal{H}_0(J)$ ($J$ is the action coordinate) has $\Omega(J) \coloneq \partial \mathcal{H}_0/\partial J = 0$, which is the appropriate generalization of the low-magnetic-shear condition \mycitep[see ][, the remark after Theorem 17, p.187]{Arnold2006}.} Interestingly, in this limit, there is no longer any difference between the drift islands for co- and counter-passing particles. In a stellarator with $s \ll \rho_\star$, it is possible to describe all passing orbits using an adiabatic invariant because, as explained in \cref{subsec:shear}, stochastic orbits due to drift-island overlap do not occur at such low values of the magnetic shear.

\section{Higher-order transit adiabatic invariant}\label{sec:higherorder}

In \cref{sec:adiabaticinvariant}, we used the transit adiabatic invariant to determine the drift-island shape. Our results agreed well with simulations at lower energies but began to deviate at higher energies. This is unsurprising because we expanded in $\rho_\star\ll 1$, which means we neglected higher-order corrections that become more important at larger energies. In this section, we derive a more accurate equation for the drift-island shape by calculating the first-order correction to the transit invariant. We use a method similar to the one described in appendix B of \citet{Cary1986}. The resulting, improved drift-island shape agrees very well with simulations, even for alpha particles at $\SI{3.5}{\mega\electronvolt}$.

\subsection{Higher-order adiabatic invariant}\label{subsec:higherorderinvariant}

The calculation in \cref{sec:adiabaticinvariant} used two small parameters: $\rho_\star\ll 1$ and ${|\iota - N/M| \ll 1}$. The higher-order calculation is more straightforward if there is only one expansion parameter. Since we are interested in the particles that move around an island of width $\sim\!(\rho_\star\delta/s)^{1/2}\? \Psi_{\mathrm{t}}$, in this section we order $s\sim \delta \sim 1$ and $\psi-\psi_\mathrm{r} \sim \sqrt{\rho_\star}\,\Psi_{\mathrm{t}}$.\footnote{Ordering $\delta\sim 1$ allows us to study both non-omnigeneous and omnigeneous stellarators, since we may take the subsidiary limit $\delta\ll 1$ to describe an optimized device.}

When this ordering is satisfied, the results of \cref{sec:adiabaticinvariant} can be summarized as follows. Both $\eta$ and $\psi$ evolve slowly, at rates ${\rmd\eta/\rmd t\sim \sqrt{\rho_\star}\?  v/L}$ and $\rmd\psi/\rmd t\sim \rho_\star \Psi_{\mathrm{t}} v/L$. The transit invariant $\mathcal{I}$ is also slowly varying (\viz $\rmd\mathcal{I}/\rmd t \sim \sqrt{\rho_\star}\? v^2$) because it depends only on the slow coordinates $\eta$ and $\psi$. The special importance of $\mathcal{I}$ lies in the fact that changes in $\mathcal{I}$ do not accumulate over long times: the transit average $\langle\rmd\mathcal{I}/\rmd t\rangle_\mathrm{t} \sim \rho_\star v^2$ is an order (in $\sqrt{\rho_\star}$) smaller than the instantaneous value of $\rmd\mathcal{I}/\rmd t$. According to the general theory of adiabatic invariants \mycitep{Lenard1959, Kruskal1962, Henrard1993}, it should be possible to construct an asymptotic series
\begin{equation}\label{eq:asymptoticseries}
    \mybar{\mathcal{I}}(\psi,\eta,\zeta,\mathcal{E},\mu,\sigma) = \mathcal{I}(\psi,\eta,\mathcal{E},\mu,\sigma) + \mathcal{I}^{(1/2)}(\psi,\eta,\zeta,\mathcal{E},\mu,\sigma) + \ldots,
\end{equation}
with $\smash{\mathcal{I}^{(n/2)}}\sim \smash{\rho_\star^{n/2}}vL$, that is conserved to all orders in $\sqrt{\rho_\star}$. Any finite truncation of this series, up to and including the term $\smash{\mathcal{I}^{(n/2)}}$, gives a quantity $\mybar{\mathcal{I}}$ that has a small time derivative, $\rmd\mybar{\mathcal{I}}/\rmd t\sim \rho_\star^{(n+1)/2}\?  v^2$, but an even smaller transit-averaged time derivative, $\langle\rmd\mybar{\mathcal{I}}/\rmd t\rangle_\mathrm{t}\sim  \rho_\star^{(n+2)/2}\?  v^2$. Therefore, knowledge of the higher-order terms in \eqref{eq:asymptoticseries} allows us to more precisely constrain the position of the particle over long times. 

In this section, we calculate $\mathcal{I}^{(1/2)}$. The resulting higher-order transit invariant changes by less than $\sim\!\rho_\star vL$ over one transit and will therefore capture order-$\rho_\star$ effects, such as tangential drifts (which cause the drift islands to shift radially, as discussed in \cref{subsec:lowshear}) and finite radial orbit widths.

\subsection{Guiding-centre equations}

For this derivation, we must ensure that our guiding-centre equations of motion are correct to a high enough order in $\rho_\star$. When $\psi - \psi_\mathrm{r}\sim\sqrt{\rho_\star}\,\Psi_\mathrm{t}$, the guiding-centre equations \eqref{eq:guidingcentre2}, including error estimates, read
\begin{subequations}\label{eq:witherrors}
\begin{align}
\label{eq:psiwitherrors} \frac{\rmd\psi}{\rmd t} &= \bb{v}_\mathrm{d}\bb{\cdot}\bb{\nabla}\psi + O\biggl(\rho_\star^2\frac{v\,\Psi_{\mathrm{t}}}{L}\biggr)\,,\\
\label{eq:alphawitherrors} \frac{\rmd\eta}{\rmd t} &= v_\parallel\biggl(\iota(\psi)-\frac{N}{M}\biggr)\,\bh \bb{\cdot}\bb{\nabla}\zeta + \bb{v}_\mathrm{d}\bb{\cdot}\bb{\nabla}\eta + O\Bigl(\rho_\star^{3/2} \frac{v}{L}\Bigr)\,,\\
\label{eq:zetawitherrors} \frac{\rmd\zeta}{\rmd t} &= v_\parallel\bh \bb{\cdot}\bb{\nabla}\zeta + O\Bigl(\rho_\star \frac{v}{L}\Bigr)\,.
\end{align}
\end{subequations}
The order-$\rho_\star^2$ error in \eqref{eq:psiwitherrors} is due to higher-order guiding-centre drifts. The order-$\rho_\star^{3/2}$ error in \eqref{eq:alphawitherrors} is due to parallel drifts projected in the $\bb{\nabla}\eta$ direction, while the order-$\rho_\star$ error in \eqref{eq:zetawitherrors} is due to the magnetic drift in the $\bb{\nabla}\zeta$ direction.
 
To find $\mathcal{I}^{(1/2)}$, we will need the equations of motion for $\psi$, $\eta$, and $\zeta$ to one order (in $\sqrt{\rho_\star}$) beyond leading order. Thus, the guiding-centre equations \eqref{eq:witherrors} are sufficiently accurate for our purposes. In \cref{app:mu}, we explain why we do not need to consider variations in the magnetic moment, despite the fact that its rate-of-change, ${\mathrm{d}\mu/\mathrm{d}t\sim \rho_\star v^3/BL}$, is as large as, for example, the higher-order corrections to $\rmd \eta / \rmd t$ that we must retain in \eqref{eq:witherrors}.

\subsection{Calculation of higher-order invariant}\label{subsec:higherordercalc}

We proceed to impose $\rmd \mybar{\mathcal{I}}/\rmd t = 0$ order-by-order on \eqref{eq:asymptoticseries}. At lowest order, we find
\begin{equation}\label{eq:lowestorder}
    \partial_\psi\mathcal{I}\,\frac{\rmd\psi}{\rmd t}  + \partial_\eta\mathcal{I}\,\frac{\rmd\eta}{\rmd t} + \partial_\zeta \mathcal{I}^{(1/2)}\,\frac{\rmd\zeta}{\rmd t} = 0
\end{equation}
because the time derivatives of the coordinates satisfy $\rmd\psi/\rmd t\sim \rho_\star \Psi_\mathrm{t}\, v/L$, $\rmd \eta/\rmd t\sim \sqrt{\rho_\star}\?  v/L$ and $\rmd\zeta/\rmd t \sim v/L$, and $\mathcal{I}$ has a large radial gradient: $\partial_\psi\mathcal{I}\sim (1/\sqrt{\rho_\star})\,vL/\Psi_\mathrm{t}$. Let
\begin{align}\label{eq:Gammadefs}
\Gamma_\psi &\coloneq \frac{\bb{v}_\mathrm{d}\bb{\cdot}\bb{\nabla}\psi}{v_\parallel \bh \bb{\cdot}\bb{\nabla}\zeta}\,, &
\Gamma_\eta &\coloneq \iota(\psi) - \frac{N}{M}  + \frac{\bb{v}_\mathrm{d}\bb{\cdot}\bb{\nabla}\eta}{v_\parallel \bh \bb{\cdot}\bb{\nabla}\zeta}\,.
\end{align}
Integrating \eqref{eq:lowestorder} to find $\mathcal{I}^{(1/2)}$ gives
\begin{equation}\label{eq:Ihalf}
    \mathcal{I}^{(1/2)}(\psi,\eta,\zeta) = -\biggl(\partial_\psi\mathcal{I}\int_0^\zeta \!\Gamma_\psi \,\rmd \zeta' + \partial_\eta\mathcal{I}\int_0^\zeta \!\Gamma_\eta \,\rmd\zeta'\biggr) + \mathcal{I}^{(1/2)}(\psi,\eta,0)\,.
\end{equation}
We no longer write out the velocity-space arguments $(\mathcal{E},\mu,\sigma)$. In \eqref{eq:Ihalf}, we have found the dependence of $\mathcal{I}^{(1/2)}$ on $\zeta$. However, we still need to determine the constant of integration $\mathcal{I}^{(1/2)}(\psi,\eta,0)$, which depends on the slow coordinates only. Its value is fixed by a solvability condition derived from the next-order analogue of \eqref{eq:lowestorder}:
\begin{equation}\label{eq:higherorder}
    \Bigl(\partial_\psi\mathcal{I} + \partial_\psi\mathcal{I}^{(1/2)}\Bigr)\?\frac{\rmd\psi}{\rmd t} + \Bigl(\partial_\eta\mathcal{I} + \partial_\eta\mathcal{I}^{(1/2)}\Bigr)\?\frac{\rmd\eta}{\rmd t} + \Bigl(\partial_\zeta\mathcal{I}^{(1/2)} + \partial_\zeta\mathcal{I}^{(1)}\Bigr)\?\frac{\rmd\zeta}{\rmd t} = 0\,.
\end{equation}
We can eliminate the unknown quantity $\mathcal{I}^{(1)}$ from this equation by dividing by $\rmd{\zeta}/\rmd t$ and integrating over $\zeta$:
\begin{equation}\label{eq:nextorder}
    \partial_\psi \mathcal{I}\oint \Gamma_\psi \,\rmd \zeta + \partial_\eta\mathcal{I}\oint \Gamma_\eta \,\rmd \zeta + \oint \?\Bigl( \Gamma_\psi\,\partial_\psi\mathcal{I}^{(1/2)} + \Gamma_\eta\,\partial_\eta\mathcal{I}^{(1/2)}\Bigr)\,\rmd \zeta = 0\,.
\end{equation}
We found, back in \eqref{eq:changeslow}, that
\begin{align}\label{eq:Gammaidentities}
    \oint \?  \Gamma_\psi \,\rmd \zeta &= \frac{mc}{Ze}\,\partial_\eta\mathcal{I}\,, &
    \oint \?  \Gamma_\eta \,\rmd \zeta &= -\frac{mc}{Ze}\,\partial_\psi\mathcal{I}\,.
\end{align}
Therefore, the first two terms in \eqref{eq:nextorder} cancel. This cancellation had to occur at order $\sqrt{\rho_\star}\?  v^2$, as in \eqref{eq:lowestorder}, since $\mathcal{I}$ is an adiabatic invariant to lowest order. The cancellation also occurs at order $\rho_\star\?  v^2$ because the transit-averaged identities \eqref{eq:psidriftav}--\eqref{eq:alphadriftav} are satisfied exactly, as we pointed out in \cref{subsec:hamiltonianderivation}. Therefore, \eqref{eq:nextorder} becomes
\begin{equation}\label{eq:prepoisson}
    0 = \oint \?\Bigl( \Gamma_\psi\?\partial_\psi\mathcal{I}^{(1/2)} + \Gamma_\eta\?\partial_\eta\mathcal{I}^{(1/2)}\Bigr)\,\rmd \zeta\,.
\end{equation}
In \cref{app:manipulations}, we show that, after substituting \eqref{eq:Ihalf} into \eqref{eq:prepoisson}, the resulting equation can be manipulated into the Poisson-bracket form
\begin{equation}\label{eq:poisson}
    \partial_\psi F\, \partial_\eta\mathcal{I} - \partial_\eta F\, \partial_\psi\mathcal{I} = 0\,,
\end{equation}
where
\begin{equation}\label{eq:Cdef}
    F(\psi,\eta) 
    \coloneq \mathcal{I}^{(1/2)}(\psi,\eta,0) - \frac{Ze}{mc}\?\iota_{\rm r}'(\psi-\psi_\mathrm{r})\oint(\zeta' - \upi M)\? \Gamma_\psi \,\rmd\zeta' + O(\rho_\star vL)\,.
\end{equation}
Equation \eqref{eq:poisson} implies that $F(\psi,\eta)$ depends on $\psi$ and $\eta$ only through $\mathcal{I}$. Thus, we can write $F(\psi,\eta) = f(\mathcal{I}(\psi, \eta))$, for some function $f$. Substituting this result into \eqref{eq:Ihalf} gives
\begin{equation}\label{eq:Jnearly}
    \mathcal{I}^{(1/2)}(\psi,\eta,\zeta) = \frac{Ze}{mc}\?\iota_{\rm r}'(\psi-\psi_\mathrm{r})\biggl(\?\int_0^{\zeta} 2\upi M\Gamma_\psi \,\rmd\zeta' + \oint\?  (\zeta' - \zeta - \upi M)\?\Gamma_\psi \,\rmd\zeta'   \biggr) + f(\mathcal{I})\,.
\end{equation}
In \eqref{eq:Jnearly}, we have determined $\mathcal{I}^{(1/2)}$ up to a function of the lowest-order adiabatic invariant. We have complete freedom to pick $f(\mathcal{I})$---we explain why in \cref{app:constant}---so we choose $f = 0$ for simplicity. Furthermore, we make a negligible error by evaluating the line integrals in \eqref{eq:Jnearly} at the rational surface $\psi_\mathrm{r}$, which, as above, we represent using a subscript `r' next to the integral sign. Overall, the higher-order transit adiabatic invariant is
\begin{align}\label{eq:higherorderinvariant}
    \mybar{\mathcal{I}} = \sigma I - \frac{2\upi M Ze}{mc}\!\int_{\psi_\mathrm{r}}^{\psi}\biggl(\iota(\psi') - \frac{N}{M}\biggr)\,\rmd\psi' &+ \frac{Ze}{mc}\?\iota_{\rm r}'(\psi-\psi_\mathrm{r})\oint_\mathrm{r}\,(\zeta' - \zeta - \upi M)\?  \Gamma_\psi\,\rmd\zeta'\nn
    & + \frac{Ze}{mc}\?\iota_{\rm r}'(\psi-\psi_\mathrm{r}) \int_{0, \mathrm{r}}^\zeta 2\upi M \Gamma_\psi\,\rmd\zeta'\,,
\end{align}
where
\begin{equation}\label{eq:Gammapsieta}
    \Gamma_\psi = \frac{\bb{v}_\mathrm{d}\bb{\cdot}\bb{\nabla}\psi}{v_\parallel \bh \bb{\cdot}\bb{\nabla}\zeta} = \frac{mc}{Ze}\?\bigl[ \partial_\eta(v_\parallel \bh \bb{\cdot}\partial_\zeta\bb{x}) - \partial_\zeta(v_\parallel \bh \bb{\cdot}\partial_\eta\bb{x}) \bigr]\,.
\end{equation}
Equation \eqref{eq:higherorderinvariant} is the main result of this section. Before considering the implications of \eqref{eq:higherorderinvariant} for the drift-island shape, we first discuss certain consistency checks that \eqref{eq:higherorderinvariant} must pass in order for the higher-order transit invariant to be well-defined.

The first requirement is that $\mybar{\mathcal{I}}$ must be a continuous function of position on each flux surface. As written, \eqref{eq:higherorderinvariant} does not manifestly possess this property. The reason is that the higher-order correction terms in \eqref{eq:higherorderinvariant} contain the toroidal angle explicitly, but the definitions we adopted for our coordinates restricted ${\eta\in[0,2\upi/M)}$ and ${\zeta\in[0,2\upi M)}$, so there are adjacent points on any given flux surface which have very different values of $\zeta$. For example, suppose we pick a point and increase its poloidal angle until the point crosses the curve $\eta = 2\upi/ M$. Then, the $\eta$ coordinate of this point will jump discontinuously from $\eta = 2\upi/ M$ to $\eta = 0$, while its $\zeta$ coordinate will change discontinuously by a multiple of $2\upi$. It is not necessarily clear that the value of $\mybar{\mathcal{I}}$ is unchanged by these discontinuous jumps.

In fact, \eqref{eq:higherorderinvariant} can be rewritten in a form which explicitly shows that $\mybar{\mathcal{I}}$ \emph{is} a continuous function of position. To achieve this, we change the dummy integration variables in the  higher-order correction terms in \eqref{eq:higherorderinvariant} to
\begin{equation}
\xi \coloneq \begin{cases}
        \zeta' - \zeta &\text{ if } \zeta\leq\zeta'\leq 2\upi M\\
        2\upi M + \zeta' - \zeta &\text{ if } 0\leq\zeta'\leq \zeta\,,
    \end{cases} 
\end{equation}
which removes any reference to an arbitrary origin for the toroidal angle coordinate. After this change of variables, \eqref{eq:higherorderinvariant} becomes
\begin{equation}\label{eq:origininvariant}
    \mybar{\mathcal{I}} = \sigma I - \frac{2\upi M Ze}{mc}\!\int_{\psi_\mathrm{r}}^{\psi}\biggl(\iota(\psi') - \frac{N}{M}\biggr)\,\rmd\psi' + \frac{Ze}{mc}\?\iota_{\rm r}'(\psi-\psi_\mathrm{r})\oint_\mathrm{r}\? (\xi - \upi M) \?  \Gamma_\psi \,\rmd\xi\,.
\end{equation}
The line integral in the last term of \eqref{eq:origininvariant} starts at $\xi = 0$ (the particle position) instead of at the arbitrary origin $\zeta = 0$, and extends along the curve of constant $\eta$ for a full transit. This geometric prescription for the integration curve, together with the fact that the integrand depends on on $\xi$ and $\eta$ in a continuous manner, ensures that this line integral does not change discontinuously anywhere on the flux surface. Therefore, the higher-order transit invariant is a continuous function of position.

The second requirement which \eqref{eq:higherorderinvariant} needs to satisfy is that the value of $\mybar{\mathcal{I}}$ must be the same, up to terms of order $\rho_\star vL$, in any set of straight-field-line coordinates. This is a form of gauge invariance. It is not obvious that $\mathcal{\mybar{I}}$ is gauge invariant in this way because of the explicit toroidal-angle dependence in the higher-order correction terms in \eqref{eq:higherorderinvariant} or \eqref{eq:origininvariant}. Furthermore, the curves in physical space corresponding to fixed $\eta = \theta - (N/M)\? \zeta$ (which we used to define the kinetic term $I$) are different in different straight-field-line coordinate systems. In \cref{app:gauge}, we prove explicitly that formula \eqref{eq:higherorderinvariant} is, nevertheless, gauge invariant.

\subsection{Higher-order island shape}\label{sec:hoislandshape}

We now use the higher-order transit invariant, defined by  \eqref{eq:higherorderinvariant} or \eqref{eq:origininvariant}, to determine the drift-island shape more accurately. Since the higher-order transit invariant depends not just on $\psi$ and $\eta$ but also on $\zeta$, the shape of the drift islands varies slightly as we move toroidally around the device. To visualize the islands as they would appear in a Poincar\'e plot at a fixed toroidal angle, we must plot the level sets of $\mybar{\mathcal{I}}(\psi,\eta,\zeta,\mathcal{E},\mu,\sigma)$ with $\zeta$ fixed; for this discussion, we fix $\zeta = 0$. 

%
%
\begin{figure}
\centering
\begin{tikzpicture}[x=\textwidth]
\node[anchor=south west,inner sep=0] (image) at (0.0,0) {\includegraphics[width=\textwidth]{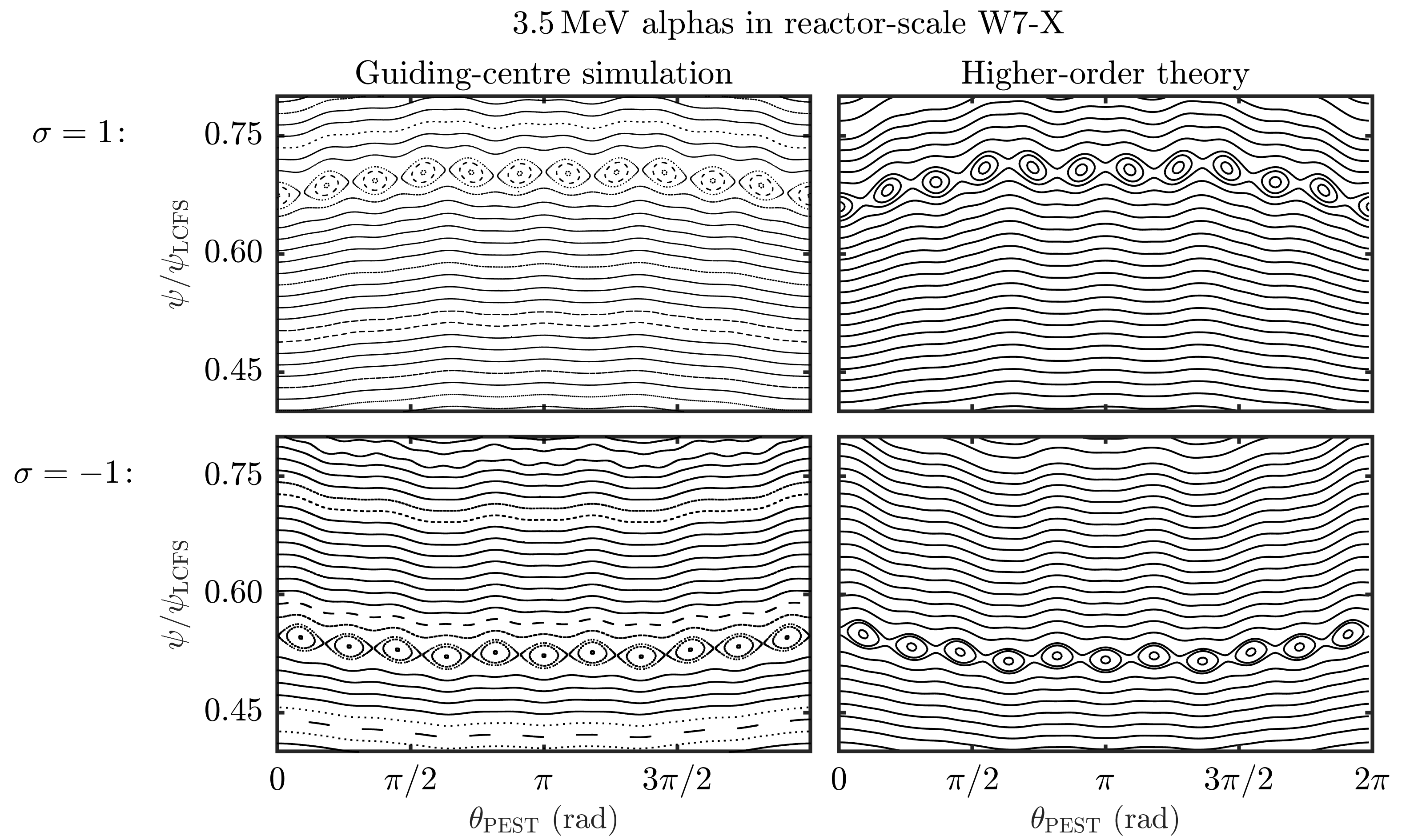}};
    \begin{scope}[x={(image.south east)},y={(image.north west)}]
    \end{scope}
\end{tikzpicture}
    \caption{(left panels) Here, we repeat the Poincar\'e plots of \cref{fig:W7X_lowestorder} for passing alpha particles with energy $\SI{3.5}{\mega\electronvolt}$ and $\mu=0$ in a reactor-scale W7-X equilibrium. (right panels) This time, we plot the level sets of the higher-order transit invariant $\mybar{\mathcal{I}}(\psi,\eta,0,\mathcal{E},\mu,\sigma)$, given by \eqref{eq:origininvariant}.}
    \label{fig:W7X_higherorder}
\end{figure}

The modifications to the drift-island shape can be most easily understood by Taylor expanding \eqref{eq:origininvariant} around the rational surface, remembering that $\psi-\psi_\mathrm{r}\sim\sqrt{\rho_\star}\,\Psi_\mathrm{t}$ and discarding terms of order $\rho_\star vL$ or smaller:
\begin{align}\label{eq:correctedshape}
    \mybar{\mathcal{I}}(\psi,\eta,0,\mathcal{E},\mu,\sigma) &\simeq \sigma I_{\rm r}(\eta,\mathcal{E},\mu) + \sigma(\psi-\psi_\mathrm{r})\,\partial_\psi I(\psi_\mathrm{r},\eta,\mathcal{E},\mu)\nn
    \al - \frac{\upi M Ze}{mc}\?\iota_{\rm r}'(\psi-\psi_\mathrm{r})^2 - \frac{\upi M Ze}{3mc}\?\iota_{\rm r}''(\psi-\psi_\mathrm{r})^3\nn
    \al + \frac{Ze}{mc}\?\iota_{\rm r}'(\psi-\psi_\mathrm{r})\oint_\mathrm{r}\? (\xi - \upi M) \?  \Gamma_\psi \,\rmd\xi\,,
\end{align}
where $\iota_{\rm r}''\coloneq (\rmd^2 \iota / \rmd\psi^2)(\psi=\psi_\mathrm{r})$. This invariant coincides with \eqref{eq:simpleinvariant} to lowest order but includes new terms that modify the drift-island shape in the following ways.

The first line of \eqref{eq:correctedshape} contains the familiar term $\sigma I_{\rm r}(\eta,\mathcal{E},\mu)$ plus a linear (in $\psi-\psi_\mathrm{r}$) correction that shifts the centre of the islands radially. This correction is proportional to the transit average of the particle's tangential magnetic drift, according to \eqref{eq:alphadriftav}. Therefore, this term allows the tangential drift to shift the position of the drift islands radially, as discussed in \cref{subsec:lowshear}.

In the second line of \eqref{eq:correctedshape}, the geometric term now has a cubic correction involving $\iota_{\rm r}''$. This shows that, at higher order, the island shape is sensitive to the curvature of the $\iota$-profile around the rational surface.

Finally, in the third line of \eqref{eq:correctedshape}, we find another linear term that shifts the islands radially. This is the only term that takes different values as we follow a curve of constant $\psi$ and $\eta$ around the stellarator: it is different at $\zeta = 0, 2\upi, \ldots, 2\upi\?(M-1)$. Therefore, this is the term that allows the higher-order transit invariant to describe deformed island chains in which each island in the chain has a slightly different shape and radial location.

In \cref{fig:W7X_higherorder}, we compare the higher-order drift-island shape of \eqref{eq:correctedshape} with the simulations presented earlier, in \cref{fig:W7X_lowestorder}. It is clear that, with the higher-order corrections, the theory does a much better job of capturing the drift-island shape and position.

\section{Hamiltonian perturbation theory for passing particles}\label{sec:irrationalsurfaces}

We have shown that the orbits of passing particles near a rational flux surface can be calculated using the transit adiabatic invariant. We derived this invariant using a local expansion in which $\iota-N/M$ was ordered as small. There are two disadvantages to such an expansion. First, given the initial coordinates of a particle, there is no clear criterion for which rational surface should play the role of $\psi_\mathrm{r}$; after all, there are infinitely many rational surfaces to choose from. Second, according to the general theory of nearly integrable Hamiltonian systems, when the drift islands from two nearby rational surfaces become large enough to overlap, the islands are replaced by a region of stochastic orbits \mycitep{Chirikov1979}. It is difficult to see that such a transition should occur when using the local expansion.

In this section, we address these issues by describing another method \Dash based on Hamiltonian perturbation theory \Dash for calculating the orbits of passing particles. This method can be applied to particles on rational or irrational surfaces. Within this framework, a finite set of `low-order' rational surfaces are naturally identified, around which the particle orbits must be described using the transit adiabatic invariant.

The version of Hamiltonian perturbation theory that we use is essentially the  \citet{Hori1966} and \citet{Deprit1969} Lie-transform approach to Poincar\'{e}--von Zeipel perturbation theory \mycitep[for reviews, see ]{Cary1981, Morbidelli, Arnold2006}. However, we avoid Lie transforms and instead use Littlejohn's phase-space Lagrangian formulation \mycitep{Littlejohn1983, CaryBrizard2009, Parra2011}. The phase-space Lagrangian makes it straightforward to use non-canonical coordinates, which we find more convenient for describing guiding-centre motion. The calculation is similar to the calculation of the distorted magnetic surfaces in a perturbed magnetic field, to which the phase-space Lagrangian method has also been applied \mycitep[\eg]{CaryLittlejohn1983, Viana2023, Escande2024, Chu2025}.

\subsection{Phase-space Lagrangian formulation}\label{subsec:lagrangianformulation}

\citet[][]{Littlejohn1983} showed that the guiding-centre equations of motion can be obtained from a phase-space Lagrangian of the form
\begin{equation}\label{eq:phasespaceL}
    \mathcal{L}_\mathrm{gc} \coloneq \Bigl( \bb{A} + \frac{v_\parallel}{\Omega}\bb{B} + \ldots \Bigr)\0\frac{\rmd\bb{X}}{\rmd t} - \Bigl(\frac{mc}{Ze}\Bigr)^{\!2}\mu\?\frac{\rmd\varphi}{\rmd t} - \frac{mc}{Ze}\? \mathcal{E}\,.
\end{equation}
The phase-space coordinates employed here are the guiding-centre position $\bb{X}$, the energy per unit mass $\mathcal{E}$, the magnetic moment $\mu$, and the gyrophase $\varphi$; these coordinates must be defined appropriately in order for the Lagrangian to take this form \mycitep[\eg]{Littlejohn1983, Parra2011}. All other symbols have the same meaning as before, and functions of position, such as the parallel velocity $v_\parallel = \sigma\sqrt{2(\mathcal{E} - \mu B)}$, are evaluated at the guiding-centre position.

In \eqref{eq:phasespaceL}, the ellipses indicate that higher-order corrections (in powers of $\rho_\star\ll 1)$ can be included so that the associated guiding-centre equations are accurate to higher order, thus describing the particle motion accurately over longer times. In this section, we ignore these higher-order corrections because we have argued (in \cref{sec:higherorder}) that, to calculate the first correction $\mathcal{I}^{(1/2)}$ to the transit adiabatic invariant, the usual guiding-centre equations \Dash without higher-order drifts \Dash are sufficiently accurate. It is worth bearing in mind that the calculation we describe in this section could be continued to arbitrarily high order in $\rho_\star$, but this would require higher-order accuracy in the starting phase-space Lagrangian and in the definitions of the guiding-centre coordinates.

We are only interested in the shape of the guiding-centre trajectory through space, not the rate at which the particle advances along this trajectory. As a result, the calculation is simpler if we parametrize position along the trajectory using the toroidal angle $\zeta$ instead of time $t$. The relevant phase-space Lagrangian is then
\begin{equation}\label{eq:fullLagrangian}
    \mathcal{L}^\zeta_\mathrm{gc} \coloneq \Bigl( \bb{A} + \frac{v_\parallel}{\Omega}\bb{B}  \Bigr)\0\frac{\rmd\bb{X}}{\rmd \zeta} - \Bigl(\frac{mc}{Ze}\Bigr)^{\!2}\mu\?\frac{\rmd\varphi}{\rmd \zeta} - \frac{mc}{Ze}\? \mathcal{E}\? \frac{\rmd t}{\rmd \zeta}\,,
\end{equation}
where $t$ is now treated as a phase-space coordinate. Since $\mathcal{E}$ and $\mu$ are both constant, and since we are not interested in calculating how $\varphi$ or $t$ evolve with $\zeta$, we can lower the dimensionality of phase space by using, instead of \eqref{eq:fullLagrangian}, the reduced Lagrangian
\begin{equation}\label{eq:simpleLagrangian}
    \mathcal{L} \coloneq \Bigl( \bb{A} + \frac{v_\parallel}{\Omega}\bb{B} \Bigr)\0\frac{\rmd\bb{X}}{\rmd \zeta}\,.
\end{equation}
In \eqref{eq:simpleLagrangian}, $\mathcal{E}$ and $\mu$ are treated as fixed parameters. This fact will be important whenever we add a total $\zeta$ derivative to the Lagrangian \eqref{eq:simpleLagrangian}. There will be no need to include $\mathcal{E}$ or $\mu$ derivatives when using the chain rule (\ie $\rmd f/\rmd\zeta = \partial_\psi f\?(\rmd\psi/\rmd\zeta) + \partial_\theta f\?(\rmd\theta/\rmd\zeta) + \partial_\zeta f$). The reduced Lagrangian \eqref{eq:simpleLagrangian} can also be obtained by enforcing $\mathcal{E} = \text{const.}$ and $\mu = \text{const.}$ as holonomic constraints in \eqref{eq:fullLagrangian} \mycitep[\eg]{Goldstein2002}. We can think of \eqref{eq:simpleLagrangian} as the field-line Lagrangian \mycitep{CaryLittlejohn1983} for the fictitious magnetic field $\bb{B}^\star$ which, as explained in \cref{subsec:magneticislands}, the guiding centre follows. We will use \eqref{eq:simpleLagrangian} to calculate the guiding-centre orbits of passing particles in a stellarator magnetic field.

\subsection{Drift surfaces}

First, we express the Lagrangian in the straight-field-line coordinates $(\psi,\theta,\zeta)$ appropriate for magnetic fields with nested toroidal flux surfaces. For the vector potential, we can take \mycitep[\eg]{Helander2014}
\begin{equation}
    \bb{A} \coloneq \psi\,\bb{\nabla}\theta - \chi(\psi)\? \bb{\nabla}\zeta\,,
\end{equation}
where $\chi(\psi) \coloneq \int_0^\psi \iota(\psi') \,\rmd\psi'$ is the poloidal flux contained inside flux surface $\psi$. Then, \eqref{eq:simpleLagrangian} becomes
\begin{equation}\label{eq:BstarLagrangian}
   \mathcal{L} =  \Bigl(\psi + \frac{v_\parallel}{\Omega}\bb{B} \0\con{\theta}\Bigr)\frac{\rmd\theta}{\rmd\zeta} + \frac{v_\parallel}{\Omega}\bb{B} \0\con{\psi}\,\frac{\rmd\psi}{\rmd\zeta} -\chi + \frac{v_\parallel}{\Omega}\bb{B} \0\con{\zeta}\,.
\end{equation}
Throughout this section, the partial derivatives $\partial_\psi$, $\partial_\theta$ and $\partial_\zeta$ are taken with respect to one of the coordinates $(\psi,\theta,\zeta)$, with the other two held fixed. Note that this changes the definition of $\partial_\zeta$, which in previous sections denoted a derivative at fixed $\psi$ and $\eta$.

It is useful to think of \eqref{eq:BstarLagrangian} as a small perturbation to the Lagrangian for the field lines of the magnetic field, which is
\begin{equation}\label{eq:BLagrangian}
    \mathcal{L}_B \coloneq \psi\?\frac{\rmd\theta}{\rmd\zeta} - \chi(\psi)\,.
\end{equation}
The Euler-Lagrange equations of motion corresponding to \eqref{eq:BLagrangian} are simply
\begin{align}\label{eq:BEL}
     \frac{\rmd\psi}{\rmd\zeta} &= 0\,, & \frac{\rmd\theta}{\rmd\zeta} &= \iota(\psi)\,.
\end{align}
The first equation states that $\bb{B}$ has flux surfaces, while the second states that $\theta$ and $\zeta$ are straight-field-line coordinates. The reason these equations are so simple is that $\mathcal{L}_B$ is independent of the angles $\theta$ and $\zeta$.\footnote{The coordinates $\psi$ and $\theta$ are action-angle coordinates for the Hamiltonian system described by phase-space Lagrangian $\mathcal{L}_B$.} By contrast, the equations of motion for the guiding-centre Lagrangian \eqref{eq:BstarLagrangian} are%
\begin{subequations}\label{eq:gcEL}
\begin{align}
    \label{eq:psizeta}\frac{\rmd\psi}{\rmd\zeta} &= \Bigl[\partial_\theta\Bigl(\frac{v_\parallel}{\Omega}\bb{B} \0\con{\zeta}\Bigr) - \partial_\zeta\Bigl(\frac{v_\parallel}{\Omega}\bb{B} \0\con{\theta}\Bigr)\Bigr] + O(\rho_\star^2\,\Psi_\mathrm{t})\,,\\
    \label{eq:thetazeta}\frac{\rmd\theta}{\rmd\zeta} &= \iota(\psi) - \Bigl[\partial_\psi\Bigl(\frac{v_\parallel}{\Omega}\bb{B} \0\con{\zeta}\Bigr) - \partial_\zeta\Bigl(\frac{v_\parallel}{\Omega}\bb{B} \0\con{\psi}\Bigr)\Bigr] + O(\rho_\star^2)\,.
\end{align}
\end{subequations}
In comparison with \eqref{eq:BEL}, each equation here has an additional, order-$\rho_\star$ term that depends on both angles $\theta$ and $\zeta$; physically, these terms represent spatially varying drifts. The key idea of Hamiltonian perturbation theory is to look for a coordinate transformation that removes the angle dependence from the Lagrangian \eqref{eq:BstarLagrangian}, so that the guiding-centre equations of motion become as simple as \eqref{eq:BEL}.

Since the angle-dependent terms in \eqref{eq:BstarLagrangian} are small in $\rho_\star$, we will look for a \emph{near-identity} coordinate transformation of the form
\begin{subequations}\label{eq:irrationaltransform}
\begin{align}
    \psi &= \mybar{\psi} + \psi^{(1)}(\mybar{\psi}, \mybar{\theta}, \zeta)\,,\\
    \theta &= \mybar{\theta} + \theta^{(1)}(\mybar{\psi}, \mybar{\theta}, \zeta)\,,
\end{align}
\end{subequations}
where $\psi^{(1)}\sim\rho_\star \Psi_\mathrm{t}$ and $\theta^{(1)}\sim \rho_\star$ are small corrections. At this stage, \eqref{eq:irrationaltransform} is a completely general near-identity coordinate transformation. Shortly, we will make a particular choice for $\psi^{(1)}$ and $\theta^{(1)}$ that eliminates any dependence on $\mybar{\theta}$ and $\zeta$ from the Lagrangian (or, at least, relegates this dependence to higher order). Equation \eqref{eq:irrationaltransform} defines the new coordinates $\mybar{\psi}$ and $\mybar{\theta}$ implicitly, analogous to how the guiding-centre position is often defined implicitly by
\begin{equation}\label{eq:gctransform}
    \bb{x} = \bb{X} + \bb{\rho}(\bb{X}, \mathcal{E}, \mu, \varphi)\,,
\end{equation}
where $\bb{x}$ is the particle position, $\bb{X}$ is the guiding-centre position, and $\bb{\rho}$ is the gyroradius vector.

Using \eqref{eq:irrationaltransform} to change coordinates in the Lagrangian gives
\begin{align}\label{eq:Lagrangianterms1}
    \mathcal{L} = \Bigl( \mybar{\psi} + \psi^{(1)} + \frac{v_\parallel}{\Omega}\bb{B} \0\con{\theta}\Bigr)\biggl(\frac{\rmd\mybar{\theta}}{\rmd\zeta} + \frac{\rmd\theta^{(1)}}{\rmd\zeta}\biggr) &+ \frac{v_\parallel}{\Omega}\bb{B} \0\con{\psi}\,\biggl(\frac{\rmd\mybar{\psi}}{\rmd\zeta}
    + \frac{\rmd\psi^{(1)}}{\rmd\zeta}\biggr)\nn
    &{}- \chi - \iota\? \psi^{(1)} + \frac{v_\parallel}{\Omega}\bb{B} \0\con{\zeta}\,,
\end{align}
where we drop terms of order $\rho_\star^2$ from the Lagrangian throughout the rest of this section.\footnote{More precisely: if we represent our Lagrangian in the form $\mathcal{L} = \Lambda_\psi (\rmd\mybar{\psi}/\rmd\zeta) + \Lambda_\theta (\rmd\mybar{\theta}/\rmd\zeta) - \mathcal{H}$, then we drop terms of order $\rho_\star^2$ from the symplectic parts $\Lambda_\psi$ and $\Lambda_\theta$, and from the Hamiltonian $\mathcal{H}$. Such terms are as small as the higher-order terms that we neglected, \textit{ab initio}, in \eqref{eq:phasespaceL}.} From here on, functions of position without arguments written explicitly are assumed to be evaluated at the point with magnetic coordinates $(\mybar{\psi}(\psi, \theta, \zeta),\mybar{\theta}(\psi, \theta, \zeta),\zeta)$, rather than at the actual guiding-centre position $(\psi, \theta, \zeta)$.

Recall that a total time derivative can be added to a Lagrangian without modifying the associated equations of motion. To remove the derivatives of the as-yet-undetermined functions $\psi^{(1)}$ and $\theta^{(1)}$ from \eqref{eq:Lagrangianterms1}, we subtract the total $\zeta$ derivative
\begin{equation}
    \frac{\rmd}{\rmd\zeta}\Bigl[ \Bigl( \mybar{\psi} + \psi^{(1)} + \frac{v_\parallel}{\Omega}\bb{B} \0\con{\theta}\Bigr)\,\theta^{(1)} + \frac{v_\parallel}{\Omega}\bb{B} \0\con{\psi}\,\psi^{(1)} \Bigr]\,.
\end{equation}
Furthermore, we anticipate that we may need to add other $\zeta$ derivatives to the Lagrangian to put it in a form that has no angle dependence. To allow for this possibility, we subtract $\rmd S^{(1)}/\rmd \zeta$ for some `generating function' $S^{(1)}\sim \rho_\star \Psi_\mathrm{t}$, which we will choose later. The result is
\begin{align}\label{eq:Lagrangianterms2}
        \mathcal{L} \equiv \Bigl( \mybar{\psi} + \psi^{(1)} + \frac{v_\parallel}{\Omega}\bb{B} \0\con{\theta} - \partial_\theta S^{(1)}\Bigr)\,\frac{\rmd\mybar{\theta}}{\rmd\zeta} &+ \Bigl(-\theta^{(1)} + \frac{v_\parallel}{\Omega}\bb{B} \0\con{\psi} - \partial_\psi S^{(1)}\Bigr)\,\frac{\rmd\mybar{\psi}}{\rmd\zeta}\nn
        &- \chi - \iota\? \psi^{(1)} + \frac{v_\parallel}{\Omega}\bb{B} \0\con{\zeta} - \partial_\zeta S^{(1)}\,,
\end{align}
where we use the symbol $\equiv$ to denote equality up to a total $\zeta$ derivative. We can remove some of the angle dependence in \eqref{eq:Lagrangianterms2} by picking
\begin{subequations}\label{eq:transformdefined}
\begin{align}
    \psi^{(1)} &\coloneq -\frac{v_\parallel}{\Omega}\bb{B} \0\con{\theta} + \partial_\theta S^{(1)}\,,\\
    \theta^{(1)} &\coloneq \frac{v_\parallel}{\Omega}\bb{B} \0\con{\psi} - \partial_\psi S^{(1)}\,,
\end{align}
\end{subequations}
for whatever choice of $S^{(1)}$ we eventually make. This leaves
\begin{equation}\label{eq:Hamiltonian}
    \mathcal{L}\equiv \mybar{\psi}\,\frac{\rmd\mybar{\theta}}{\rmd\zeta} - \chi + V - \bigl(\partial_\zeta S^{(1)} + \iota\? \partial_\theta S^{(1)}\bigr)\,,
\end{equation}
where, as in \cref{subsec:resonance},
\begin{equation}
    V = \frac{v_\parallel}{\Omega}\bb{B} \0(\con{\zeta} + \iota\? \con{\theta}) = \frac{mc}{Ze} \frac{v_\parallel}{\bh \0\bb{\nabla}\zeta}\,.
\end{equation}
The remaining angle dependence in \eqref{eq:Hamiltonian} would be eliminated if we could choose $S^{(1)}$ to satisfy the magnetic differential equation
\begin{equation}\label{eq:R}
    \partial_\zeta S^{(1)} + \iota\?  \partial_\theta S^{(1)} =  V - \langle V \rangle\,.
\end{equation}
Here, we have introduced the average over the poloidal and toroidal angles
\begin{equation}
    \langle \? \ldots\?  \rangle  \coloneq \frac{1}{(2\upi)^2}\int_0^{2\upi}\int_0^{2\upi} (\ldots) \,\rmd\theta\,\rmd\zeta\,.
\end{equation}
We included the term $\langle V \rangle$ in \eqref{eq:R} because the left-hand side clearly vanishes under this average, so the right-hand side must as well. Unfortunately, we will find that \eqref{eq:R} is not always solvable for $S^{(1)}$. Furthermore, even when \eqref{eq:R} is solvable, the solutions have certain deficiencies that make them unsuitable for completing the construction of our near-identity coordinate transformation. In the next subsection, we discuss these issues with \eqref{eq:R} in more detail.

\subsection{Solvability of the magnetic differential equation}\label{subsec:mdes}

For a general inhomogeneous term $V$, magnetic differential equations of the form \eqref{eq:R} have been extensively studied \mycitep[\eg]{Kruskal1958, Newcomb1959, Hamada1962, Russmann1975, Imbert2024}. In this subsection, we investigate when \eqref{eq:R} can be solved in the specific case where $V = (mc/Ze)(v_\parallel/\bh \0\bb{\nabla}\zeta)$.

Equation \eqref{eq:R} specifies the rate of change of $S^{(1)}$ along magnetic field lines, since it can be recast as
\begin{equation}\label{eq:Udef}
    \bh \0\bb{\nabla}S^{(1)} = \frac{mc}{Ze}\,\biggl(v_\parallel - \bh \0\bb{\nabla}\zeta\,\biggl\langle \frac{v_\parallel}{\bh \0\bb{\nabla}\zeta}\biggr\rangle \biggr) \coloneq U\,.
\end{equation}
Symbolically, the solution to this equation is
\begin{equation}\label{eq:Rfieldline}
    S^{(1)} \stackrel{?}{=} \int^l U\,\rmd l'\,,
\end{equation}
where the integral is taken along field lines. However, it is unclear whether \eqref{eq:Rfieldline} defines a continuous function of position on each flux surface. On an irrational surface, two arbitrarily close points could be connected by an arbitrarily large distance along a single field line, meaning their values of $S^{(1)}$ \Dash calculated by integrating $U$ along this line \Dash could be very different. The problem with \eqref{eq:Rfieldline} is that it does not `know about' the toroidal periodicity of the domain.

This observation suggests we look for a periodic solution to \eqref{eq:R} using a Fourier expansion. Let
\begin{align}
    S^{(1)} &= \mspace{-8mu}\sum_{(p,q)\in \mathbb{Z}^2}\mspace{-8mu} S^{(1)}_{pq}\,\mathrm{e}^{\mathrm{i}(p\theta-q\zeta)}\,, & V &= \mspace{-8mu}\sum_{(p,q)\in \mathbb{Z}^2}\mspace{-8mu} V_{pq}\,\mathrm{e}^{\mathrm{i}(p\theta-q\zeta)}\,.
\end{align}
In terms of these Fourier coefficients, \eqref{eq:R} becomes
\begin{equation}\label{eq:R3}
    -\mathrm{i}(q - \iota\?  p)S^{(1)}_{pq} = \begin{cases}
        V_{pq} &\text{if $(p,q)\neq(0,0)\,$,}\\
        0 &\text{if $(p,q)=(0,0)\,$.}
    \end{cases}
\end{equation}
Formally, this means
\begin{equation}\label{eq:Rsum}
    S^{(1)} \stackrel{?}{=} \mspace{-8mu}\sum_{(p,q) \in \,\mathbb{Z}^2\backslash\{(0,0)\}}\frac{\mathrm{i}V_{pq}}{q-\iota\?  p}\,\mathrm{e}^{\mathrm{i}(p\theta-q\zeta)}\,,
\end{equation}
where we have chosen the solution with zero average over the poloidal and toroidal angles ($S^{(1)}$ is determined up to an arbitrary additive constant).

Now, the question of whether the magnetic differential equation \eqref{eq:R} is solvable becomes the question of whether the Fourier series in \eqref{eq:Rsum} converges. This series will not, typically, converge if the rotational transform $\iota$ is rational: in this case, there are combinations of $p$ and $q$ that make the denominator vanish, which is called a `resonant' denominator. It turns out that the series also diverges at certain irrational values of $\iota$ \mycitep{Russmann1975, Yoccoz1992}.

The series \eqref{eq:Rsum} converges in two physically relevant scenarios, which are summarized below.
\begin{enumerate}
\item The series converges for certain `highly irrational' values of $\iota$. For example, a sufficient condition for the series to converge for all analytic $V$ is that $\iota$ satisfies the following Diophantine condition: there are constants $C>0$ and $\varsigma > 0$ such that
\begin{equation}\label{eq:Diophantine}
    \Bigl| \iota - \frac{N}{M} \Bigr| > \frac{C}{|M|^{2+\varsigma}}\,,
\end{equation}
for all integers $N$ and $M$ \mycitep[see ][, \S3]{Arnol1963}. This condition prevents $\iota$ from being too close to any rational number. When \eqref{eq:Diophantine} is satisfied, the terms in the series \eqref{eq:Rsum} decay fast enough for convergence because the numerators $V_{pq}$ decay exponentially for analytic $V$ while the denominators decay at most algebraically.

\item \hypertarget{item:omnigeneity} The series converges if the numerators vanish whenever the denominators do. This is clearly a necessary condition for convergence; interestingly, it is also sufficient, according to the following result of \citet{Newcomb1959}. Let $(\psi_1, \psi_2)$ be an interval in which the shear $\iota'(\psi)$ never vanishes. The series \eqref{eq:Rsum} converges for all $\psi\in (\psi_1, \psi_2)$ if (1) $V$ is sufficiently smooth and (2) for every rational surface ${\psi_{M\!N}\in (\psi_1, \psi_2)}$ and for all $k\in \mathbb{Z}$, we have ${V_{kM\!,\? kN}(\psi_{M\!N}) = 0}$, where $N$ and $M>0$ are coprime integers defined by ${\iota(\psi_{M\!N}) = N/M}$. When condition (2) is satisfied, certain terms in the series in \eqref{eq:Rsum} have an indeterminate `$0/0$' form at rational surfaces. This indeterminacy arises because \eqref{eq:R3} does not fix the quantities $S^{(1)}_{kM\!,\? kN}(\psi_{M\!N})$; as a result, we can freely choose their values. To ensure continuity with the irrational surfaces, we must choose
\begin{equation}
    S^{(1)}_{kM\!,\? kN}(\psi_{M\!N}) = -\frac{\mathrm{i}\? \partial_\psi V_{kM\!,\? kN}(\psi_{M\!N})}{kM\iota'(\psi_{M\!N})}\,,
\end{equation}
and we assume that this choice has been made when interpreting \eqref{eq:Rsum} on rational surfaces. 

The condition $V_{kM\!,\? kN}(\psi_{M\!N}, \mathcal{E}, \mu, \sigma) = 0$ (where we have temporarily restored the velocity-space arguments) arises naturally in the context of passing-particle orbits in stellarators. We showed in \cref{subsec:resonance} that this condition is equivalent to the absence of drift islands around rational surface $\psi_{M\!N}$, for particles of energy $\mathcal{E}$ and magnetic moment $\mu$. If $V_{kM\!,\? kN}(\psi_{M\!N}, \mathcal{E}, \mu, \sigma) = 0$ for all passing $\mathcal{E}$ and $\mu$, then the magnetic field must be cyclometric on $\psi_{M\!N}$, as discussed in \cref{subsec:omnigeneity}. Since omnigeneous stellarators are always cyclometric, this condition is automatically satisfied in omnigeneous devices. The fact that $S^{(1)}$ \Dash and, therefore, $\mybar{\psi}$ \Dash is always well-defined in perfectly omnigeneous stellarators is unsurprising, as it is well-known that $(\psi, \alpha, B)$ coordinates may be used to derive a formula equivalent to \eqref{eq:psistar} for $\mybar{\psi}$ which does not involve a sum and which, therefore, never suffers from convergence issues \mycitep[\eg]{Helander2009, Landreman2012}.
\end{enumerate}

To summarize, the Fourier series \eqref{eq:Rsum} sometimes converges, for example when $\iota$ is `highly irrational' or when the stellarator is perfectly cyclometric in a neighbourhood of the flux surface of interest. Unfortunately, even in these cases, the resulting $S^{(1)}$ cannot be used to complete the construction of our near-identity coordinate transformation, for the following reasons.
\begin{enumerate}
    \item In the case of highly irrational flux surfaces, the Fourier series for $S^{(1)}$ converges on these surfaces. However, it still (generically) diverges on the dense set of surfaces that do not satisfy \eqref{eq:Diophantine}; this set includes, for example, all rational surfaces. Therefore, $S^{(1)}$ must be discontinuous in the radial direction. Our perturbative change of coordinates required $S^{(1)}$ to have a well-defined radial derivative \Dash in particular, the definition of $\theta^{(1)}$ involved $\partial_\psi S^{(1)}$ \Dash so this is a problem.
    \item In the case of cyclometric stellarators, the Fourier series for $S^{(1)}$ diverges if any harmonic $V_{kM\!,\?kN}(\psi_{M\!N})$ is non-zero, regardless of how small the harmonic may be. A formalism that breaks down in the presence of an infinitesimal deviation from cyclometry is inadequate for describing real devices.
\end{enumerate}
In the next subsection, we explain how these problems associated with the resonant denominators may be circumvented.

\subsection{Dealing with resonances}\label{subsec:weakresonances}

Our approach is based on two key ideas \mycitep[other methods exist; see \eg]{Morbidelli}. The first idea is that we do not need to eliminate the angle dependence in the Lagrangian completely; we only need to eliminate it to order $\rho_\star$. To use this idea, we follow \citet{Arnold1963} and introduce an `ultraviolet cutoff' by splitting the perturbation into ${V = V^{<K} + V^{\geq K}}$, where $V^{<K}$ and $V^{\geq K}$ are defined as follows. The first term, $V^{<K}$, contains all harmonics of $V$ that have $|p|+|q|<K$ for some cutoff $K>0$. The second term, $V^{\geq K}$, contains all the remaining harmonics. Crucially, we define the cutoff $K$ to be large enough that $V^{\geq K}$ is an order (in $\rho_\star$) smaller than $V^{<K}$; thus $V^{\geq K} \sim \rho_\star^2 \? \Psi_\mathrm{t}$. For a sufficiently smooth $V$, we can always find a $K$ large enough to satisfy this requirement. The ultraviolet cutoff is useful because $V^{\geq K}$ is small enough that it can be neglected in the Lagrangian, so we only need to eliminate the angle dependence contained in $V^{<K}$. 

Resuming our calculation from \eqref{eq:Hamiltonian}, we can write
\begin{equation}\label{eq:Lcutoff}
    \mathcal{L}\equiv \mybar{\psi}\,\frac{\rmd\mybar{\theta}}{\rmd\zeta} - \chi + V^{<K} - \bigl(\partial_\zeta S^{(1)} + \iota\? \partial_\theta S^{(1)}\bigr)\,,
\end{equation}
since we are dropping terms of order $\rho_\star^2$ or smaller from the Lagrangian. The calculation now proceeds as before, but with $V$ replaced by the smoother function $V^{<K}$. Thus, to eliminate the angle dependence in \eqref{eq:Lcutoff}, we take 
\begin{equation}\label{eq:Rwithoutres}
        S^{(1)} \coloneq \mspace{-8mu}\sum_{(p,q) \in \,\mathbb{I}^{<K}}\mspace{-8mu} \frac{\mathrm{i}(V^{<K})_{pq}}{q-\iota\? p}\,\mathrm{e}^{\mathrm{i}(p\theta-q\zeta)}\,,
\end{equation}
where the index set is $\mathbb{I}^{<K} \coloneq \{ (p, q)\in \mathbb{Z}^2: |p|+|q| < K \text{ and } (p,q) \neq (0,0)\}$. Equation \eqref{eq:Rwithoutres} replaces the problematic definition \eqref{eq:Rsum}. The series in \eqref{eq:Rwithoutres} converges on all flux surfaces except a finite set of rational surfaces: those that resonate with one of the finitely many harmonics of $V^{<K}$. We will refer to these surfaces as `low-order' rational surfaces.

To avoid these divergences on low-order rational surfaces, we introduce the second key idea: our perturbative construction does not need to be valid for all passing particles simultaneously. Instead, we can restrict attention to particles in a region that does not contain any low-order rational surfaces; in such a region, the rotational transform is, in some sense, `sufficiently irrational'. From now on, we assume $\psi\in\mathfrak{R}_0$, where $\mathfrak{R}_0$ is a spatial region that does not contain any of the low-order rational surfaces. Note that we will slightly refine the definition of $\mathfrak{R}_0$ below, in \cref{def:R0}.

Within $\mathfrak{R}_0$, the series \eqref{eq:Rwithoutres} converges and the near-identity coordinate transformation specified by \eqref{eq:irrationaltransform}, \eqref{eq:transformdefined} and \eqref{eq:Rwithoutres} is well-defined. The Lagrangian in these new coordinates is
\begin{equation}\label{eq:barLagrangian}
    \mathcal{L} \equiv \mybar{\psi}\,\frac{\rmd\mybar{\theta}}{\rmd\zeta} - \mybarup{\chi}(\mybar{\psi})\,,
\end{equation}
where $\mybarup{\chi} \coloneq \chi - \langle V \rangle$. We have achieved our aim of removing angle dependence from the Lagrangian; any angle dependence is now hidden in higher-order terms that have been suppressed in \eqref{eq:barLagrangian}.

According to \eqref{eq:barLagrangian}, the new coordinates $\mybar{\psi}$ and $\mybar{\theta}$ evolve with $\zeta$ in a trivial way. One equation of motion simple states that $\mybar{\psi}$ is constant (to leading order; its $\zeta$ derivative can be non-zero at order $\rho_\star^2\,\Psi_\mathrm{t}$ due to the suppressed higher-order terms). Physically, this means passing particles on irrational surfaces do not experience a secular radial drift over long times of order $(1/\rho_\star)(L/v)$, as stated in \cref{sec:basictheory}. The other equation of motion is
\begin{equation}\label{eq:kiota}
    \frac{\rmd\mybar{\theta}}{\rmd\zeta} = \iota(\mybar{\psi}) + \frac{mc}{Ze}\? \partial_\psi\biggl\langle \frac{v_\parallel}{\bh \0\bb{\nabla}\zeta} \biggr\rangle \coloneq \mybarup{\iota}(\mybar{\psi})\,,
\end{equation}
which means $\mybar{\theta}$ changes with $\zeta$ at a constant rate (again, there may be deviations from this behaviour at order $\rho_\star^2$). This rate $\mybarup{\iota}$ is the kinetic rotational transform of the orbit; it is determined by a combination of the rotational transform of the field and the flux-surface-averaged tangential drift \Dash which is a small correction \Dash as discussed in \cref{subsec:lowshear}.

Upon inverting the coordinate transformation \eqref{eq:irrationaltransform} to first order in $\rho_\star$, we find that
\begin{equation}\label{eq:psistar}
    \mybar{\psi}(\psi, \theta, \zeta) = \psi + \Bigl(\frac{v_\parallel}{\Omega}\bb{B} \0\con{\theta}\Bigr)(\psi, \theta, \zeta) - \bigl(\partial_\theta S^{(1)}\bigr)(\psi,\theta,\zeta) + O(\rho_\star^2\? \Psi_\mathrm{t})
\end{equation}
is a conserved quantity. This is an equation for the passing-particle drift surfaces in regions of sufficiently irrational rotational transform in any stellarator. It is consistent with a less general result derived in \citet[equation (12)]{Tykhyy2006} and \citet[appendix A]{Kolesnichenko2006} for a magnetic field whose field strength is equal to a constant plus a small angle-dependent term, provided an ultraviolet cutoff is used to make the Fourier series in these references well-behaved.

\begin{figure}
\vspace{3mm}
\centering
\includegraphics[width=\textwidth]{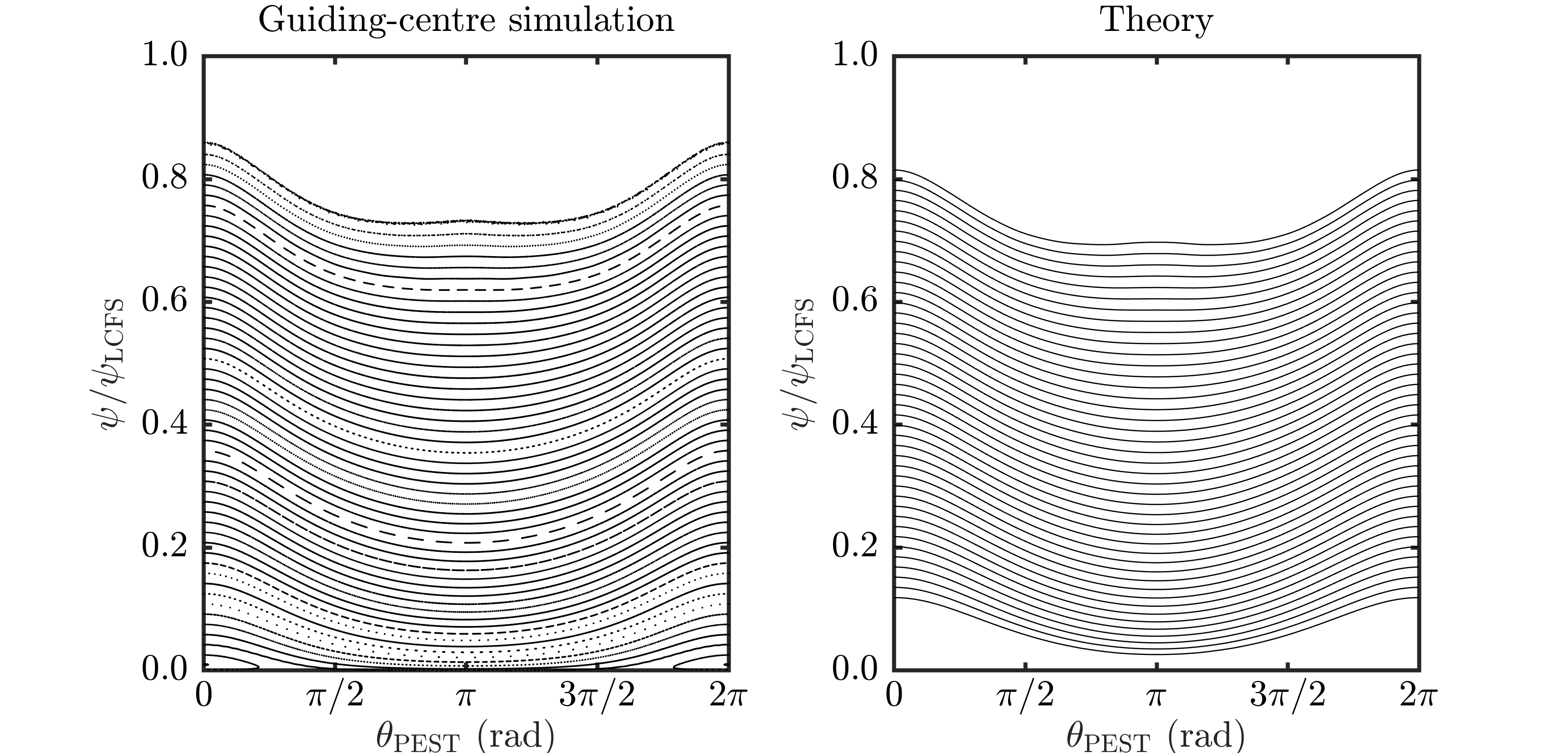}
    \caption{$(a)$ Poincar\'e plot of the guiding-centre orbits of passing alpha particles in a reactor-scale LHD equilibrium, simulated using ASCOT5. The alphas have energy $\SI{3.5}{\mega\electronvolt}$, $\mu/\mathcal{E} = \SI{0.111}{\per\tesla}$, and $\sigma = 1$. Drift surfaces beyond $\psi/\psi_{\rm LCFS}\approx 0.8$ are missing because we have not plotted orbits that crossed the last closed flux surface. $(b)$ Level sets of $\mybar{\psi}$, defined by \eqref{eq:psistar}. The plotted contour levels were chosen as follows: a contour level was computed for each particle by averaging $\mybar{\psi}$ over the points plotted in $(a)$ for that particle.}
    \label{fig:LHD}
\end{figure}

\begin{figure}
\vspace{4mm}
\centering
\includegraphics[width=\textwidth]{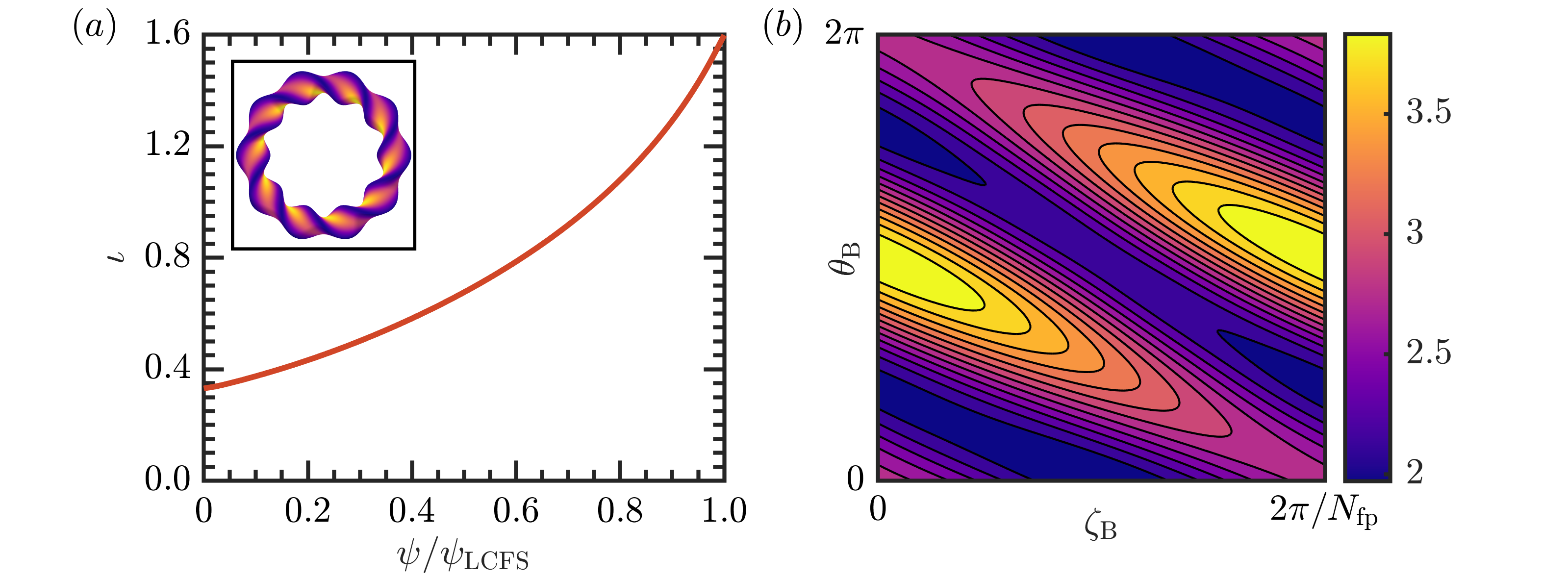}
    \caption{$(a)$ Rotational transform of the LHD equilibrium we use to trace alpha orbits in \cref{fig:LHD}. $(b)$ Plot of the magnetic field strength in Boozer coordinates on the $\iota = 1$ flux surface.}
    \label{fig:LHDeqm}
\end{figure}

In \cref{fig:LHD}, we compare this calculation of particle orbits in a region of sufficiently irrational rotational transform with an ASCOT5 simulation for a reactor-scale LHD equilibrium. More detail about the equilibrium is provided in \cref{fig:LHDeqm}. Despite the fact that the rotational transform passes through many rational values, there are no visible drift islands; the same is true at all other pitch angles. This must be a consequence of the high shear and smallness of any resonant harmonics of $V$.

%
%

%
%

\begin{figure}
\vspace{3mm}
\centering
\includegraphics[width=\textwidth]{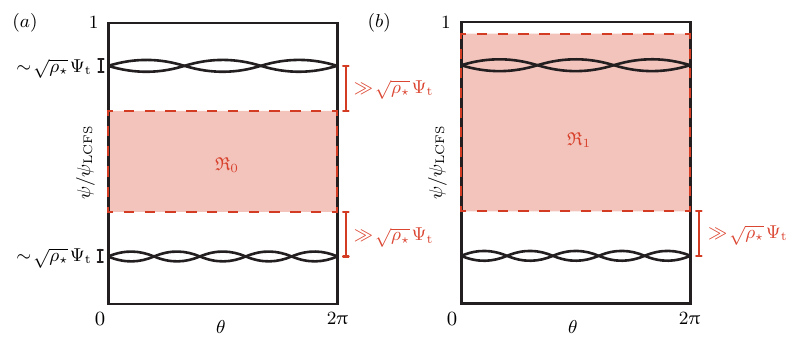}
    \caption{Diagrams showing the regions of validity for the orbit calculations presented in \crefrange{subsec:weakresonances}{subsec:oneres}. In each case, a Poincar\'e plot is drawn with drift-island chains around two low-order rational surfaces. $(a)$ Region $\mathfrak{R}_0$ must remain much further than $\sim\!\sqrt{\rho_\star}\?\Psi_{\rm t}$ from either low-order rational surface. $(b)$ Region $\mathfrak{R}_1$ contains one of the low-order rational surfaces and must remain much further than $\sim\!\sqrt{\rho_\star}\?\Psi_{\rm t}$ from the other.}
    \label{fig:regions}
\end{figure}

Before moving on, it is instructive to see how the above calculation fails close to a low-order rational surface. Let the nearest low-order rational surface to $\mathfrak{R}_0$ be $\psi_\mathrm{r}$. The calculation can fail because, when $|\psi-\psi_\mathrm{r}|$ is too small for some $\psi\in \mathfrak{R}_0$, $S^{(1)}$ becomes large enough to violate our orderings. For the relations $\psi^{(1)}\ll\Psi_\mathrm{t}$ and $\theta^{(1)}\ll 1$ to hold, we need $S^{(1)}\ll \Psi_\mathrm{t}$ and $\partial_\psi S^{(1)}\ll 1$. From \eqref{eq:Rwithoutres}, $S^{(1)}$ diverges as ${S^{(1)} \sim \rho_\star \Psi_\mathrm{t}/\iota_{\rm r}'(\psi-\psi_\mathrm{r})}$ (we show this explicitly in \cref{app:limits} \Dash see \eqref{eq:RRdiff2}), which means ${\partial_\psi S^{(1)} \sim \rho_\star \Psi_\mathrm{t}/ \iota_{\rm r}' (\psi-\psi_\mathrm{r})^2}$. Thus, for our construction to be valid, we need ${|\psi-\psi_\mathrm{r}| \gg (\rho_\star/s)^{1/2}\,\Psi_\mathrm{t}}$. The lower bound in this inequality is the width of the drift islands around $\psi_\mathrm{r}$. Therefore, for our coordinate transformation to be valid throughout $\mathfrak{R}_0$, we need $\mathfrak{R}_0$ to stay much further than a drift-island width away from any low-order rational surfaces, as shown in \crefsub{fig:regions}{a}. For reference, we now state the complete definition of $\mathfrak{R}_0$.
\begin{definition}\label{def:R0}
     $\mathfrak{R}_0$ is an interval $(\psi_1, \psi_2)$ whose distance to any low-order rational surface, $\psi_{\rm r}$, satisfies $\mathrm{dist}(\psi_{\rm r}, \mathfrak{R}_0)\gg (\rho_\star/s)^{1/2}\,\Psi_\mathrm{t}$.\footnote{Note that, as $\rho_\star \to 0$, the condition $V^{\geq K}\sim \rho_\star V^{<K}$ means we must have $K\to\infty$. As $K$ increases, the number of low-order rational surfaces grows without bound. Thus, any region $\mathfrak{R}_0$ that satisfies \cref{def:R0} must decrease in size as $\rho_\star\to 0$. In order to construct a theory that remains valid within a finite region as $\rho_\star\to 0$, we could fix the value of $K$ and use $K\gg 1$ as an additional expansion parameter, independent of $\rho_\star$.}
\end{definition}

Of course, we would also like to be able to describe orbits near the low-order rational surfaces, since this is where drift islands can appear. We consider this problem in the next subsection. 

\subsection{Drift surfaces in the presence of a single resonance}\label{subsec:oneres}

By modifying the method above, it is possible to calculate orbits in a region $\mathfrak{R}_1$ that contains a single low-order rational surface. Let this surface be $\psi_\mathrm{r}$, with $\iota(\psi_\mathrm{r}) = N/M$ as usual. The key idea here is that we can still eliminate most of the angle dependence from the Lagrangian by following the same steps as before. However, to avoid divergences, we must exclude all harmonics that are resonant at $\psi_\mathrm{r}$ from the Fourier expansion of $S^{(1)}$.

First, we split the perturbation as $V = V_\mathrm{r} + V_\mathrm{nr}^{<K} + {V_\mathrm{nr}^{\geq K}}$, where
\begin{equation}
        V_\mathrm{r} \coloneq  \sum_{k \in \mathbb{Z}}V_{kM\!,\? kN}\,\mathrm{e}^{\mathrm{i}k(M\theta-N\zeta)}
\end{equation}
contains all resonant harmonics of $V$, $V^{<K}_\mathrm{nr}$ contains all non-resonant harmonics with $|p|+|q|<K$, and $V^{\geq K}_\mathrm{nr}$ contains the non-resonant harmonics with $|p|+|q|\geq K$. As before, we choose $K$ large enough that $V^{\geq K}_\mathrm{nr}\sim \rho_\star^2\? \Psi_\mathrm{t}$. Then, the Lagrangian in \eqref{eq:Hamiltonian} becomes
\begin{equation}\label{eq:newnot}
    \mathcal{L}\equiv \mybar{\psi}_\mathrm{nr}\?\frac{\rmd\mybar{\theta}_\mathrm{nr}}{\rmd\zeta} - \chi + V_\mathrm{r} + V_\mathrm{nr}^{<K} - \bigl(\partial_\zeta S^{(1)}_\mathrm{nr} + \iota\? \partial_\theta S^{(1)}_\mathrm{nr}\bigr)\,.
\end{equation}
In moving from \eqref{eq:Hamiltonian} to \eqref{eq:newnot}, we have introduced new notation because we are now dealing with particles in $\mathfrak{R}_1$ instead of $\mathfrak{R}_0$. To avoid confusion, when we are working in $\mathfrak{R}_1$ we denote the generating function by $S^{(1)}_\mathrm{nr}$ and the corresponding barred coordinates by $\mybar{\psi}_\mathrm{nr}$ and $\mybar{\theta}_\mathrm{nr}$. To be explicit, $\mybar{\psi}_\mathrm{nr}$ and $\mybar{\theta}_\mathrm{nr}$ are defined by
\begin{subequations}
\begin{align}
    \psi &= \mybar{\psi}_\mathrm{nr} + \psi^{(1)}(\mybar{\psi}_\mathrm{nr}, \mybar{\theta}_\mathrm{nr}, \zeta)\,,\\
    \theta &= \mybar{\theta}_\mathrm{nr} + \theta^{(1)}(\mybar{\psi}_\mathrm{nr}, \mybar{\theta}_\mathrm{nr}, \zeta)\,,
\end{align}
\end{subequations}
where
\begin{subequations}\label{eq:nrtransforms}
\begin{align}
    \label{eq:transformpsi}\psi^{(1)}_\mathrm{nr} &\coloneq -\frac{v_\parallel}{\Omega}\bb{B} \0\con{\theta} + \partial_\theta S^{(1)}_\mathrm{nr}\,,\\
    \label{eq:transformtheta}\theta^{(1)}_\mathrm{nr} &\coloneq \frac{v_\parallel}{\Omega}\bb{B} \0\con{\psi} - \partial_\psi S^{(1)}_\mathrm{nr}\,,
\end{align}
\end{subequations}
analogous to \eqref{eq:irrationaltransform} and \eqref{eq:transformdefined}. This time, however, we choose
\begin{equation}\label{eq:Rnonres}
    S^{(1)}_\mathrm{nr} \coloneq \mspace{-8mu}\sum_{(p,q) \in \mathbb{I}^{<K}_\mathrm{nr}}\mspace{-8mu} \frac{\mathrm{i}(V_\mathrm{nr}^{<K})_{pq}}{q-\iota\? p}\,\mathrm{e}^{\mathrm{i}(p\theta-q\zeta)}\,,
\end{equation}
where ${\mathbb{I}^{<K}_\mathrm{nr} \coloneq \{ (p, q)\in \mathbb{Z}^2: |p|+|q| < K \text{ and } (p,q) \neq (kM,kN) \text{ for any }k\in\mathbb{Z}\}}$ contains only non-resonant harmonics. This choice of generating function satisfies
\begin{equation}\label{eq:Rnrdef}
    \partial_\zeta S^{(1)}_\mathrm{nr} + \iota\? \partial_\theta S^{(1)}_\mathrm{nr} = V^{<K}_\mathrm{nr}\,,
\end{equation}
so the Lagrangian \eqref{eq:newnot} becomes 
\begin{equation}\label{eq:resonantLagrangian}
    \mathcal{L}\equiv \mybar{\psi}_\mathrm{nr}\?\frac{\rmd\mybar{\theta}_\mathrm{nr}}{\rmd \zeta} - \chi + \sum_{k\in\mathbb{Z}} V_{kM\!,\? kN}\,\mathrm{e}^{\mathrm{i}k(M\bar{\theta}_\mathrm{nr} - N\zeta)}\,.
\end{equation}
We can write \eqref{eq:resonantLagrangian} in a simple form by introducing the $\zeta$-average
\begin{equation}\label{eq:flav}
	\langle \?  \ldots \?  \rangle_\zeta \coloneq \frac{1}{2\upi M} \oint (\ldots)\, \rmd \zeta\,,
\end{equation}
where we integrate over a curve of constant $\psi$ and $\eta$ as usual. This average is useful because it extracts Fourier modes with the helicity of the rational surface:
\begin{equation}
		\langle V \rangle_\zeta = \sum_{k\in\mathbb{Z}} V_{kM\!,\? kN}\,\mathrm{e}^{\mathrm{i}kM\eta} = V_\mathrm{r}\,.
\end{equation}
Then, using $\mybarup{\eta}_\mathrm{nr} \coloneq \mybar{\theta}_\mathrm{nr} - (N/M)\? \zeta$ as a coordinate instead of $\mybar{\theta}_\mathrm{nr}$, the Lagrangian \eqref{eq:resonantLagrangian} becomes
\begin{equation}\label{eq:finalLagrangian}
    \mathcal{L}\equiv \mybar{\psi}_\mathrm{nr}\?\frac{\rmd\mybarup{\eta}_\mathrm{nr}}{\rmd\zeta} - \mathcal{H}_{\rm r}(\mybar{\psi}_\mathrm{nr}, \mybarup{\eta}_\mathrm{nr})\,,
\end{equation}
where $\mathcal{H}_{\rm r} \coloneq \chi_\mathrm{r} - \langle V \rangle_\zeta$ and $\chi_\mathrm{r}$ is the helical flux introduced back in \eqref{eq:helicalflux}. This Lagrangian is the generalization of \eqref{eq:barLagrangian} to the case where there is a single low-order rational surface in the domain of interest.

In \eqref{eq:finalLagrangian}, the dependence on the fast angle $\zeta$ has been eliminated. Thus, in the presence of a single resonance, although we could not eliminate both angular coordinates as in \cref{subsec:weakresonances}, it is still possible to significantly simplify the Lagrangian. The $\zeta$ independence of this Lagrangian implies conservation of
\begin{equation}\label{eq:Hbar}
    \mathcal{H}_{\rm r}(\mybar{\psi}_\mathrm{nr},\mybarup{\eta}_\mathrm{nr}) = \int_{\psi_\mathrm{r}}^{\bar{\psi}_\mathrm{nr}}\biggl(\iota(\psi') -\frac{N}{M}\biggr)\,\rmd\psi' - \frac{1}{2\upi M}\frac{mc}{Ze}\oint v_\parallel\bh \0(\con{\zeta}+\iota\? \con{\theta})\,\rmd\zeta'\,.
\end{equation}
Here, as in the Lagrangian, all functions are evaluated at a $\psi$ coordinate of $\mybar{\psi}_\mathrm{nr}(\psi, \eta, \zeta)$ and an $\eta$ coordinate of $\mybarup{\eta}_\mathrm{nr}(\psi, \eta, \zeta)$. To express the conserved quantity \eqref{eq:Hbar} in the original guiding-centre coordinates, let $\mybarup{\chi}_{\rm r}(\psi, \eta, \zeta) \coloneq \mathcal{H}_{\rm r}(\mybar{\psi}_\mathrm{nr}(\psi, \eta, \zeta),\mybarup{\eta}_\mathrm{nr}(\psi, \eta, \zeta))$. Substituting the definitions of $\mybar{\psi}_\mathrm{nr}$ and $\mybarup{\eta}_\mathrm{nr}$ into \eqref{eq:Hbar} and Taylor expanding, we find
\begin{align}\label{eq:H}
    \mybarup{\chi}_{\rm r}(\psi,\eta,\zeta) &= \int_{\psi_\mathrm{r}}^{\psi}\biggl(\iota(\psi') - \frac{N}{M}\biggr)\,\rmd\psi' - \frac{1}{2\upi M}\frac{mc}{Ze}\oint v_\parallel\bh \0(\con{\zeta}+\iota\? \con{\theta})\rmd\zeta'\nn
    \al + \biggl(\iota(\psi) - \frac{N}{M}\biggr)\Bigl(\frac{mc}{Ze}v_\parallel\bh \0\con{\theta} -\partial_\theta S^{(1)}_\mathrm{nr}\Bigr) + O(\rho_\star^2\?  \Psi_\mathrm{t})\,,
\end{align}
where all functions on the right are now evaluated at $(\psi, \eta, \zeta)$. This conserved quantity $\mybarup{\chi}_{\rm r}$ determines the orbits of all passing particles in $\mathfrak{R}_1$, just as $\mybar{\psi}$ did for particles in $\mathfrak{R}_0$. Region $\mathfrak{R}_1$ contains both the low-order rational surface $\psi_\mathrm{r}$, around which the orbits trace out drift islands, and irrational surfaces away from $\psi_\mathrm{r}$, where the drift surfaces are intact (to this order). Both types of orbit are described by a single invariant $\mybarup{\chi}_{\rm r}$, whose definition \eqref{eq:H} is uniformly valid throughout $\mathfrak{R}_1$. 

In \cref{app:limits}, we prove that, in two opposite limits, \eqref{eq:H} is consistent with results from earlier in the paper. For particles close to $\psi_\mathrm{r}$, conservation of $\mybarup{\chi}_{\rm r}$ is equivalent to conservation of the transit adiabatic invariant, including the higher-order corrections derived in \cref{sec:higherorder}. For particles on irrational surfaces far from $\psi_\mathrm{r}$, conservation of $\mybarup{\chi}_{\rm r}$ is equivalent to conservation of $\mybar{\psi}$, defined by \eqref{eq:psistar} for particles in a region, $\mathfrak{R}_0$, of sufficiently irrational rotational transform.

In \cref{subsec:weakresonances}, we discussed a restriction on region $\mathfrak{R}_0$: it needed to remain much more than a drift-island width away from any low-order rational surfaces. Similar reasoning shows that $\mathfrak{R}_1$ must remain much more than a drift-island width away from any low-order rational surfaces other than $\psi_\mathrm{r}$, as shown in \crefsub{fig:regions}{b}.

These restrictions on $\mathfrak{R}_0$ and $\mathfrak{R}_1$ can only be satisfied if the drift-island width is much smaller than the distance between low-order rational surfaces. This condition may not be satisfied if $\rho_\star$ is insufficiently small; then, our perturbative construction would fail. Such a failure could be anticipated on physical grounds because, once the drift-island width is comparable to the distance between rational surfaces, drift-island overlap can occur. Overlapping islands are known to give rise to stochastic orbits \mycitep{Chirikov1979}, which cannot be described by conservation of an invariant such as $\mybar{\psi}$ or $\mybarup{\chi}_{\rm r}$.

%
%
\begin{figure}
\vspace{2mm}
\centering
\includegraphics[width=32pc]{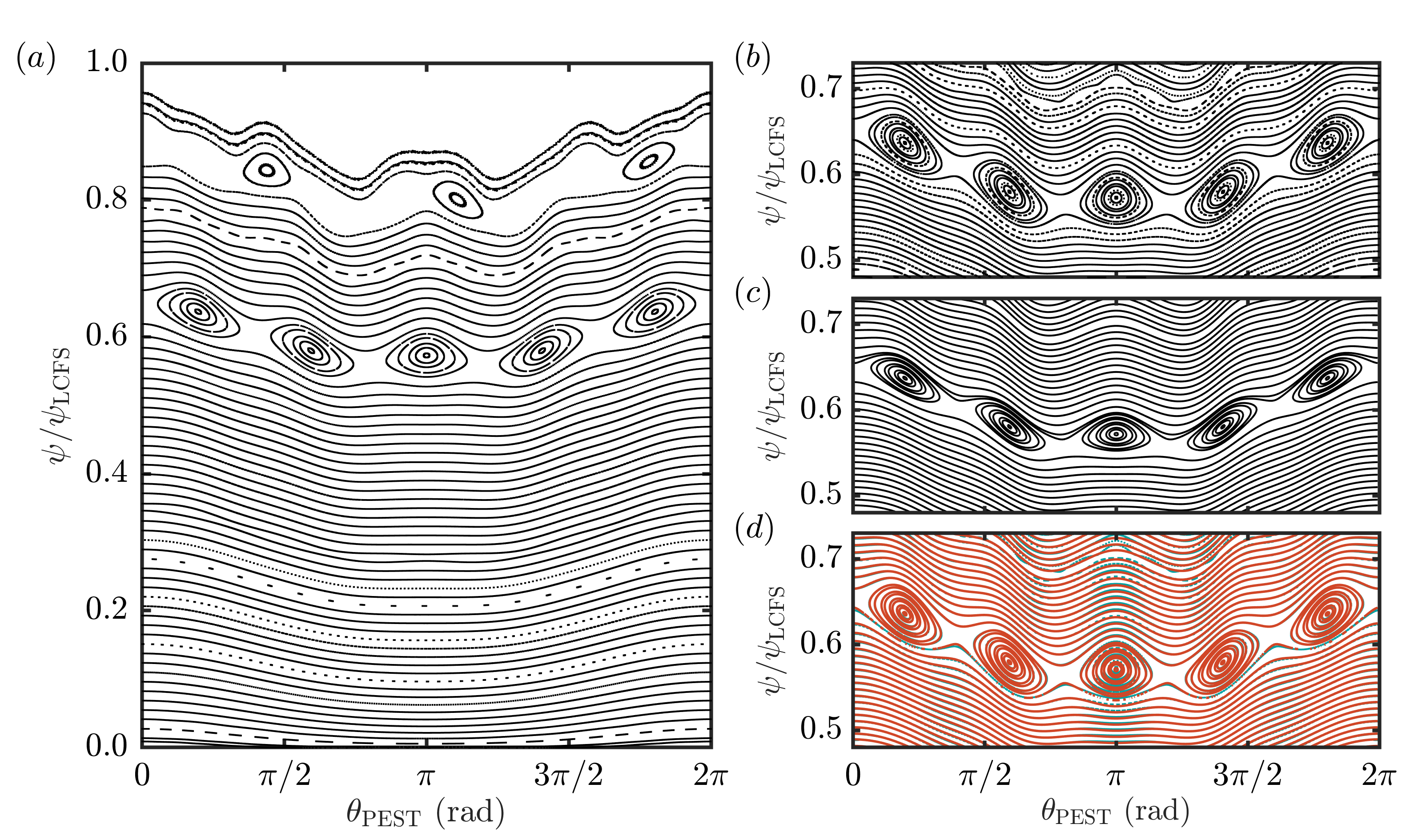}
    \caption{$(a)$ Poincar\'e plot, at cylindrical angle $\phi = 0$, of passing alpha orbits in a reactor-scale NCSX equilibrium, simulated using ASCOT5. The alphas have energy $\SI{3.5}{\mega\electronvolt}$, $\mu = 0$, and $\sigma = -1$. Multiple chains of drift islands are visible. $(b)$ Zoomed-in view of the $(3,5)$ drift-island chain. $(c)$ Level sets of $\mybarup{\chi}_{\rm r}(\psi,\eta,\zeta=0)$, computed using \eqref{eq:H} and plotted for particles in a region containing the $\iota = 3/5$ rational surface. $(d)$ Similar to $(c)$, except we plot the level sets of the higher-order invariant $\mathcal{H}_{2, \mathrm{r}}(\mybardown{\mybar{\psi}}_\mathrm{nr}(\psi,\eta,\zeta=0), \mybardown{\mybar{\eta}}_\mathrm{nr}(\psi,\eta,\zeta=0))$, defined by \eqref{eq:hoconserved}--\eqref{eq:hoS1}, in red. We also underlay the simulation data from $(b)$ in blue; the close agreement between the theory and simulation results is evident. In $(c)$ and $(d)$, the contour levels are equal to the average value of the plotted invariant for each particle in the ASCOT5 simulation shown in $(b)$, as in \cref{fig:LHD}.}
    \label{fig:NCSX}
\end{figure}

%
%
\begin{figure}
\vspace{4mm}
\centering
\includegraphics[width=\textwidth]{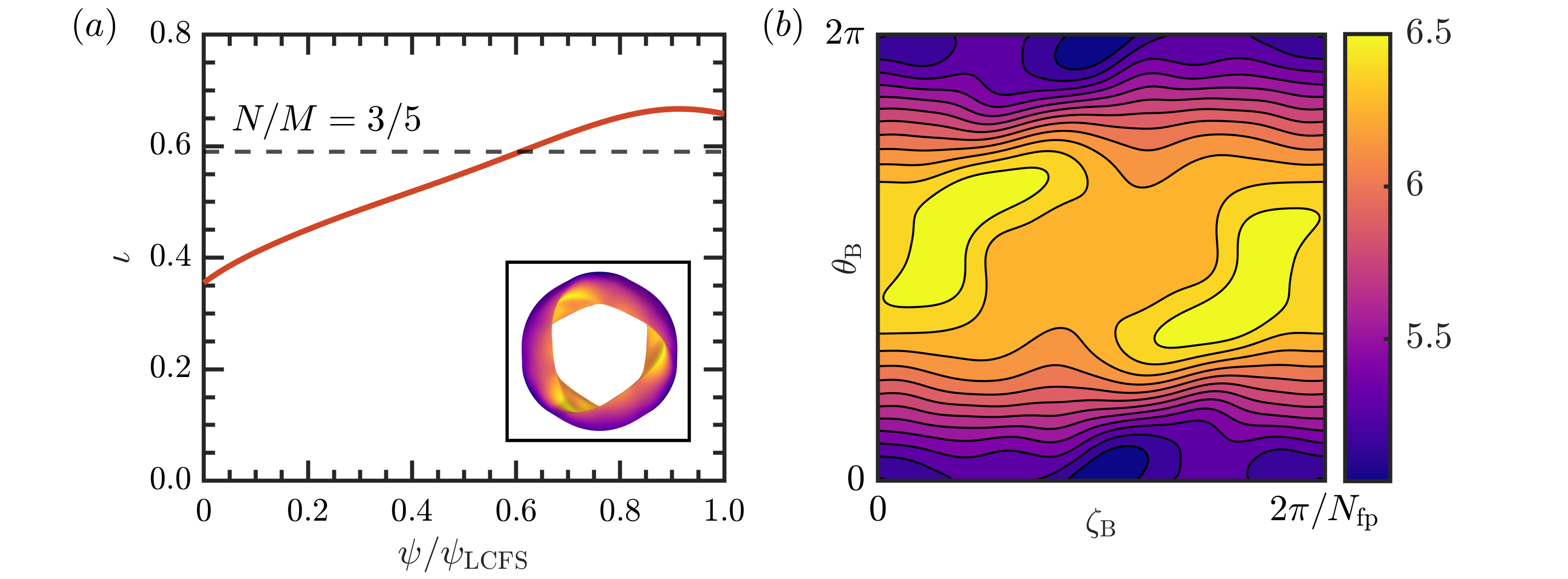}
    \caption{$(a)$ Rotational transform of the NCSX equilibrium we use to trace alpha orbits in \cref{fig:NCSX}. $(b)$ Plot of the magnetic field strength in Boozer coordinates on the rational surface with $\iota = N/M = 3/5$.}
    \label{fig:NCSXeqm}
\end{figure}

%
%
\begin{figure}
\vspace{2mm}
\centering
\includegraphics[width=32pc]{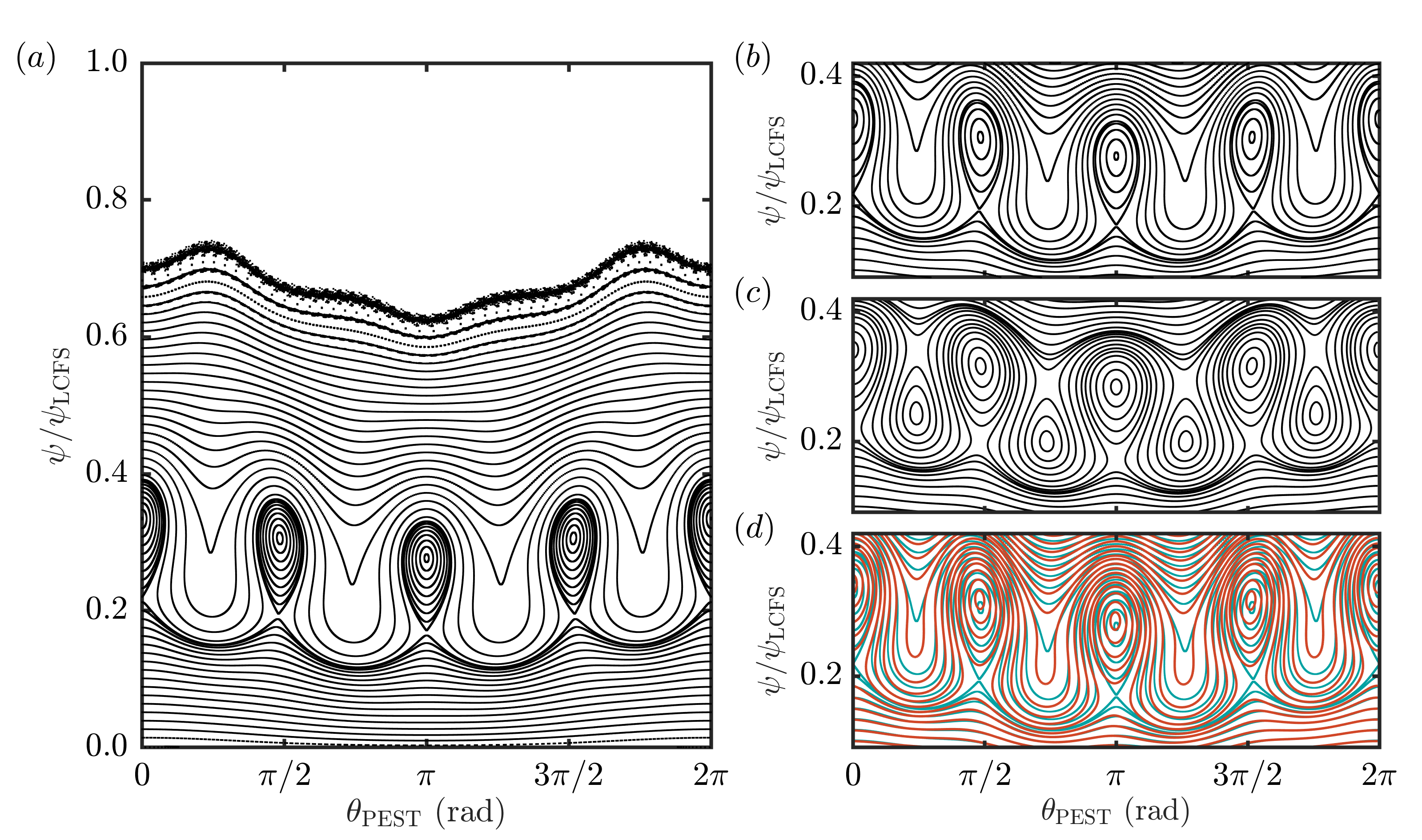}
    \caption{Similar to \cref{fig:NCSX} but for a reactor-scale WISTELL-A equilibrium. $(a)$ Poincar\'e plot, at cylindrical angle $\phi = 0$, for passing alpha particles with energy $\SI{3.5}{\mega\electronvolt}$, $\mu/\mathcal{E} = \SI{0.1425}{\per\tesla}$, and $\sigma = 1$. $(b)$ Zoomed-in view of the prominent drift-island chain. $(c)$ Level sets of $\mybarup{\chi}_{\rm r}(\psi,\eta,\zeta=0)$, plotted for particles in a region containing the $N=1$, $M=1$ resonance. $(d)$ Similar to $(c)$, except we plot the level sets of $\mathcal{H}_{2, \mathrm{r}}(\mybardown{\mybar{\psi}}_\mathrm{nr}(\psi,\eta,\zeta=0), \mybardown{\mybar{\eta}}_\mathrm{nr}(\psi,\eta,\zeta=0))$, in red. We also replot the simulation data from $(b)$ in blue to allow easy comparison. In $(c)$ and $(d)$, the contour levels are equal to the average value of the plotted invariant for each particle in the ASCOT5 simulation shown in $(b)$, as in \cref{fig:LHD}.}
    \label{fig:WISTELL}
\end{figure}

%
%
\begin{figure}
\vspace{4mm}
\centering
\includegraphics[width=\textwidth]{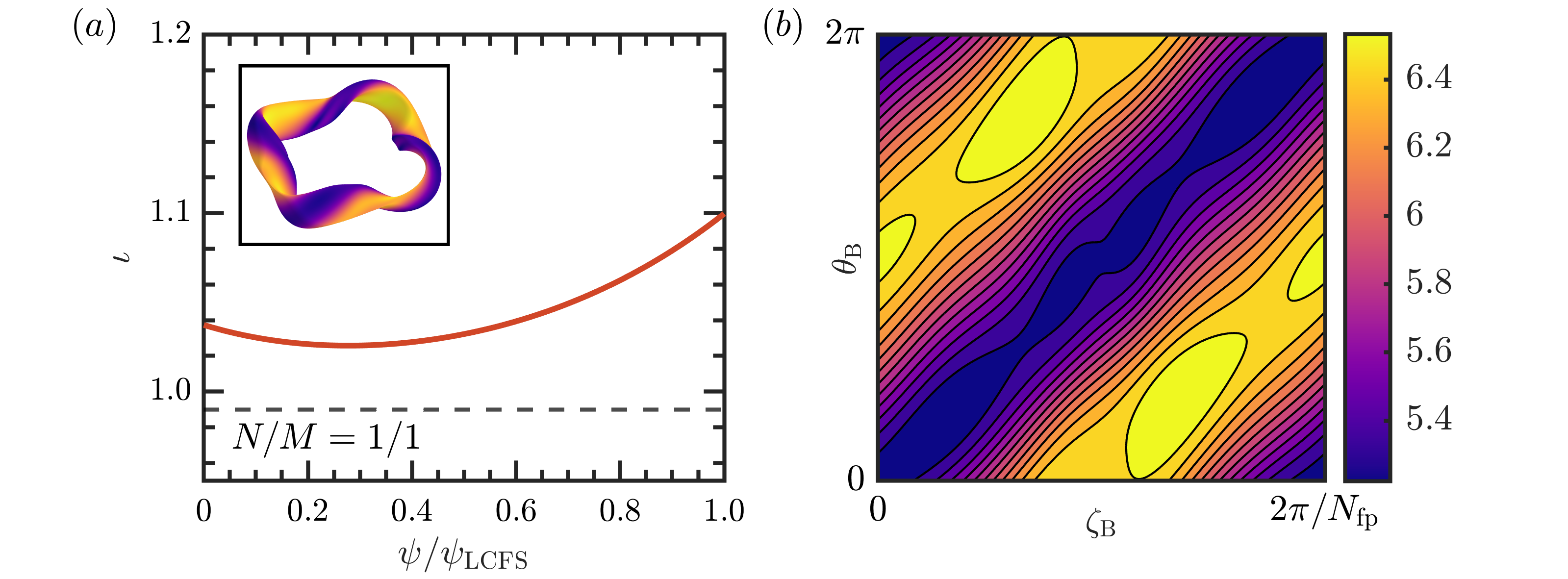}
    \caption{$(a)$ Rotational transform of the WISTELL-A equilibrium we use to trace alpha orbits in \cref{fig:WISTELL}. $(b)$ Plot of the magnetic field strength in Boozer coordinates on the flux surface $\psi/\psi_{\rm LCFS} = 0.3$.}
    \label{fig:WISTELLeqm}
\end{figure}

In \crefrange{fig:NCSX}{fig:WISTELLeqm}, we compare the analytical results of this subsection with ASCOT5 simulations. \Crefsuband{fig:NCSX}{a}{fig:NCSX}{b} show a Poincar\'e plot of simulated guiding-centre orbits of freely passing, $\SI{3.5}{\mega\electronvolt}$ alpha particles with $\mu = 0$ and $\sigma=-1$ in a reactor-scale NCSX equilibrium. The equilibrium was taken from the DESC examples directory; more information about it is provided in \cref{fig:NCSXeqm}. There is a rational surface with $\iota = N/M = 3/5$ in the plasma \Dash note that this rational surface has $N = N_{\rm fp}$, which, according to the discussion in \cref{subsec:resonance}, promotes large drift islands. In \crefsub{fig:NCSX}{c}, we have plotted the level sets of $\mybarup{\chi}_{\rm r}(\psi,\eta,\zeta=0)$, taking the low-order resonance in $\mathfrak{R}_1$ to be $(N,M) = (3,5)$. \Crefsub{fig:NCSX}{d} shows the level sets of a higher-order version of the $\mybarup{\chi}_{\rm r}$ invariant, which will be introduced in \cref{subsec:evenhigherorder}. We see that conservation of $\mybarup{\chi}_{\rm r}$ provides a very accurate prediction of the simulated alpha-particle orbits.

Similarly, \crefsuband{fig:WISTELL}{a}{fig:WISTELL}{b} show a Poincar\'e plot, constructed from an ASCOT5 simulation, for barely passing, $\SI{3.5}{\mega\electronvolt}$ alpha particles with $\mu/\mathcal{E} = \SI{0.1425}{\per\tesla}$ and $\sigma=1$ in a reactor-scale WISTELL-A quasi-helically symmetric equilibrium \mycitep{Bader2020}. This equilibrium was taken from the DESC examples directory. Despite the fact that the rotational-transform profile shown in \cref{fig:WISTELLeqm} avoids the low-order rational value $\iota = 1$, wide drift islands corresponding to an $N=M$ resonance are visible in the Poincar\'e plot. This occurs because the kinetic-rotational-transform profile of the co-passing orbits is lowered \Dash relative to the magnetic rotational-transform profile shown in \cref{fig:WISTELLeqm} \Dash by the tangential magnetic drift, as discussed in \cref{subsec:lowshear}. This allows the kinetic rotational transform $\mybarup{\iota}$ to pass through $\mybarup{\iota} = 1$; equivalently, it means the $N=M$ resonance shifts into the plasma. In \crefsub{fig:WISTELL}{c}, we plot contours of $\mybarup{\chi}_{\rm r}(\psi,\eta,\zeta=0)$, taking the low-order resonance in $\mathfrak{R}_1$ to be $(N,M) = (1,1)$. \Crefsub{fig:WISTELL}{d} shows contours of the higher-order version of $\mybarup{\chi}_{\rm r}$ that will be discussed in \cref{subsec:evenhigherorder}. Once again, the agreement between theory and simulations is excellent.

Before we move on to discuss higher-order corrections, we explain some interesting features of the drift islands in \cref{fig:WISTELL}. First, these drift islands are particularly large. This may be surprising, since \cref{fig:WISTELLeqm} shows that WISTELL\nobreakdash-A has a relatively small quasisymmetry error. One reason for their size is that these drift islands are situated near a local minimum of the rotational-transform profile (around $\psi/\psi_{\rm LCFS} = 0.3$; see \cref{fig:WISTELLeqm}). This means they are located in a region where the magnetic shear is small, and low shear tends to produce wider drift islands. A second reason why these islands are large is that they correspond to an $N = M = 1$ resonance. This resonance is very low order and has $\widetilde{N} = N_{\rm fp}$ (in the notation of \cref{subsec:resonance}), which promotes wide drift islands for the reasons set out in \cref{subsec:resonance}.

The drift islands in \cref{fig:WISTELL} also have an unusual shape. They do not resemble the phase portrait of a pendulum, as might be expected from the simplified lowest-order transit invariant formula \eqref{eq:lowshearI} that we derived for stellarators with $s\sim\rho_\star$ (this is the appropriate ordering to use here because the drift islands in \cref{fig:WISTELL} are not located close to their associated rational surface; in fact, there is no rational surface with $\iota = N/M = 1$ in the plasma). Their unusual shape is caused by the magnetic shear vanishing near the resonance, so the linear approximation ${\iota(\psi) - N/M \simeq \iota'_{\rm s}(\psi - \psi_{\rm s})}$ \Dash which we used to derive \eqref{eq:lowshearI} for drift islands near a shifted resonant surface $\psi_{\rm s}$ \Dash is not appropriate here. Nevertheless, provided this approximation is not used, the transit invariant \eqref{eq:invariant}, or the more complicated invariants plotted in \crefsuband{fig:WISTELL}{c}{fig:WISTELL}{d}, still correctly describe the shape of these drift islands.\footnote{More generally, when an integrable Hamiltonian map is perturbed, unusual island shapes like those of \cref{fig:WISTELL} are known to arise if the unperturbed map is a `nontwist' map \mycitep{dCN1996}.} The fact that $\iota'_{\rm{s}} \approx 0$ at the location of the drift islands means the usual island-width estimate $\Delta\psi \sim (\rho_\star\delta/s)^{1/2}\?\Psi_{\rm t}$ does not apply in this case. However, we can still estimate the island width using the method of \cref{subsec:orbitwidth} if we employ $\Delta\theta \sim \iota_{\rm s}'' (\psi - \psi_{\rm s})^2$, where $\iota_{\rm s}''\coloneq (\rmd^2\iota/\rmd\psi^2)(\psi = \psi_{\rm s})$, for the amount by which a near-resonant orbit fails to close after one transit. With this modification, we obtain a corrected island-width scaling of $\Delta\psi \sim (\rho_\star\delta\? \Psi_{\rm t} / \iota_{\rm s}'')^{1/3}$. The fact that this width scales with $\rho_\star^{1/3}$, instead of the typical scaling with $\sqrt{\rho_\star}$, is another way of understanding that the drift islands are wider if they arise near a local minimum of $\iota(\psi)$.

A final noteworthy feature of the drift islands in \cref{fig:WISTELL} is that they only exist for a very narrow range of pitch angles. This is because the pitch angle must be close to the trapped--passing boundary, where the tangential distance through which a particle drifts in a single transit is largest (this tangential distance diverges logarithmically as $\mu$ approaches the trapped--passing boundary). A large tangential drift is needed so that the kinetic-rotational-transform profile is lowered enough \Dash relative the the magnetic rotational-transform profile shown in \cref{fig:WISTELLeqm} \Dash to reach the low-order rational value $N/M = 1$. In the case shown in \cref{fig:WISTELL}, reducing $\mu$ by a few percent removes the drift islands, since $\mu$ is then too far from the trapped--passing boundary for the $N/M = 1$ resonance to be present in the plasma. The fact that wide drift islands can arise at extremely specific pitch angles suggests that a scan across many different pitch angles could be necessary to check that a given stellarator design does not contain any large drift islands like these.

\subsection{Higher-order Hamiltonian perturbation theory}\label{subsec:evenhigherorder}

To conclude this section, we explain how these Hamiltonian calculations may be systematically extended to higher order. We begin with the higher-order calculation of orbits in $\mathfrak{R}_0$. The Lagrangian \eqref{eq:barLagrangian} that we derived in \cref{subsec:weakresonances} for particles in $\mathfrak{R}_0$ takes the form
\begin{equation}\label{eq:R0W}
    \mathcal{L} \equiv \bigl(\mybar{\psi} + \Lambda_{\theta 2}(\mybar{\psi}, \mybar{\theta}, \zeta)\bigr)\frac{\rmd\mybar{\theta}}{\rmd \zeta} + \Lambda_{\psi 2}(\mybar{\psi}, \mybar{\theta}, \zeta)\,\frac{\rmd\mybar{\psi}}{\rmd\zeta} - \mybarup{\chi}(\mybar{\psi}) + \mathcal{H}_2(\mybar{\psi}, \mybar{\theta}, \zeta)\,,
\end{equation}
where $\Lambda_{\theta 2}$, $\Lambda_{\psi 2}$, and $\mathcal{H}_2$ are order $\rho_\star^2$ and contain all higher-order terms that we previously suppressed in this Lagrangian. These terms contain angle dependence which complicates the higher-order equations of motion.

We can now repeat the procedure that lead to \eqref{eq:R0W}. That is, we look for a near-identity change of coordinates from $(\mybar{\psi}, \mybar{\theta})$ to a new set $(\mybardown{\mybar{\psi}}, \mybardown{\mybar{\theta}})$, defined implicitly by
\begin{subequations}\label{eq:nearidentity2}
\begin{align}
    \mybar{\psi} &= \mybardown{\mybar{\psi}} + \psi^{(2)}(\mybardown{\mybar{\psi}}, \mybardown{\mybar{\theta}}, \zeta)\,,\\
    \mybar{\theta} &= \mybardown{\mybar{\theta}} + \theta^{(2)}(\mybardown{\mybar{\psi}}, \mybardown{\mybar{\theta}}, \zeta)\,,
\end{align}
\end{subequations}
where $\psi^{(2)}\sim \rho_\star^2\?  \Psi_\mathrm{t}$ and $\theta^{(2)}\sim\rho_\star^2$. We will choose $\psi^{(2)}$ and $\theta^{(2)}$ so that any dependence on $\mybardown{\mybar{\theta}}$ and $\zeta$ in the Lagrangian is removed \Dash or rather, pushed to even higher order. We will need to add the total $\zeta$ derivative of a new generating function $S^{(2)}\sim\rho_\star^2\?  \Psi_\mathrm{t}$ to achieve this. Then, with a suitable choice of $\psi^{(2)}$ and $\theta^{(2)}$, the Lagrangian can be put in the form
\begin{equation}\label{eq:secondorderL}
    \mathcal{L}\equiv \mybardown{\mybar{\psi}}\,\frac{\rmd\mybardown{\mybar{\theta}}}{\rmd\zeta} - \mybarup{\chi} + V_2 - \bigl(\partial_\zeta S^{(2)} + \iota\? \partial_\theta S^{(2)}\bigr)\,,
\end{equation}
for some $V_2(\mybardown{\mybar{\psi}}, \mybardown{\mybar{\theta}}, \zeta)\sim \rho_\star^2\? \Psi_\mathrm{t}$ whose precise form is not important for understanding the general method. In \eqref{eq:secondorderL}, we are suppressing angle-dependent terms of order $\rho_\star^3\?\Psi_{\rm t}$.

As before, in order to avoid resonant denominators in $\mathfrak{R}_0$, we can split $V_2 = V_2^{<K_2} + V_2^{\geq K_2}$, for some new cutoff $K_2$ that makes $V_2^{\geq K_2}\sim \rho_\star^3\? \Psi_\mathrm{t}$ (typically, $K_2$ will be larger than the cutoff $K$ that we introduced at the previous order). Then, we choose $S^{(2)}$ to solve the magnetic differential equation
\begin{equation}
    \partial_\zeta S^{(2)} + \iota\?  \partial_\theta S^{(2)} =  V_2^{<K_2} - \bigl\langle V_2^{<K_2} \bigr\rangle\,.
\end{equation}
We can solve for $S^{(2)}$ as a Fourier series as long as $\mathfrak{R}_0$ does not contain any resonances with harmonics of $V_2^{<{K_2}}$. The final result is a Lagrangian of the form
\begin{equation}\label{eq:barbarLag}
    \mathcal{L}\equiv \mybardown{\mybar{\psi}}\,\frac{\rmd\mybardown{\mybar{\theta}}}{\rmd\zeta} - \mybardown{\mybarup{\chi}}(\mybardown{\mybar{\psi}})\,,
\end{equation}
where $\mybardown{\mybarup{\chi}}\coloneq \mybarup{\chi} - \langle V_2 \rangle$. The equations of motion tell us that $\mybardown{\mybar{\psi}}(\psi, \theta, \zeta)$ is approximately conserved; its $\zeta$ derivative is $\rmd\mybardown{\mybar{\psi}}/\rmd\zeta\sim \rho_\star^3\?  \Psi_\mathrm{t}$, so $\mybardown{\mybar{\psi}}$ is conserved more accurately than the quantity $\mybar{\psi}$ that we constructed at first order in \eqref{eq:psistar}. Therefore, $\mybardown{\mybar{\psi}}(\psi, \theta, \zeta) = \text{const.}$ is a more accurate equation for the drift surfaces of passing particles in a region of sufficiently irrational rotational transform.

This procedure can be iterated to obtain more and more accurate equations for the drift surfaces. However, it cannot be continued indefinitely. Eventually, for any given $\mathfrak{R}_0$, we will reach an order at which the angle-dependent perturbation $V_n\sim\rho_\star^n\?  \Psi_\mathrm{t}$ in the Lagrangian contains a harmonic that resonates with a rational surface in $\mathfrak{R}_0$ and has an amplitude of order $\rho_\star^n\?\Psi_\mathrm{t}$. Such a harmonic is too large to be `hidden' in a higher-order remainder term. From this point onwards, we need to use the approach of \cref{subsec:oneres} to describe motion in a region containing a single resonant surface. If we wish to keep going to higher order beyond this point, then we need to know how to continue the calculation of particle orbits in $\mathfrak{R}_1$ to higher order.\footnote{For completeness, we point out another issue that may be encountered at higher orders. At each order, we calculate the corrections $\psi^{(n)}$, $\theta^{(n)}$, and $S^{(n)}$, which are assumed to be smaller, by a factor of $\rho_\star \ll 1$, than $\psi^{(n-1)}$, $\theta^{(n-1)}$, and $S^{(n-1)}$, respectively. These corrections typically contain coefficients that grow rapidly with $n$. Therefore, $n$ will eventually become large enough that the $n$th-order corrections are no smaller than the corrections from the previous order, violating our orderings.}

The higher-order calculation of orbits in $\mathfrak{R}_1$ proceeds in a similar manner: we iteratively look for near-identity coordinate transformations that simplify the Lagrangian. The difference is that we look for transformations that eliminate all angle dependence except for dependence on the resonant combination $\eta = \theta - (M/N)\? \zeta$. At every order, we will obtain an integrable Lagrangian of the form \eqref{eq:finalLagrangian}, which gives a higher-order conserved quantity generalizing \eqref{eq:H}. 

The calculation of orbits in $\mathfrak{R}_1$ is carried out to next order in \cref{app:higherorder}. In this calculation, we neglect the order-$\rho_\star^2$ correction to Littlejohn's guiding-centre Lagrangian \eqref{eq:phasespaceL}. Neglecting this term means we are calculating higher-order corrections to the orbit that would be observed in a guiding-centre code, such as ASCOT5, that uses equations derived from the usual guiding-centre Lagrangian (without higher-order corrections). The results of this calculation are summarized below.

The higher-order conserved quantity can be expressed in the form
\begin{equation}\label{eq:hoconserved}
    \mathcal{H}_{2, \mathrm{r}}(\mybardown{\mybar{\psi}}_\mathrm{nr}, \mybardown{\mybar{\eta}}_\mathrm{nr})\coloneq \chi_\mathrm{r}(\mybardown{\mybar{\psi}}_\mathrm{nr}) - \langle V \rangle_\zeta(\mybardown{\mybar{\psi}}_\mathrm{nr}, \mybardown{\mybar{\eta}}_\mathrm{nr}) - \langle V_2 \rangle_\zeta(\mybardown{\mybar{\psi}}_\mathrm{nr}, \mybardown{\mybar{\eta}}_\mathrm{nr})=\text{const.}\,,
\end{equation}
where $\mybardown{\mybar{\eta}}_\mathrm{nr} = \mybardown{\mybar{\theta}}_\mathrm{nr} - (N/M)\? \zeta$ and $V_2$ is defined by
\begin{equation}\label{eq:hoV1}
    V_2 \coloneq V^{\geq K}_\mathrm{nr} + \bigl(\psi^{(1)}_\mathrm{nr}\partial_\psi V_\mathrm{r} + \theta^{(1)}_\mathrm{nr}\partial_\theta V_\mathrm{r}\bigr) + \frac{1}{2}\iota' \bigl(\psi^{(1)}_\mathrm{nr}\bigr)^{2} - \psi^{(1)}_\mathrm{nr}\bigl(\partial_\zeta \theta^{(1)}_\mathrm{nr} + \iota\? \partial_\theta \theta^{(1)}_\mathrm{nr}\bigr)\,.
\end{equation}
The new coordinates $\mybardown{\mybar{\psi}}_\mathrm{nr}$ and $\mybardown{\mybar{\eta}}_\mathrm{nr}$ appearing in \eqref{eq:hoconserved} are defined by
\begin{subequations}
\begin{align}
    \mybar{\psi}_{\rm nr} &= \mybardown{\mybar{\psi}}_{\rm nr} + \psi^{(2)}_{\rm nr}(\mybardown{\mybar{\psi}}_{\rm nr}, \mybardown{\mybar{\theta}}_{\rm nr}, \zeta)\,,\\
    \mybar{\theta}_{\rm nr} &= \mybardown{\mybar{\theta}}_{\rm nr} + \theta^{(2)}_{\rm nr}(\mybardown{\mybar{\psi}}_{\rm nr}, \mybardown{\mybar{\theta}}_{\rm nr}, \zeta)\,,
\end{align}
\end{subequations}
where
\begin{subequations}\label{eq:hocoords1}
\begin{align}
    \psi^{(2)}_\mathrm{nr} &\coloneq \psi^{(1)}_\mathrm{nr}\Bigl[\partial_\theta \Bigl( \frac{v_\parallel}{\Omega}\bb{B} \0\con{\psi} \Bigr) - \partial_\psi\Bigl( \frac{v_\parallel}{\Omega}\bb{B} \0\con{\theta} \Bigr) \Bigr] + \theta^{(1)}_\mathrm{nr}\partial_\theta\psi^{(1)}_\mathrm{nr} + \partial_\theta S^{(2)}_\mathrm{nr}\,,\\
    \theta^{(2)}_\mathrm{nr} &\coloneq \theta^{(1)}_\mathrm{nr}\Bigl[\partial_\theta \Bigl( \frac{v_\parallel}{\Omega}\bb{B} \0\con{\psi} \Bigr) - \partial_\psi\Bigl( \frac{v_\parallel}{\Omega}\bb{B} \0\con{\theta} \Bigr) \Bigr] - \theta^{(1)}_\mathrm{nr}\partial_\psi\psi^{(1)}_\mathrm{nr} - \partial_\psi S^{(2)}_\mathrm{nr}\,,
\end{align}
\end{subequations}
and the generating function is
\begin{equation}\label{eq:hoS1}
    S^{(2)}_\mathrm{nr} \coloneq \mspace{-8mu}\sum_{(p,q) \in \mathbb{I}^{<K_2}_\mathrm{nr}}\mspace{-8mu} \frac{\mathrm{i}(V_{2,\mathrm{nr}}^{<K_2})_{pq}}{q-\iota\? p}\,\mathrm{e}^{\mathrm{i}(p\theta-q\zeta)}\,,
\end{equation}
for $V_{2,\mathrm{nr}}^{<K_2}$ defined by $V_2 = \langle V_2 \rangle_\zeta + V_{2,\mathrm{nr}}^{<K_2} + V_{2,\mathrm{nr}}^{\geq K_2}$.

We compare this higher-order theory with ASCOT5 simulations in \crefsuband{fig:NCSX}{d}{fig:WISTELL}{d}. The agreement between the theory and simulations is striking, and consistent across all orbits in all equilibria that we have examined. Thus, perturbative calculations based on a asymptotic expansion in $\rho_\star$ are able to describe the orbits of passing alpha particles in reactor-scale stellarators with impressive accuracy, even at $\SI{3.5}{\mega\electronvolt}$.

\subsection{Discussion}

The calculation in this section was motivated, in part, by the need to find a clear criterion for which flux surfaces count as `rational' for the purposes of calculating passing-particle orbits; that is, around which rational surfaces must particle orbits be described using the transit adiabatic invariant? There are infinitely many rational surfaces, so it cannot be the case that we can choose any one of them to be $\psi_\mathrm{r}$.

The answer obtained in this section is that the transit adiabatic invariant describes particle orbits around those rational surfaces that resonate with low-order harmonics of ${V = (mc/Ze)(v_\parallel/\bh \0\bb{\nabla}\zeta)}$. These harmonics depend on the velocity-space coordinates $\mathcal{E}$ and $\mu$, so it may be possible to treat a flux surface as `irrational' for one particle despite the fact that it must be treated as `rational' for another. As we continue the calculation to higher order in $\rho_\star\ll 1$, the set of rational surfaces that give rise to a resonant denominator gets larger and larger; higher-order calculations are able to detect smaller and smaller drift islands. Therefore, the answer to which flux surfaces count as `rational' also depends on the order of accuracy of our calculation. This makes sense physically: at higher order, we compute the particle orbits more accurately and we can, therefore, follow a given particle for a longer time. As a result, higher-order calculations can detect orbits that close on themselves after longer times, which occur on higher-order rational surfaces.

\section{Conclusions}\label{sec:summary}

In this paper, we have calculated the guiding-centre orbits of passing particles in a general stellarator. These orbits contain drift islands around rational flux surfaces, as observed in recent simulations \mycitep{White2022, White2022b, White2025}. We derived simple equations for the drift-island shape and used them to investigate how the stellarator magnetic field might be designed to ensure that these islands are small. Our key findings are as follows.

\begin{enumerate}

    \item We characterized stellarator magnetic fields in which the drift-island width vanishes for all passing particles, finding that they must be \emph{cyclometric} (see \cref{subsec:cyclometry}). Cyclometry is similar to the well-known \citet{Cary1997} condition for omnigeneity, but is less restrictive. Therefore, to remove a set of drift islands around a rational surface, optimizing for cyclometry on that surface may be more compatible with other objectives than simply optimizing for better~omnigeneity.

    \item The drift-island width usually scales as $w \sim (\rho_\star \delta /s)^{1/2}\?a$, where $\rho_\star$ is the energetic-particle gyroradius divided by a typical length scale of the field, $\delta$ is the deviation from cyclometry, $s$ is the magnetic shear, and $a$ is the minor radius. Therefore, wide drift islands may arise in stellarators that are insufficiently cyclometric and have low magnetic shear. The drift islands can shift away from their associated rational surface by a radial distance $\sim\!(\rho_\star/s)\?a$. Therefore, when there is a low-order rational surface at the plasma edge, as is the case in stellarators with an island divertor, wide drift islands may shift into the plasma for either co- or counter-passing alphas.

    \item Passing particles near a low-order rational surface conserve the \emph{transit adiabatic invariant}, defined in \eqref{eq:invariant}. We calculated higher-order corrections to the transit invariant, which become more important at higher energies. The resulting theory agrees extremely well with guiding-centre and full-orbit simulations conducted using ASCOT5. This suggests that the transit invariant and orbit-averaged guiding-centre equations can accurately describe the motion of passing alpha particles in reactor-scale stellarators, even at $\SI{3.5}{\mega\electronvolt}$.

    \item We showed that the orbits of passing particles in a general stellarator, on both rational and irrational surfaces, can be systematically calculated using Hamiltonian perturbation theory. These calculations match simulations almost perfectly (\eg \cref{fig:NCSX}), indicating that passing alpha orbits in stellarator reactors can be modelled using an asymptotic expansion in $\rho_\star\ll 1$, even at the birth energy.
\end{enumerate}

\begin{acknowledge}
We are grateful to Richard~Nies, Xu~Chu and Matt~Kunz for many helpful conversations. 
\end{acknowledge}

\begin{funding}
This manuscript has been authored by Princeton University/Princeton Plasma Physics Laboratory under contract number DE\nobreakdash-AC02\nobreakdash-09CH11466 with the U.S. Department of Energy. The United States Government retains a non-exclusive, paid-up, irrevocable, world-wide licence to publish or reproduce the published form of this manuscript, or allow others to do so, for United States Government purpose(s). This research was supported in part by Grants No. PID2021\nobreakdash-123175NB\nobreakdash-I00 and PID2024\nobreakdash-155558OB\nobreakdash-I00, funded by Ministerio de Ciencia, Innovaci\'on y Universidades/Agencia Estatal de Investigaci\'on/10.13039/501100011033 and by ERDF/EU.
\end{funding}


\appendix

\section{Comparison of guiding-centre and full-orbit simulations}\label{app:GCvsFO}

\begin{figure}
\vspace{3mm}
\centering
\includegraphics[width=32pc]{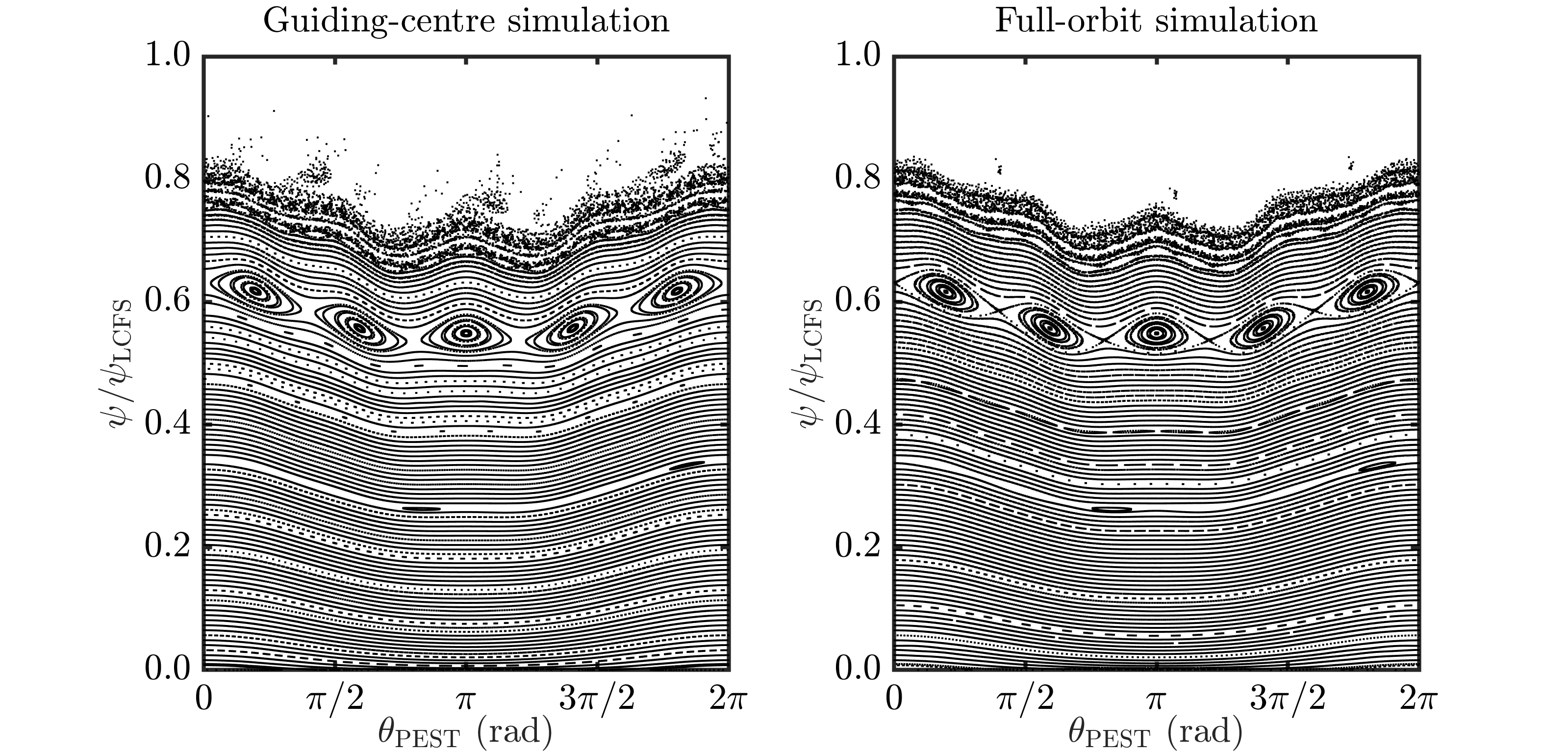}
    \caption{Comparison of guiding-centre and full-orbit simulations for passing alpha particles with energy $\SI{3.5}{\mega\electronvolt}$, $\mu/\mathcal{E} = \SI{0.096}{\per\tesla}$, and $\sigma = 1$, in a reactor-scale NCSX equilibrium. We provide more detail about this equilibrium in \cref{fig:NCSXeqm}.}
    \label{fig:GCFO1}
\end{figure}
\begin{figure}
\vspace{1mm}
\centering
\includegraphics[width=32pc]{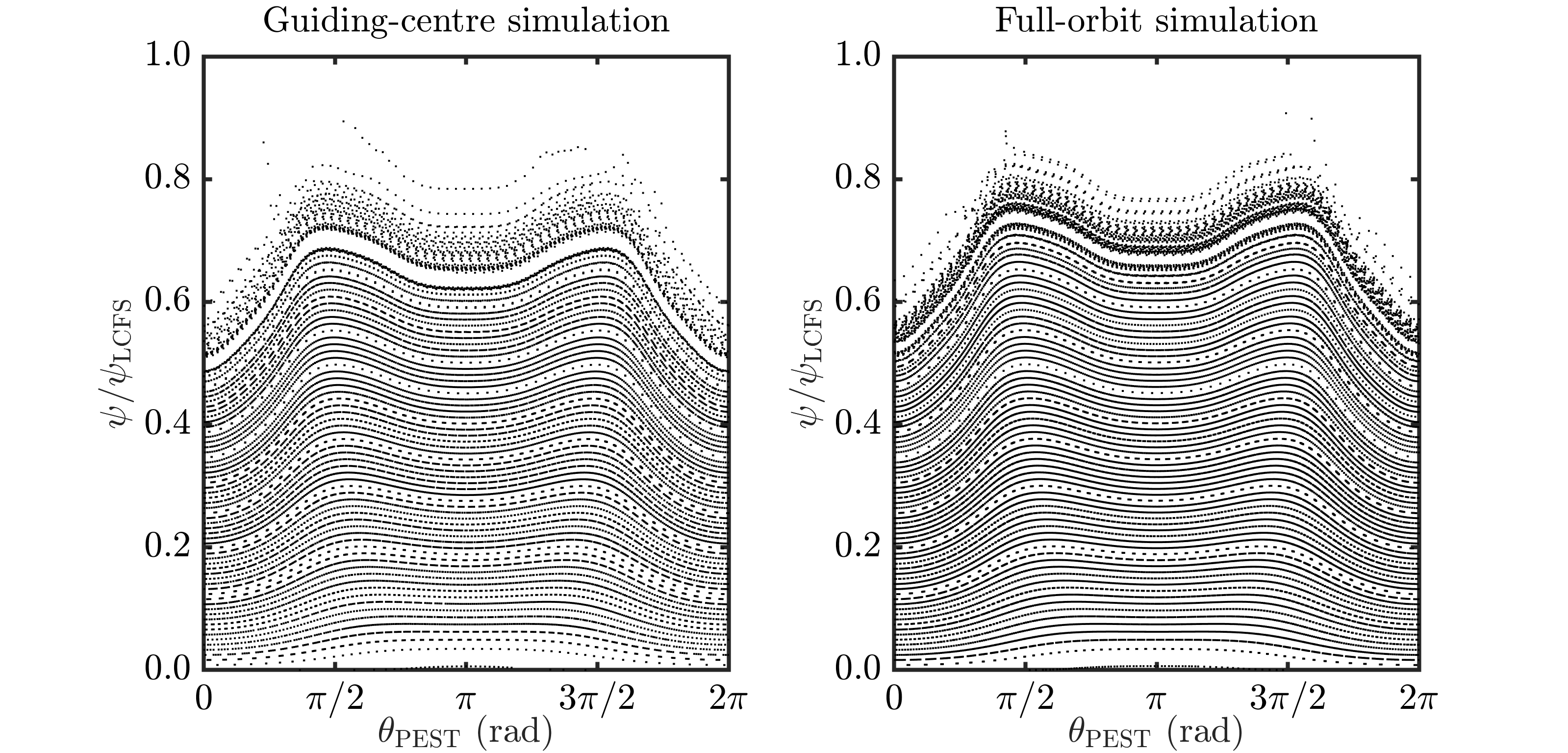}
    \caption{Comparison of guiding-centre and full-orbit simulations for trapped alpha particles with energy $\SI{3.5}{\mega\electronvolt}$ and $\mu/\mathcal{E} = \SI{0.177}{\per\tesla}$, in a CIEMAT-QI equilibrium with $N_{\rm fp} = 4$ and $\beta \sim 4\%$. CIEMAT-QI is a modern quasi-isodynamic stellarator design \protect{\citep{Sanchez2023, Velasco2023, Garcia2024}}; we used it for the trapped-particle comparison because trapped alphas are rapidly lost from the other, less well-optimized reactor-scale equilibria considered in this article. To construct these Poincar\'e plots for trapped particles, we recorded a point whenever the guiding centre crossed the $\phi = \upi/4$ plane ($\phi$ is cylindrical angle; this plane is roughly midway between the particles' bounce points) with $\vp > 0$.}
    \label{fig:GCFO2}
\end{figure}

To verify that the guiding-centre model accurately describes the orbits of alpha particles in a stellarator reactor, we compared some of our guiding-centre simulations with full-orbit ASCOT5 simulations. This is particularly important in light of recent concerns that the guiding-centre model may not be valid for energetic alphas \mycitep{Burby2025}. \crefand{fig:GCFO1}{fig:GCFO2} show side-by-side Poincar\'e plots created from the guiding-centre and full-orbit simulations. To construct the Poincar\'e plot in the full-orbit cases, we used ASCOT5 to compute the coordinates of the guiding centre from the particle coordinates at each timestep and recorded a point in the plot whenever the guiding centre crossed the Poincar\'e plane. These figures show excellent agreement between the guiding-centre and full-orbit simulations. The agreement was equally good in other equilibria and for other particle energies and pitch angles (for passing particles and trapped particles away from the trapped-passing boundary) that we checked. 

These results indicate that observed discrepancies between the alpha-particle losses in guiding-centre and full-orbit simulations \mycitep[\eg]{deAssuncao2023, Rodrigues2024, Carbajal2025} are not caused by breakdown of the perturbative guiding-centre expansion. Instead, our results support the explanation proposed by \citet{Rodrigues2024} for the primary cause of these discrepancies: full-orbit simulations should lose additional particles through the last closed flux surface (LCFS) because the guiding centre of a particle can remain inside the LCFS even though the particle itself crosses the LCFS as it completes its gyro-orbit. The additional lost particles are those whose orbits carry their guiding centre within roughly a gyroradius of the LCFS; note that such particles can be born on flux surfaces far from the edge if their orbit width is large. Although the number of alpha particles that satisfy this condition is small when $\rho_\star\ll 1$ (relative to the total number of alphas in the device), the alpha losses could still be enhanced by a considerable factor in cases where the losses in a guiding-centre simulation are very small to begin with, as might be the case for a well-optimized stellarator.

\section{Relaxing the assumption of flux surfaces}\label{app:magneticislands}

Throughout the paper, we assume that the magnetic field possesses nested toroidal flux surfaces. The reader may be concerned that this assumption is inappropriate for particles near rational surfaces, since rational surfaces can easily be broken into magnetic islands by small magnetic perturbations. In this appendix, we argue that the assumption of nested flux surfaces is not a problem for sufficiently energetic particles.

Our orbit calculations remain accurate in the presence of magnetic islands around the rational surface $\psi_\mathrm{r}$ provided the magnetic-island width is much smaller than the drift-island width. The physical reason is that particles circulating around a drift island spend most of their orbit in regions where the flux surfaces are intact. To make this argument more rigorous, we proceed to explain how our calculations can be modified to include the effect of magnetic islands whose width is comparable to the drift-island width.

Suppose the magnetic field is $\bb{B} = \bb{B}_0 + \bb{B}_1$, where $\bb{B}_0$ possesses nested toroidal flux surfaces and $\bb{B}_1$ is a small island-generating perturbation. To make the magnetic-island width comparable to the drift-island width, we order $ \bb{B}_1 \sim \rho_\star \bb{B}_0$. Then, for particles a distance $\psi - \psi_{\rm r}\sim (\rho_\star/s)^{1/2}\,\Psi_\mathrm{t}$ from the rational surface, the equations of motion for the guiding centre, \eqref{eq:guidingcentre2}, become
\begin{subequations}\label{eq:dmislands}
\begin{align}
    \frac{\rmd\psi}{\rmd t} &= v_\parallel\frac{\bb{B}_1}{B_0}\bb{\cdot}\bb{\nabla}\psi + \bb{v}_\mathrm{d}\bb{\cdot}\bb{\nabla}\psi\,,\\
    \frac{\rmd\eta}{\rmd t} &= v_\parallel\bh\bb{\cdot}\bb{\nabla}\eta\,,\\
    \frac{\rmd\zeta}{\rmd t} &= v_\parallel\bh\bb{\cdot}\bb{\nabla}\zeta\,,
\end{align}
to leading order. Here, the magnetic coordinates $(\psi,\theta,\zeta)$ are constructed from the unperturbed field $\bb{B}_0$, as are $B_0 = |\bb{B}_0|$,  $\bh = \bb{B}_0/B_0$, and
\begin{align}
    &v_\parallel(\bb{x},\mathcal{E},\mu,\sigma) = \sigma\sqrt{2\?(\mathcal{E}-\mu B_0)}\,,\\
    &\bb{v}_\mathrm{d}(\bb{x},\mathcal{E},\mu) = \frac{v_\parallel}{B_0}\,\bb{\nabla}\bb{\times}\biggl(\frac{v_\parallel\bb{B}_0}{\Omega_0}\biggr)\,,
\end{align}
\end{subequations}
where $\Omega_0 = ZeB_0/mc$. To this order, the same equations would arise if the guiding centre moved along field lines of
\begin{equation}\label{eq:Bstar2}
	\bb{B}^\star = \bb{B}_0 + \underbrace{\bb{B}_1}_{\sim\rho_\star B_0} + \underbrace{\bb{\nabla}\bb{\times}\biggl( \frac{v_\parallel\bb{B}_0}{\Omega_0} \biggr)}_{\sim\rho_\star B_0}\,.
\end{equation}
We see that the drift-island-generating perturbation in $\bb{B}^\star$ now consists of two parts: a part which is simply the magnetic perturbation $\bb{B}_1$, and a part due to the magnetic drift caused by $\bb{B}_0$.

We can calculate the drift-island shape from \eqref{eq:dmislands} by following the same steps as in \cref{sec:adiabaticinvariant}. The resulting island-shape is the same as \eqref{eq:simpleinvariant} but with $I_\mathrm{r}(\eta,\mathcal{E},\mu)$ replaced by
\begin{equation}\label{eq:js2}
	I^\mathrm{dm}_\mathrm{r}(\eta,\mathcal{E},\mu,\sigma) \coloneq \frac{Ze\sigma}{mc}\oint_\mathrm{r}\, \bb{A}_1\bb{\cdot}\partial_\zeta\bb{x}\, \rmd \zeta + \oint_\mathrm{r}\, |v_\parallel|\?\bh\bb{\cdot}\partial_\zeta\bb{x}\, \rmd \zeta\,,
\end{equation}
where $\bb{B}_1 = \bb{\nabla}\bb{\times}\bb{A}_1$ . Therefore, the particle orbit traces out islands whose shape is determined by a combination of drifts and the magnetic perturbation; we call these `drift-magnetic' islands. Higher-order corrections to the shape of the drift-magnetic islands may be derived using the methods of \cref{sec:higherorder} and \cref{sec:irrationalsurfaces}.

Using \eqref{eq:js2}, we can investigate when it is legitimate to assume that the field has nested toroidal flux surfaces for the purposes of calculating particle orbits around a rational surface $\psi_\mathrm{r}$. This assumption is legitimate when the second term in \eqref{eq:js2} dominates the dependence of $I^\mathrm{dm}_\mathrm{r}$ on $\eta$. This occurs when ${|\bb{B}_1|\ll \rho_\star |\bb{B}_0|}$ or, more precisely, when the harmonics of $\bh\?\0\bb{A}_1/\?\bh\?\0\grad{\zeta}$ that are resonant at $\psi_\mathrm{r}$ are much smaller than the corresponding harmonics of $(mc/Ze)(v_\parallel/\bh\0\grad{\zeta})$.

\section{Proof that field-strength contours cannot intersect field lines tangentially in cyclometric magnetic fields}\label{app:topology}

In \cref{subsec:bulk}, we state that, if the magnetic field is cyclometric on a rational surface $\psi_\mathrm{r}$, then there cannot be tangential intersections between field-strength contours and magnetic field lines on $\psi_\mathrm{r}$. In this appendix, we provide a proof of this claim, assuming the field strength is a smooth, well-behaved function.

We will use the following coordinates. All quantities will be evaluated at the rational surface $\psi_\mathrm{r}$; therefore, we suppress any $\psi$ arguments. We describe position on $\psi_\mathrm{r}$ using coordinates $\eta$ and $l$, where $l$ is arc length along the field lines. We use $\partial_l$ and $\partial_\eta$ to denote partial derivatives with respect to one of $l$ or $\eta$, respectively, with the other held fixed. 

\begin{figure}
\vspace{4mm}
\centering
\includegraphics[width=\textwidth]{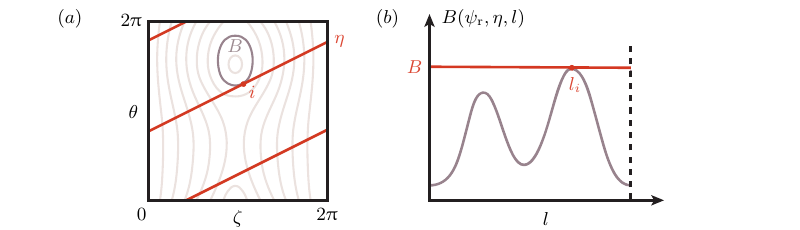}
    \caption{$(a)$~Plot, in straight-field-line coordinates, of a field line (shown in red) that has a tangential intersection with a field-strength contour. At the intersection point $i$, we must have $\partial_l B = 0$. $(b)$~Sketch of the field strength as a function of distance along this field line.}
    \label{fig:tangential}
\end{figure}

Suppose a field-strength contour and a field line intersect tangentially, as shown in \cref{fig:tangential}. At the intersection point, we have (1) $\partial_l B = 0$ and, provided the intersection point does not happen to be a stationary point of $B$ on the flux surface, (2) $\partial_\eta B \neq 0$. We will prove that, in a cyclometric field, $\partial_\eta B = 0$ at any point where $\partial_l B = 0$. This means there cannot be tangential intersections between field-strength contours and field lines in a cyclometric field, since it is not possible to have tangential intersections only at the stationary points of $B$.

To prove that $\partial_\eta B = 0$ wherever $\partial_l B = 0$, our strategy will be as follows. We will consider two rational-surface field lines labeled by $\eta = \eta_*$ and $\eta = \eta_*+\delta\eta$, where $|\delta \eta|\ll 1$, as well as two values attained by the magnetic field strength on these field lines, namely $B = B_*$ and $B = B_*+\delta B$, where $\delta B$ is small. We will derive an expression for ${\delta L \coloneq L(\eta_*+\delta\eta, B_*+\delta B) - L(\eta_*, B_*+\delta B)}$. In a cyclometric field, $\delta L$ must be independent of $\delta \eta$, for all $\delta B$. We will find that this implies $\partial_\eta B = 0$ at each point where $\partial_l B = 0$ along $\eta_*$.

First, we introduce some helpful notation. Along any rational-surface field line $\eta$, there will be a number of points at which the magnetic field strength takes a given value $B$. As in \cref{subsec:bulk}, we write $l_i(\eta, B)$ for the $l$-coordinates of these points, which we label using a discrete index $i$. Also, we set $l_{i*} \coloneq l_i(\eta_*, B_*)$.

Suppose $\partial_l B(\eta_*, l_{i*})$ is non-zero for some $i$. Then, for sufficiently small $\delta\eta$ and $\delta B$, every point $l_i(\eta_*, B_*+\delta B)$ on field line $\eta_*$ where the field strength is $B_*+\delta B$ is accompanied by a nearby point $l_i(\eta_*+\delta\eta, B_*+\delta B)$ on field line $\eta_*+\delta \eta$ where the field strength is also $B_*+\delta B$; we assume these nearby points are indexed by the same $i$. Then, their $l$-coordinates are related by
\begin{equation}\label{eq:dl}
l_i(\eta_*+\delta\eta, B_*+\delta B) - l_i(\eta_*, B_*+\delta B) \simeq -\frac{\partial_\eta B(\eta_*, l_{i*})}{\partial_l B(\eta_*, l_{i*})}\,\delta\eta\,.
\end{equation}
When $\partial_l B(\eta_*, l_{i*})$ is non-zero for all $l_{i*}$, we can compute $\delta L$ by summing $\eqref{eq:dl}$ over $i$, and accounting for whether the change in $l_i$ increases the size of a well (contributing positively to $\delta L$) or decreases the size of a well (contributing negatively to $\delta L$). We find
\begin{equation}\label{eq:simpleL}
    \delta L = -\mspace{-8mu}\sum_{i \in \mathcal{P}(\eta_*, B_*)} \frac{\partial_\eta B(\eta_*, l_{i*})}{|\partial_l B(\eta_*, l_{i*})|}\,\delta\eta\,,
\end{equation}
where $\mathcal{P}(\eta_*, B_*)$ denotes the set of points along field line $\eta_*$ where the field strength is $B_*$.

If $\partial_l B(\eta_*, l_{i*}) = 0$ at any of the points $l_{i*}$, then these points will need to be treated separately, since \eqref{eq:dl} is incorrect for them. These points are the locations of local minima or maxima when $B$ is plotted against $l$ along field line $\eta_*$. For simplicity, we do not consider magnetic fields with points of inflection when $B$ is plotted against $l$, although it is not difficult to generalize \eqref{eq:deltaL} below to include such points.

\begin{figure}
\vspace{2mm}
\centering
\includegraphics[width=\textwidth]{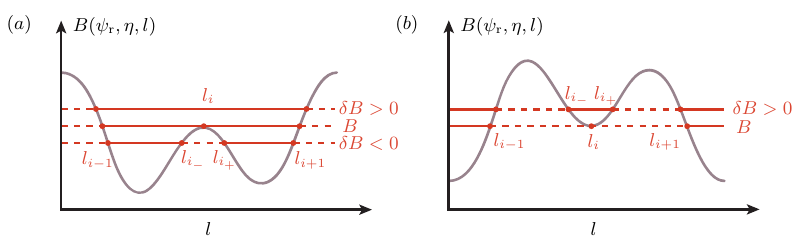}
    \caption{Sketch of how the wells (indicated by solid red lines) corresponding to a field strength $B$ change when $B$ is slightly increased ($\delta B >0$) or decreased ($\delta B<0$). In each case, there is a point $l_i$ at which the field strength is $B$ and $\partial_l B = 0$. $(a)$ In this case, $\partial^2_l B < 0$ at $l_i$. If $\delta B >0$, the wells on either side of $l_i$ merge. If $\delta B <0$, these wells move apart. $(b)$ In this case, $\partial^2_l B > 0$ at $l_i$. If $\delta B >0$, a new well opens up. If $\delta B < 0$, there will be no well here.}
    \label{fig:specialpoints}
\end{figure}

Consider a point $l_{i*}$ with $\partial_l B(\eta_*, l_{i*}) = 0$ and suppose that it is a local maximum, meaning ${\partial_l^2B(\eta_*, l_{i*})<0}$. Then, there are two wells on either side of $l_{i*}$ that touch at $l_{i*}$. If the field strength is changed from $B_*$ to $B_*+
\delta B$, there are two possibilities, as shown in figure \cref{fig:specialpoints}. If $\delta B > 0$, then the two wells on either side of $l_{i*}$ merge into one larger well. In this case, the point $l_{i*}$ does not smoothly connect to another point in $\mathcal{P}(\eta_*, B_*+\delta B)$. On the other hand, if $\delta B < 0$, then the two wells on either side of $l_{i*}$ move apart. In this case, the point $l_{i*}$ splits into two points in $\mathcal{P}(\eta_*, B_*+\delta B)$, separated by a region where $B(\eta_*, l) > B_*+\delta B$; we will use $i_-$ and $i_+$ to index these two points. The distance between them is
\begin{equation}\label{eq:sqrtl}
    l_{i_+}(\eta_*, B_*+\delta B) - l_{i_-}(\eta_*, B_*+\delta B) \simeq 2\sqrt{\frac{2\?\delta B}{\partial_l^2 B(\eta_*, l_{i*})}}\,.
\end{equation}
Now suppose $l_{i*}$ is a local minimum of $B$ along $\eta_*$. We can think of such a point as being the location of an infinitesimal well. Then, when $\delta B < 0$, the well disappears. When $\delta B > 0$, the well widens and $l_{i*}$ splits into two points $i_-$ and $i_+$, which also satisfy \eqref{eq:sqrtl}. 

The local minima or maxima behave in a similar way when we simultaneously change the magnetic field by $\delta B$ and the field-line label by $\delta \eta$. Therefore, the generalization of \eqref{eq:simpleL} to the case where some $l_i$ are local maxima or minima is
\begin{align}\label{eq:deltaL}
     \delta L &\simeq - \mspace{-8mu}\sum_{i\in \mathcal{Q}(\eta_*, B_*)} \frac{\partial_\eta B(\eta_*, l_{i*})}{|\partial_l B(\eta_*, l_{i*})|}\,\delta\eta\nn
     \al - \mspace{-8mu}\sum_{i\in \mathcal{P}_M(\eta_*, B_*)} \Biggl[2\sqrt{\Xi\biggl(\frac{2\?[\delta B
    - \delta\eta\,\partial_\eta B(\eta_*, l_{i*})]}{\partial_l^2 B(\eta_*, l_{i*})}\biggr)} - 2\sqrt{\Xi\biggl(\frac{2\?\delta B}{\partial_l^2 B(\eta_*, l_{i*})}\biggr)}\?\Biggr]\nn
    \al + \mspace{-8mu}\sum_{i\in \mathcal{P}_m(\eta_*, B_*)} \Biggl[2\sqrt{\Xi\biggl(\frac{2\?[\delta B - \delta\eta\,\partial_\eta B(\eta_*, l_{i*})]}{\partial_l^2 B(\eta_*, l_{i*})}\biggr)} - 2\sqrt{\Xi\biggl(\frac{2\?\delta B}{\partial_l^2 B(\eta_*, l_{i*})}\biggr)}\?\Biggr]\,.
\end{align}
In \eqref{eq:deltaL}, $\Xi(x)\coloneq x\? \Theta(x)$, where $\Theta(x)$ is the Heaviside step function, and we write $\mathcal{P}_M(\eta, B)$ for the subset of $\mathcal{P}(\eta, B)$ that are local maxima, $\mathcal{P}_m(\eta, B)$ be the subset of $\mathcal{P}(\eta, B)$ that are local minima, and $\mathcal{Q}(\eta, B)$ for all other points in $\mathcal{P}(\eta, B)$.

In \eqref{eq:deltaL}, for sufficiently small $\delta\eta$ and $\delta B$, the terms linear in $\delta \eta$ are negligible compared to the terms containing square roots. Assuming $B_* < B_M$, we can choose $\delta B$ to be a large enough positive number that ${\delta B - \delta\eta\,\partial_\eta B(\eta_*, l_{i*})>0}$ for all $i$ in ${\mathcal{P}_M(\eta_*, B_*)\cup\mathcal{P}_m(\eta_*, B_*)}$, over a range of $\delta \eta$. Then, 
\begin{equation}
    \delta L \simeq \mspace{-8mu}\sum_{i\in \mathcal{P}_m(\eta_*, B_*)} \Biggl[ 2\sqrt{\frac{2\?[\delta B - \delta\eta\,\partial_\eta B(\eta_*, l_{i*})]}{\partial_l^2 B(\eta_*, l_{i*})}} - 2\sqrt{\frac{2\delta B}{\partial_l^2 B(\eta_*, l_{i*})}} \,\Biggr]\,.
\end{equation}
For $\delta L$ to be independent of $\delta \eta$ within this range, we need
\begin{equation}\label{eq:changeindeltaL}
    \frac{\partial\?\delta L}{\partial \?\delta\eta}\biggl|_{\delta B} = \mspace{-4mu}\sum_{i\in \mathcal{P}_m(\eta_*, B_*)} \frac{-\sqrt{2}\,\partial_\eta B(\eta_*, l_{i*})}{\sqrt{[\partial_l^2 B(\eta_*, l_{i*})][\delta B - \delta\eta\,\partial_\eta B(\eta_*, l_{i*})]}} = 0\,.
\end{equation}
Since $1/\sqrt{\delta B - \delta\eta\,\partial_\eta B(\eta_*, l_{i*})}$ are linearly independent functions of $\delta \eta$ and $\delta B$ when all the values of $\partial_\eta B(\eta_*, l_{i*})$ are different from each other, \eqref{eq:changeindeltaL} requires $\partial_\eta B(\eta_*, l_{i*}) = 0$ at every local minimum.

Similar reasoning, with $\delta B$ chosen to be sufficiently negative, allows us to conclude that $\partial_\eta B(\eta_*, l_{i*}) = 0$ at every local maximum. Therefore, $\partial_\eta B = 0$ at all the points where $\partial_l B = 0$, which is what we set out to prove.

\section{Proof that low-shear ($s \sim \rho_\star$) stellarators without drift islands must be cyclometric}\label{app:lowshearcyclometry}

In \cref{subsec:cyclometry}, we prove that cyclometry is the necessary and sufficient condition for there to be no drift islands, for all passing particles, at a given rational surface $\psi_{\rm r}$. Our proof uses the fact that a particle orbit contains no drift islands if and only if its transit-averaged radial drift vanishes on $\psi_{\rm r}$. This follows from \eqref{eq:simpleinvariant}, which is valid as long as any resonant (closed) passing orbits, around which drift islands could form, are located near $\psi_{\rm r}$.

In stellarators with a very low magnetic shear, however, the drift islands can shift significant radial distances away from $\psi_{\rm r}$, as discussed in \cref{subsec:lowshear}. When the shear is as low as $s \sim \rho_\star$, the shifted position of the drift islands depends on $\mathcal{E}$ and $\mu$. In order to remove the drift islands at a given flux surface $\psi$, it is only necessary to ensure that the transit-averaged radial drift vanishes \emph{for those passing particles that have closed orbits on this flux surface}. This is less stringent than requiring that the transit-averaged radial drift vanishes on this flux surface for all passing particles. Therefore, one might wonder whether cyclometry is a stronger condition than necessary for the elimination of drift islands from a stellarator with $s \sim \rho_\star$.

In this appendix, we extend the arguments of \cref{subsec:cyclometry} to stellarators with $s \sim \rho_\star$ and show that, to avoid drift islands in these devices, cyclometry is still required. In fact, stellarators with $s \sim \rho_\star$ need to be cyclometric throughout a volume \Dash on all flux surfaces that the drift islands could shift to \Dash rather than on a single rational surface.

\begin{figure}
\vspace{4mm}
\centering
\includegraphics[width=\textwidth]{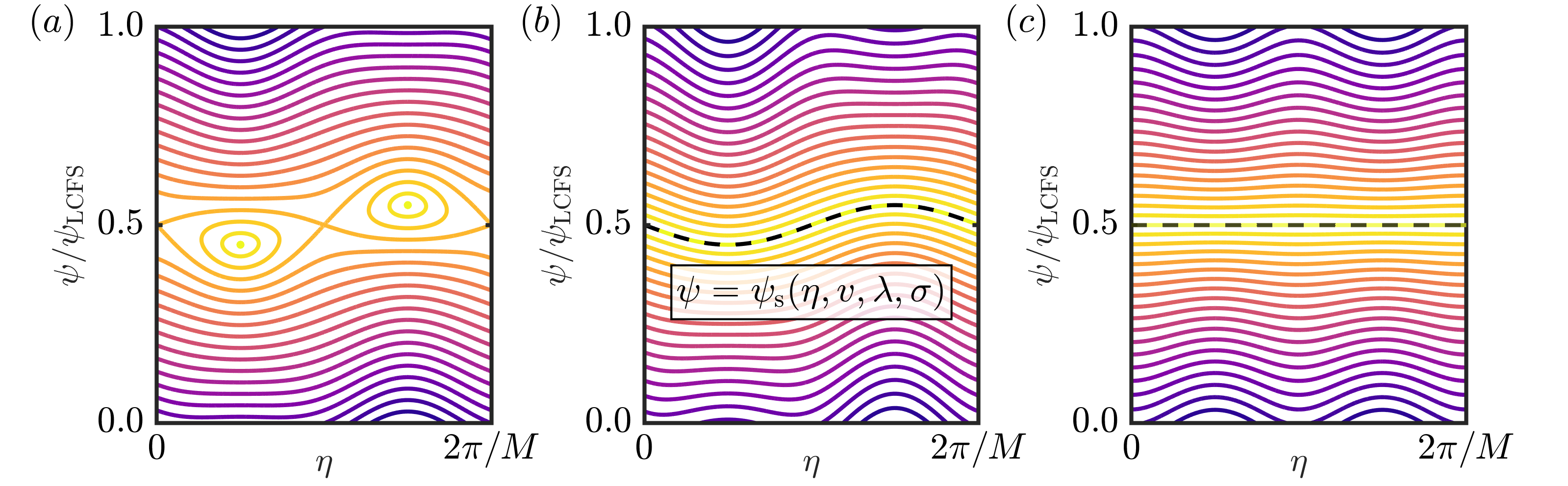}
    \caption{Three possible topologies of the level sets of $\mathcal{I}$ in a contour plot in the $(\eta,\psi)$ plane. $(a)$ Typically, $\mathcal{I}$ will have isolated extrema, which represent the O-points of the drift islands. $(b)$ A case where $\mathcal{I}$ reaches its extremal value along a continuous curve. $(c)$ Here, $\mathcal{I}$ attains its extremal value on the flux surface $\psi/\psi_\mathrm{LCFS} = 0.5$.}
    \label{fig:low_shear_cyclometry}
\end{figure}

As usual, we assume the rotational transform is close to $N/M$. Then, as discussed in \cref{subsec:islandplots}, passing orbits trace out level sets of the transit invariant $\mathcal{I}(\psi, \eta, \mathcal{E}, \mu, \sigma)$. Drift islands are present when $\mathcal{I}$ has isolated local extrema in the $(\eta,\psi)$ plane, which represent drift-island O-points, as shown in \crefsub{fig:low_shear_cyclometry}{a}. Therefore, we must ensure that either
\begin{enumerate}
    \item \label{item:1} $\mathcal{I}$ has no local extrema in the plasma, or
    \item \label{item:2} $\mathcal{I}$ only attains its extremal values on continuous curves in the $(\eta,\psi)$ plane, as shown in \crefsub{fig:low_shear_cyclometry}{b}, and never at isolated points.
\end{enumerate}
Local extrema of $\mathcal{I}$ represent resonant passing orbits that close on themselves after precisely $M$ toroidal and $N$ poloidal turns (this can be seen from \eqref{eq:changeslow}). Condition (\ref{item:1}) ensures that there are no such closed orbits in the plasma, which is a trivial way to avoid any associated drift islands. We are interested in stellarators that do contain such closed orbits. In such devices, condition (\ref{item:2}) must be satisfied, which ensures the closed orbits are surrounded by topologically toroidal drift surfaces instead of drift islands. We proceed to investigate what this implies for the magnetic field.

For this discussion, it will be convenient to change our velocity-space coordinates: instead of $\mathcal{E}$ and $\mu$, we will employ the particle speed $v$, defined by $\mathcal{E} = v^2/2$, and the pitch-angle coordinate $\lambda \coloneq \mu / \mathcal{E}$. The key idea is that, whereas in \cref{subsec:cyclometry} the cyclometry condition follows from the fact that the drift-island width vanishes for a range of values of $\lambda$, here (for stellarators with $s \sim \rho_\star$) we must use the fact that the drift-island width vanishes even as we vary $\lambda$ and $v$ simultaneously.

%
%
Let $\psi = \psi_{\rm s}(\eta, v, \lambda, \sigma)$ be the curve in the $(\eta,\psi)$ plane on which $\mathcal{I}$ attains its extremal value (see \crefsub{fig:low_shear_cyclometry}{b}). We begin by proving that $\psi_{\rm s}(\eta, v, \lambda, \sigma)$ must be independent of $\eta$ \Dash that is, $\mathcal{I}$ attains its extremal value on a flux surface, as shown in \crefsub{fig:low_shear_cyclometry}{c} \Dash for all passing particles. By definition of an extremum, every point along the curve satisfies
\begin{subequations}
\begin{equation}
\label{eq:cond1old}\partial_\eta \mathcal{I} = \sigma\?\partial_\eta I =  0
\end{equation}
and
\begin{equation}
\label{eq:cond2old}\partial_\psi \mathcal{I} = \sigma\? \partial_\psi I - \frac{2\upi M Ze}{mc}\biggl(\iota - \frac{N}{M}\biggr) = 0\,.
\end{equation}
\end{subequations}
It will be useful to work with these conditions in Boozer coordinates because the definition of cyclometry on a general flux surface, given in \cref{subsec:Boozing}, was formulated in Boozer coordinates. Conditions \eqref{eq:cond1old} and \eqref{eq:cond2old} are equivalent to
\begin{subequations}
\begin{equation}
\label{eq:cond1}\partial_\eta\biggl(G\oint \frac{\sqrt{1 - \lambda B}}{B}\,\rmd \zeta_{\rm B}\biggr) = 0
\end{equation}
and
\begin{equation}
\label{eq:cond2}\sigma v\?\partial_\psi\biggl(G\oint \frac{\sqrt{1 - \lambda B}}{B}\,\rmd \zeta_{\rm B}\biggr) - \frac{2\upi M Z e}{mc}\biggl(\iota - \frac{N}{M}\biggr) = 0\,,
\end{equation}
\end{subequations}
respectively. We should think of \eqref{eq:cond2} as defining the curve $\psi = \psi_{\rm s}(\eta, v, \lambda, \sigma)$, and \eqref{eq:cond1} as an additional condition that must be enforced at each point along this curve. The reason \eqref{eq:cond1} should not be used to define the curve is that this equation cannot always be inverted for $\psi$; it may, for all passing $\lambda$, be satisfied throughout a finite area in $(\eta,\psi)$ space. For example, \eqref{eq:cond1} holds everywhere in a stellarator that is cyclometric on every flux surface. Equation \eqref{eq:cond2}, on the other hand, does not have this problem. While it is \Dash in principle \Dash possible that \eqref{eq:cond2} might hold throughout a finite area for a particular value of $v$, this cannot be the case for all $v$. Thus, for generic $v$, we can invert \eqref{eq:cond2} to determine $\psi = \psi_{\rm s}(\eta, v, \lambda, \sigma)$.

We now use the fact that \eqref{eq:cond1} and \eqref{eq:cond2} hold for particles with different speeds $v$ to prove that $\psi_{\rm s}(\eta, v, \lambda, \sigma)$ must be independent of $\eta$ \Dash that is, $\mathcal{I}$ attains its extremal value on a flux surface, as shown in \crefsub{fig:low_shear_cyclometry}{c} \Dash for all passing particles. Differentiating \eqref{eq:cond1} with respect to $v$ and \eqref{eq:cond2} with respect to $\eta$, we find that
\begin{subequations}
\begin{equation}
\label{eq:cond1new}(\partial_v\psi_{\rm s})\,\partial_\psi\partial_\eta\biggl(G\oint \frac{\sqrt{1 - \lambda B}}{B}\,\rmd \zeta_{\rm B}\biggr) = 0
\end{equation}
and
\begin{equation}
\label{eq:cond2new}
\sigma v\?\partial_\eta\partial_\psi\biggl(G\oint \frac{\sqrt{1 - \lambda B}}{B}\,\rmd \zeta_{\rm B}\biggr) + \sigma v\?(\partial_\eta\psi_{\rm s})\,\partial_\psi^2\biggl(G\oint \frac{\sqrt{1 - \lambda B}}{B}\,\rmd \zeta_{\rm B}\biggr) = 0\,,
\end{equation}
\end{subequations}
are satisfied at each point along the curve $\psi = \psi_{\rm s}(\eta, v, \lambda, \sigma)$. Since $\partial_v\psi_{\rm s}$ is generically nonzero, \eqref{eq:cond1new} implies
\begin{equation}
\partial_\psi\partial_\eta\biggl(G\oint \frac{\sqrt{1 - \lambda B}}{B}\,\rmd \zeta_{\rm B}\biggr) = 0\,.
\end{equation}
Then, using the commutativity of partial derivatives, \eqref{eq:cond2new} becomes
\begin{equation}\label{eq:cond2newnew}
(\partial_\eta\psi_{\rm s})\,\partial_\psi^2\biggl(G\oint \frac{\sqrt{1 - \lambda B}}{B}\,\rmd \zeta_{\rm B}\biggr) = 0\,.
\end{equation}
Generically, the integral in \eqref{eq:cond2newnew} has a nonzero second derivative with respect to $\psi$, so $\partial_\eta\psi_{\rm s} = 0$. Hence, $\psi_{\rm s}(\eta, v, \lambda, \sigma)$ must be independent of $\eta$.

From here on, we assume that, whenever $\mathcal{I}$ has an extremum in the plasma for some $v$, $\lambda$, and $\sigma$, the extremal value is attained everywhere on a flux surface $\psi_{\rm s}(v, \lambda, \sigma)$, as shown in \crefsub{fig:low_shear_cyclometry}{c}, on which \eqref{eq:cond1} and \eqref{eq:cond2} are satisfied. Then, we can fix our attention on a single flux surface, $\psi$, and, as we vary $\lambda$, we can imagine simultaneously adjusting $v$ in such a way that $\psi_{\rm s}(v, \lambda, \sigma) = \psi$ remains constant.\footnote{This reasoning is valid unless the stellarator has $|\iota - N/M| \ll \rho_\star$ everywhere. In such a device, the second term in \eqref{eq:cond2} can be neglected and the resulting equation cannot be solved for $v_{\rm s}$. We do not consider this possibility further because, for $|\iota - N/M| \ll \rho_\star$ to hold everywhere, the magnetic shear must satisfy $s \ll \rho_\star$, which is much smaller than the already low $s \sim \rho_\star$ ordering adopted here.} In other words, we are considering the passing particles whose orbits are resonant on flux surface $\psi$; their speed $v_{\rm s}(\psi, \lambda, \sigma)$ is given by
\begin{equation}\label{eq:vsdef}
 v_{\rm s} = \frac{2\upi M Ze}{mc}\biggl(\iota - \frac{N}{M}\biggr)\bigg/ \partial_\psi \biggl( \sigma G\oint \frac{\sqrt{1 - \lambda B}}{B}\,\rmd \zeta_{\rm B} \biggr)\,.
\end{equation}
Only one sign of $\sigma$ gives a positive $v_{\rm s}$, which means resonant orbits only occur at $\psi$ for particles circulating in one direction. 

For all possible passing $\lambda$, we need to impose condition \eqref{eq:cond1} on flux surface $\psi$. In fact, as in \cref{subsec:cyclometry}, we will only need \eqref{eq:cond1} to hold for all $\lambda$ in some open neighbourhood of zero, provided the denominator in \eqref{eq:vsdef} does not vanish at $\lambda = 0$; this assumption is satisfied on generic flux surfaces. From this point onwards, we can follow essentially the same steps that we took in \cref{subsec:cyclometry} to obtain the cyclometry constraint for stellarators with higher shear. Wherever we Taylor expanded in $\mu/\mathcal{E}$ in \cref{subsec:cyclometry}, we now Taylor expand in $\lambda$. This leads to the result that $Z_{\rm B}(\psi, \eta, B)$, defined in \eqref{eq:ZBdef}, must be independent of $\eta$, which means flux surface $\psi$ must be cyclometric. Identical reasoning could be used for any flux surface $\psi$ on which $v_{\rm s}$, defined in \eqref{eq:vsdef}, at $\lambda = 0$ is below the maximum alpha-particle speed. Physically, this means that cyclometry is necessary on all flux surfaces on which resonant passing orbits are possible for alpha particles with energies below $\SI{3.5}{\mega\electronvolt}$.

\section{Why $\mu$-variation can be neglected when calculating $\mathcal{I}^{(1/2)}$}\label{app:mu}

In our calculation of $\mathcal{I}^{(1/2)}$ in \cref{sec:higherorder}, we neglect changes in $\mu$. This is inconsistent because ${\rmd\mu/\rmd t}$ (gyroaveraged) is non-zero at order $\rho_\star\? v^3/BL$, so a term of the form $\partial_\mu \mathcal{I}\?(\rmd \mu / \rmd \zeta)$ should appear in \eqref{eq:higherorder}. In this appendix, we explain why neglecting changes in $\mu$ in this way still leads to the right formula for $\mathcal{I}^{(1/2)}$ at order $\sqrt{\rho_\star}\? vL$. 

The main idea is that $\mu$ is the lowest-order term in a series, analogous to \eqref{eq:asymptoticseries}, for an improved adiabatic invariant $\mybar{\mu}$ that is conserved to all orders in $\rho_\star$:
\begin{equation}\label{eq:mubar}
    \mybar{\mu}\?(\bb{X},\mathcal{E},\mu,\varphi,\sigma) = \mu + \mu^{(1)}(\bb{X},\mathcal{E},\mu,\varphi,\sigma) + \mu^{(2)}(\bb{X},\mathcal{E},\mu,\varphi,\sigma) + \ldots,
\end{equation}
where $\mu^{(n)}\sim\rho_\star^n\? v^2/B$ and $\varphi$ is the gyrophase. For example, the first correction is
\begin{equation}\label{eq:mucorrection}
    \mu^{(1)} = \frac{v_\parallel^2}{B}(\bh\0\bb{\nabla}\bh\0\bb{\rho}) + \frac{Ze}{2mc}v_\parallel(\bb{\rho}\0\bb{\nabla}\bh \0(\bh\bb{\times}\bb{\rho}) ) - \frac{mc}{2Ze}\frac{v_\parallel\mu}{B}\bh\0(\bb{\nabla}\bb{\times}\bh) + O\biggl(\frac{\rho_\star^2 v^2}{B}\biggr)\,,
\end{equation}
where $\bb{\rho} = (mc/ZeB)(\bh\bb{\times}\bb{v})$ is the gyroradius vector and all functions of position are evaluated at the guiding-centre position $\bb{X}$ \mycitep[\eg]{Parra2011}. 
Imagine using $\bar{\mu}$ as a coordinate instead of $\mu$ in \cref{sec:higherorder}. Then, it would be completely rigorous to neglect $\rmd\bar{\mu}/\rmd t$ to the relevant order. Thus, the calculation would proceed exactly as in \cref{sec:higherorder}, but with $\mybar{\mu}$ replacing $\mu$ everywhere. We would obtain the same formula \eqref{eq:higherorderinvariant} for the higher-order transit invariant, but with $\mybar{\mu}$ replacing $\mu$. This replacement only modifies \eqref{eq:higherorderinvariant} at order $\rho_\star vL$, which is higher-order than $\mathcal{I}^{(1/2)}$. Therefore, our formula for $\mathcal{I}^{(1/2)}$ is correct to the order of interest.

There is a subtlety in this argument: the higher-order magnetic moment $\mybar{\mu}$ in \eqref{eq:mubar} contains a term proportional to ${\bh\0(\bb{\nabla}\bb{\times}\bh) \propto j_\parallel/B}$, where $j_\parallel$ is the parallel current density. The equations of ideal MHD with nested flux surfaces can make this current density singular around rational surfaces \mycitep{Helander2014}, which would cause $\mu^{(1)}$ to diverge. This divergence indicates that the guiding-centre ordering is broken, which is expected because a sufficiently narrow current sheet creates a magnetic field with sharp gradients. In a real stellarator, singular currents will be regulated by the formation of magnetic islands \mycitep[\eg]{Reiman1983, Cary1985, Hegna1989} or by transport processes \mycitep[\eg]{Boozer1981, Hazeltine2015}, which add non-ideal terms to the MHD equations. Nevertheless, layers of finite but enhanced current density may still arise around rational surfaces. It is therefore an assumption in this work that any such layers are insignificant enough that the guiding-centre expansion remains valid as particles move across rational surfaces.

\section{Manipulations required for the higher-order correction to the transit invariant}\label{app:manipulations}

In this appendix, we show how to obtain \eqref{eq:poisson}. We will repeatedly use the fact that radial derivatives can make certain functions an order larger in $\sqrt{\rho_\star}$; in particular, ${\partial_\psi\mathcal{I}\simeq -(2\upi M Ze/mc)\?\iota'_{\rm r}(\psi - \psi_{\rm r}) \sim vL/\sqrt{\rho_\star}\?\Psi_{\rm t}}$ and ${\partial_\psi^2\mathcal{I}\simeq -(2\upi M Ze/mc)\?\iota'_{\rm r} \sim vL/\rho_\star \Psi_{\rm t}^2}$. Note, however, that $\partial_\eta\mathcal{I}$ does not have a large radial gradient: ${\partial_\psi\partial_\eta\mathcal{I}\sim vL/\Psi_{\rm t} \ll \partial_\psi \mathcal{I}}$.

First, substituting \eqref{eq:Ihalf} into \eqref{eq:prepoisson}, we find
\begin{align}\label{eq:poissonbracket}
    0 = \frac{mc}{Ze}\Bigl[ \partial_\psi\mathcal{I}^{(1/2)}(\psi,\eta,0) \, \partial_\eta\mathcal{I} &- \partial_\eta\mathcal{I}^{(1/2)}(\psi,\eta,0) \,\partial_\psi\mathcal{I} \Bigr]\nn
    &- \oint \Gamma_\psi\?\partial_\psi\biggl( \partial_\psi\mathcal{I}\int_0^\zeta \!\Gamma_\psi' \,\rmd\zeta' + \partial_\eta\mathcal{I}\int_0^\zeta  \!\Gamma_\eta' \,\rmd\zeta' \biggr)\, \rmd\zeta\nn
    &- \oint  \Gamma_\eta\?\partial_\eta\biggl( \partial_\psi\mathcal{I}\int_0^\zeta \!\Gamma_\psi' \,\rmd\zeta' + \partial_\eta\mathcal{I}\int_0^\zeta  \!\Gamma_\eta' \,\rmd\zeta' \biggr)\, \rmd\zeta\,.
\end{align}
Here, we write $\Gamma_\psi'$ and $\Gamma_\eta'$ for the values of $\Gamma_\psi$ and $\Gamma_\eta$ at toroidal angle $\zeta'$ (this notation is only necessary to avoid ambiguity in certain double integrals in this appendix, so we do not use this notation in the main text). The last two lines of \eqref{eq:poissonbracket} contain four double integrals, which we simplify as follows. For the first integral, we have
\begin{equation}
    \oint \Gamma_\psi\?\partial_\psi\biggl( \partial_\psi\mathcal{I}\int_0^\zeta \!\Gamma_\psi'\,\rmd\zeta'\biggr)\,\rmd\zeta \simeq \partial_\psi^2\mathcal{I} \oint \Gamma_\psi\biggl(\?\int_0^\zeta \!\Gamma_\psi'\,\rmd\zeta'\biggr)\, \rmd\zeta\,,
\end{equation}
due to the large radial gradient of $\mathcal{I}$. Realizing that 
\begin{equation}
    2\Gamma_\psi \int_0^\zeta \!\Gamma_\psi'\,\rmd\zeta' = \partial_\zeta\Biggl[ \biggl(\? \int_0^\zeta \!\Gamma_\psi'\,\rmd\zeta' \biggr)^{\!\!2} \Biggr]\,,
\end{equation}
we find
\begin{equation}\label{eq:integrationbyparts}
    \oint \Gamma_\psi\biggl(\int_0^\zeta \!\Gamma_\psi'\,\rmd\zeta'\biggr)\, \rmd\zeta = \frac{1}{2} \frac{mc}{Ze}\? \partial_\eta\mathcal{I} \oint \Gamma_\psi\,\rmd\zeta\,,
\end{equation}
where we have employed \eqref{eq:Gammaidentities} to write this expression in a form that will be convenient later. Then, using ${\partial_\psi^2\mathcal{I}\simeq -(2\upi M Ze/mc)\?\iota'_{\rm r}}$, the first integral becomes
\begin{align}\label{eq:int1}
    \oint \Gamma_\psi\?\partial_\psi\biggl( \partial_\psi\mathcal{I}\int_0^\zeta \!\Gamma_\psi'\,\rmd\zeta'\biggr)\,\rmd\zeta \simeq - (\partial_\eta\mathcal{I})\?\upi M\iota'_{\rm r} \oint \Gamma_\psi\,\rmd\zeta\,.
\end{align}
The second integral on the middle line of \eqref{eq:poissonbracket} is
\begin{align}\label{eq:int2}
    \oint \Gamma_\psi\?\partial_\psi\biggl( \partial_\eta\mathcal{I}\int_0^\zeta \!\Gamma_\eta'\,\rmd\zeta'\biggr)\,\rmd\zeta &\simeq (\partial_\eta\mathcal{I})\?\iota_{\rm r}' \oint \zeta\Gamma_\psi\,\rmd\zeta\,,
\end{align}
where we have used $\Gamma_\eta\simeq \iota_{\rm r}'(\psi - \psi_{\rm r})$ (by \eqref{eq:Gammadefs}) and the fact that $\partial_\eta\mathcal{I}$ does not have a large radial gradient. 

For the first integral on the last line of \eqref{eq:poissonbracket}, we have
\begin{equation}
    \oint \Gamma_\eta\?\partial_\eta\biggl( \partial_\psi\mathcal{I}\int_0^\zeta \!\Gamma_\psi'\,\rmd\zeta'\biggr)\, \rmd\zeta \simeq (\partial_\psi\mathcal{I})\? \iota'_{\rm r}(\psi - \psi_{\rm r})\? \partial_\eta \biggl[\? \oint \biggl(\? \int_0^\zeta \!\Gamma_\psi'\, \rmd\zeta'\biggr)\, \rmd\zeta \biggr]\,,
\end{equation}
because $\partial_\psi\partial_\eta\mathcal{I}$ is small. Integrating by parts gives
\begin{equation}
    \oint \biggl(\? \int_0^\zeta \!\Gamma_\psi'\,\rmd\zeta'\biggr)\, \rmd\zeta = 2\upi M \oint \Gamma_\psi\,\rmd\zeta - \oint \zeta\Gamma_\psi\,\rmd\zeta\,,
\end{equation}
from which we obtain
\begin{align}\label{eq:int3}
    \oint \Gamma_\eta\?\partial_\eta\biggl(\? \partial_\psi\mathcal{I}\int_0^\zeta \!\Gamma_\psi'\,\rmd\zeta'\biggr)\, \rmd\zeta \simeq -(\partial_\psi\mathcal{I})\?\iota'_{\rm r}(\psi - \psi_{\rm r})\?\partial_\eta \biggl(\?\oint (\zeta - 2\upi M)\Gamma_\psi \,\rmd\zeta \biggr)\,.
\end{align}
Finally, the last integral in \eqref{eq:poissonbracket} can be simplified using
\begin{align}\label{eq:int4}
    \oint \Gamma_\eta\?\partial_\eta\biggl( \partial_\eta\mathcal{I}\int_0^\zeta \!\Gamma_\eta'\,\rmd\zeta'\biggr)\, \rmd\zeta &\simeq {\iota_{\rm r}'}^2 (\psi-\psi_{\rm r})^2 (\partial_\eta^2\mathcal{I})\oint \zeta\,\rmd\zeta\nn
    &\simeq - (\partial_\psi\mathcal{I})\? \upi M \iota'_{\rm r}(\psi - \psi_{\rm r})\oint \Gamma_\psi\,\rmd\zeta\,,
\end{align}
where the last line follows from \eqref{eq:Gammaidentities}.

Replacing the integrals in \eqref{eq:poissonbracket} with the four expressions \eqref{eq:int1}, \eqref{eq:int2}, \eqref{eq:int3} and \eqref{eq:int4} gives \eqref{eq:poisson}.

\section{Origin of the arbitrary constant in $\mathcal{I}^{(1/2)}$}\label{app:constant}

In this appendix, we explain why an undetermined function $f(\mathcal{I})$ arises, in \eqref{eq:Jnearly}, in our calculation of the higher-order transit adiabatic invariant, and why we are free to set this function to zero. The fact that arbitrary choices like this one need to be made in the construction of higher-order adiabatic invariants is well known \mycitep[\eg]{Lenard1959}.

In \cref{sec:higherorder}, we find $\mathcal{I}^{(1/2)}$ by enforcing the condition
\begin{equation}\label{eq:appdef}
\biggl\langle \frac{\rmd (\mathcal{I} + \mathcal{I}^{(1/2)})}{\rmd t}\biggr\rangle_{\!\!\mathrm{t}} \sim \rho_\star^{3/2}v^2\,,
\end{equation}
which makes $\mathcal{I}+\mathcal{I}^{(1/2)}$ an approximately conserved quantity. For any small, but otherwise arbitrary, function ${f(\mathcal{I}) \sim \rho_\star^{1/2}\mathcal{I}}$, we also have
\begin{equation}
\biggl\langle \frac{\rmd (\mathcal{I} + \mathcal{I}^{(1/2)} + f(\mathcal{I}))}{\rmd t}\biggr\rangle_{\!\!\mathrm{t}} \sim \rho_\star^{3/2}v^2\,,
\end{equation}
which means $\mathcal{I}^{(1/2)} + f(\mathcal{I})$ is an equally valid choice for the higher-order correction to the transit adiabatic invariant. Since $f$ is arbitrary, we can choose $f=0$.

This freedom in the definition of $\mathcal{I}^{(1/2)}$ is related to the fact that a function of an adiabatic invariant is also an adiabatic invariant. Equation \eqref{eq:appdef} states that $\mathcal{I}+\mathcal{I}^{(1/2)}$ is a approximately conserved quantity. For any function $g$, $g(\mathcal{I}+\mathcal{I}^{(1/2)})$ must also be approximately conserved. Suppose $g(\mathcal{I}) = \mathcal{I} + f(\mathcal{I})$, for some small ${f(\mathcal{I}) \sim \smash{\rho_\star^{1/2}\mathcal{I}}}$. Expanding $g(\mathcal{I}+\mathcal{I}^{(1/2)})\simeq \mathcal{I} + \mathcal{I}^{(1/2)} + f(\mathcal{I})$, we learn that $\mathcal{I} + \mathcal{I}^{(1/2)} + f(\mathcal{I})$ is approximately conserved, so $\mathcal{I}^{(1/2)} + f(\mathcal{I})$ is an equally valid choice for the higher-order correction to the transit adiabatic invariant.

	\section{Gauge invariance of the transit adiabatic invariant}\label{app:gauge}

In this appendix, we verify explicitly that formula \eqref{eq:higherorderinvariant} gives the same value for $\mybar{\mathcal{I}}$ in any straight-field-line coordinate system (up to terms of order $\rho_\star vL$). We will use a new set of straight-field-line coordinates $(\psi,\tilde{\theta},\tilde{\zeta})$, defined by
\begin{subequations}
    \begin{align}
        \tilde{\zeta} &= \zeta + \nu(\psi, \theta,\zeta)\,,\label{eq:thetatilde}\\
        \tilde{\theta} &= \theta + \iota(\psi)\?\nu(\psi, \theta,\zeta)\,,\label{eq:zetatilde}
    \end{align}
\end{subequations}
where $\nu(\psi, \theta,\zeta)$ is $2\upi$-periodic in both its $\theta$ and $\zeta$ arguments. This is not the most general transformation between two sets of straight-field-line coordinates: we could also add two flux functions whose ratio is not the rotational transform $\iota(\psi)$ to the right sides of \eqref{eq:thetatilde} and \eqref{eq:zetatilde}. We ignore this possibility because adding a flux function to either $\theta$ or $\zeta$ is equivalent to changing the location where that angle is zero, and \eqref{eq:origininvariant} shows that this would not change the value of $\mybar{\mathcal{I}}$ (as discussed in \cref{subsec:higherordercalc}).

To begin, we introduce some helpful notation. We write $(\psi,\tilde{\eta}(\psi,\eta,\zeta),\tilde{\zeta}(\psi,\eta,\zeta))$ for the new coordinates describing the point $(\psi, \eta, \zeta)$. Similarly, we write $(\psi,\eta(\psi,\tilde{\eta},\tilde{\zeta}),\zeta(\psi,\tilde{\eta},\tilde{\zeta})$ for the old coordinates corresponding to $(\psi, \tilde{\eta}, \tilde{\zeta})$. For any other function $f(\psi, \eta, \zeta)$ of the old coordinates, we define
\begin{equation}\label{eq:astdef}
    \tilde{f}(\psi,{\tilde{\eta}},{\tilde{\zeta}}) \coloneq f(\psi,\eta(\psi,\tilde{\eta},\tilde{\zeta}),\zeta(\psi,\tilde{\eta},\tilde{\zeta}))\,.
\end{equation}
Then, the definition of the new coordinates can be expressed in two equivalent ways: 
\begin{subequations}
    \begin{align}
        \tilde{\zeta} &= \zeta + \nu(\psi, \eta,\zeta) & &\text{or} &\zeta &= \tilde{\zeta} - \tilde{\nu}(\psi, \tilde{\eta},\tilde{\zeta})  \,,\\
        \tilde{\eta} &= \eta + \biggl(\iota(\psi) - \frac{N}{M}\biggr)\,\nu(\psi, \eta,\zeta) & &\text{or} &\eta &= \tilde{\eta} - \biggl(\iota(\psi) - \frac{N}{M}\biggr)\,\tilde{\nu}(\psi, \tilde{\eta},\tilde{\zeta}) \,.
    \end{align}
\end{subequations}
For the rest of this appendix, we suppress the ${(\mathcal{E}, \mu, \sigma)}$ arguments of any functions that depend on the particle velocity.

We write $\tilde{\mybar{\mathcal{K}}}(\psi, \tilde{\eta}, \tilde{\zeta})$ for the value of the higher-order transit invariant that someone working exclusively in the new coordinates would obtain using \eqref{eq:higherorderinvariant}. To be explicit, this means $\tilde{\mybar{\mathcal{K}}}(\psi, \tilde{\eta}, \tilde{\zeta})$ is defined by
\begin{align}\label{eq:ast1}
\tilde{\mybar{\mathcal{K}}}(\psi,\tilde{\eta},\tilde{\zeta}) \coloneq &\oint \stackon[-8pt]{$(v_\parallel{\bh}\0\partial_{\tilde{\zeta}}\bb{x})$}{\vstretch{1.5}{\hstretch{2.4}{\widetilde{\phantom{\;\;\;\;\;\;\;}}}}}(\psi,\tilde{\eta},\tilde{\zeta}')\,\rmd\tilde{\zeta}' - \frac{2\upi M Ze}{mc}\!\int_{\psi_\mathrm{r}}^{\psi}\biggl(\iota(\psi') - \frac{N}{M}\biggr)\,\rmd\psi'\nn
&+ \frac{Ze}{mc}\iota_{\rm r}'(\psi-\psi_\mathrm{r})\Biggl[ \int_{0}^{{\tilde{\zeta}}} 2\upi M \stackon[-8pt]{$\displaystyle \biggl(\frac{\bb{v}_\mathrm{d}\0\bb{\nabla}\psi}{v_\parallel \bh\0\bb{\nabla}\tilde{\zeta}}\biggr)$}{\vstretch{1.5}{\hstretch{2.4}{\widetilde{\phantom{\;\;\;\;\;\;\;}}}}}(\psi_\mathrm{r},{\tilde{\eta}},\tilde{\zeta}')\,\rmd\tilde{\zeta}'\nn
&+ \oint\, (\tilde{\zeta}' - \tilde{\zeta} - \upi M) \stackon[-8pt]{$\displaystyle \biggl(\frac{\bb{v}_\mathrm{d}\0\bb{\nabla}\psi}{v_\parallel \bh\0\bb{\nabla}\tilde{\zeta}}\biggr)$}{\vstretch{1.5}{\hstretch{2.4}{\widetilde{\phantom{\;\;\;\;\;\;\;}}}}}(\psi_\mathrm{r},{\tilde{\eta}},\tilde{\zeta}')\,\rmd\tilde{\zeta}'\Biggr]\,.
\end{align}
To prove the gauge invariance of \eqref{eq:higherorderinvariant}, our aim will be to show that, in fact, ${\tilde{\mybar{\mathcal{K}}}(\psi,\tilde{\eta},\tilde{\zeta}) = \mybar{\mathcal{I}}(\psi,\eta,\zeta)}$ (up to terms of order $\rho_\star vL$). Using \eqref{eq:astdef}, we find
\begin{align}\label{eq:ast2}
		\nonumber\tilde{\mybar{\mathcal{K}}}(\psi,\tilde{\eta},\tilde{\zeta}) = &\overbrace{\oint\, (v_\parallel{\bh}\0\partial_{\tilde{\zeta}}\bb{x})(\psi,\eta(\psi,\tilde{\eta},\tilde{\zeta}'),\zeta(\psi,\tilde{\eta},\tilde{\zeta}'))\,\rmd\tilde{\zeta}'}^{\circled{1}} - \overbrace{\frac{2\upi M Ze}{mc}\!\int_{\psi_\mathrm{r}}^{\psi}\biggl(\iota(\psi') - \frac{N}{M}\biggr)\,\rmd\psi'}^{\circled{2}}\\
		\nonumber&+ \frac{Ze}{mc}\iota_{\rm r}'(\psi-\psi_\mathrm{r})\Biggl[ \int_{0}^{{\tilde{\zeta}}} 2\upi M \biggl(\frac{\bb{v}_\mathrm{d}\0\bb{\nabla}\psi}{v_\parallel \bh\0\bb{\nabla}\tilde{\zeta}}\biggr)(\psi_\mathrm{r},\eta(\psi_\mathrm{r},\tilde{\eta},\tilde{\zeta}'),\zeta(\psi_\mathrm{r},\tilde{\eta},\tilde{\zeta}'))\,\rmd\tilde{\zeta}'\\
        &\underbrace{\quad+ \oint\, (\tilde{\zeta}' - \tilde{\zeta} - \upi M) \biggl(\frac{\bb{v}_\mathrm{d}\0\bb{\nabla}\psi}{v_\parallel \bh\0\bb{\nabla}\tilde{\zeta}}\biggr)(\psi_\mathrm{r},\eta(\psi_\mathrm{r},\tilde{\eta},\tilde{\zeta}'),\zeta(\psi_\mathrm{r},\tilde{\eta},\tilde{\zeta}'))\,\rmd\tilde{\zeta}'\Biggr]\,.\quad}_{\circled{3}}
\end{align}
In the arguments of the integrands here, $\eta(\psi,\tilde{\eta},\tilde{\zeta}')$ is \textit{not} the value of $\eta$ corresponding to the guiding-centre position: for that we would write $\eta(\psi,\tilde{\eta},\tilde{\zeta})$, or simply $\eta$. The difference between $\eta(\psi,\tilde{\eta},\tilde{\zeta}')$ and $\eta$ is order $\sqrt{\rho_\star}$ and, therefore, cannot be neglected in \circled{1}. This difference is given by
	\begin{equation}\label{eq:etaredef}
		\eta(\psi,\tilde{\eta},\tilde{\zeta}') = \eta + \biggl(\iota(\psi)-\frac{N}{M}\biggr)\bigl[\tilde{\nu}(\psi,\tilde{\eta},\tilde{\zeta}) - \tilde{\nu}(\psi,\tilde{\eta},\tilde{\zeta}')\bigr]\,.
	\end{equation}
Thus, we should Taylor expand the integrand of \circled{1} in its $\eta$ argument using \eqref{eq:etaredef}. The result is
\begin{align}\label{eq:1}
		 \circled{1} &\simeq \oint\,(v_\parallel{\bh}\0\partial_\zeta\bb{x})(\psi,\eta,\zeta')\,\rmd\zeta'\nn
		\al + \iota_{\rm r}'(\psi-\psi_\mathrm{r})\oint\, [\nu(\psi_{\rm r},\eta,\zeta) - \nu(\psi_{\rm r},\eta,\zeta')]\,\partial_\eta(v_\parallel{\bh}\0\partial_\zeta\bb{x})(\psi_{\rm r},\eta,\zeta')\,\rmd\zeta'\nn
        \al - \iota_{\rm r}'(\psi-\psi_\mathrm{r})\oint\,(\partial_{\zeta}\nu\,v_\parallel{\bh}\0\partial_\eta\bb{x})(\psi_{\rm r},\eta,\zeta')\,\rmd\zeta' + O(\rho_\star vL)\,,
\end{align}
where we have also employed
	\begin{equation}
		\partial_{\tilde{\zeta}}\bb{x} = \partial_{\tilde{\zeta}}\zeta\, [ \partial_{\zeta}\bb{x} - \iota_{\rm r}'(\psi-\psi_\mathrm{r})\?\partial_\zeta \nu\, \partial_\eta\bb{x} ] + O(\rho_\star L)\,
	\end{equation}
and changed the integration variable from $\tilde{\zeta}'$ to $\zeta' = \zeta(\psi,\tilde{\eta},\tilde{\zeta}')$ (which means ${\rmd\zeta' = \partial_{\tilde{\zeta}}\zeta\,\rmd\tilde{\zeta}'}$).

We also need to manipulate the terms labelled \circled{3} (the higher-order correction terms). In these terms, we can use $\eta(\psi,\tilde{\eta},\tilde{\zeta})\simeq \eta$ and $
    {\bh}\0\bb{\nabla}\zeta \simeq \partial_{\tilde{\zeta}}\zeta\,{\bh}\0\bb{\nabla}\tilde{\zeta}$, which are both correct up to relative errors of order $\sqrt{\rho_\star}$. Changing integration variables to ${\zeta' = \zeta(\psi,\tilde{\eta},\tilde{\zeta}')}$, as above, we obtain
\begin{align}\label{eq:circ3}
		\circled{3} \simeq \frac{Ze}{mc}\iota_{\rm r}'(\psi-\psi_\mathrm{r})\Biggl[ &\int_{\zeta_0}^{{\zeta}} 2\upi M \biggl(\frac{\bb{v}_\mathrm{d}\0\bb{\nabla}\psi}{v_\parallel \bh\0\bb{\nabla}\zeta}\biggr)(\psi_\mathrm{r},\eta,\zeta')\,\rmd\zeta'\nn
        & \begin{aligned}[b] + \int_{\zeta_0}^{\zeta_0 + 2\upi M}(&\zeta' - \zeta + \nu(\psi_\mathrm{r},\eta,\zeta')\\
        &- \nu(\psi_\mathrm{r},\eta,\zeta) - \upi M)\biggl(\frac{\bb{v}_\mathrm{d}\0\bb{\nabla}\psi}{v_\parallel \bh\0\bb{\nabla}\zeta}\biggr)(\psi_\mathrm{r},\eta,\zeta')\,\rmd\zeta'\Biggr]
        \end{aligned}
\end{align}
(up to an error of order $\rho_\star vL$), where $\zeta_0 = \zeta(\psi_\mathrm{r},\tilde{\eta},\tilde{\zeta}\!=\!0)$. In fact, we can replace $\zeta_0$ with $0$ in \eqref{eq:circ3}. This can be seen by performing, in \eqref{eq:circ3}, the following manipulations:
\begin{align}
&\int_{\zeta_0}^{\zeta_0 + 2\upi M}(\zeta' - \zeta + \nu(\psi_\mathrm{r},\eta,\zeta') - \nu(\psi_\mathrm{r},\eta,\zeta) - \upi M)\biggl(\frac{\bb{v}_\mathrm{d}\0\bb{\nabla}\psi}{v_\parallel \bh\0\bb{\nabla}\zeta}\biggr)(\psi_\mathrm{r},\eta,\zeta')\,\rmd\zeta' \nn
&\begin{aligned}[t]= \biggl( \?\int_{0}^{2\upi M} - \int_{0}^{\zeta_0} + \int_{2\upi M}^{2\upi M + \zeta_0} \biggr)\biggl[(&\zeta' - \zeta + \nu(\psi_\mathrm{r},\eta,\zeta')\\
&- \nu(\psi_\mathrm{r},\eta,\zeta) - \upi M)\biggl(\frac{\bb{v}_\mathrm{d}\0\bb{\nabla}\psi}{v_\parallel \bh\0\bb{\nabla}\zeta}\biggr)(\psi_\mathrm{r},\eta,\zeta')\biggr]\,\rmd\zeta' \\
\end{aligned}\nn
&= \int_{0}^{2\upi M}(\zeta' - \zeta + \nu(\psi_\mathrm{r},\eta,\zeta') - \nu(\psi_\mathrm{r},\eta,\zeta) - \upi M)\biggl(\frac{\bb{v}_\mathrm{d}\0\bb{\nabla}\psi}{v_\parallel \bh\0\bb{\nabla}\zeta}\biggr)(\psi_\mathrm{r},\eta,\zeta')\,\rmd\zeta' \nn
\al + \int_{0}^{\zeta_0}2\upi M\biggl(\frac{\bb{v}_\mathrm{d}\0\bb{\nabla}\psi}{v_\parallel \bh\0\bb{\nabla}\zeta}\biggr)(\psi_\mathrm{r},\eta,\zeta')\,\rmd\zeta'\,.
\end{align}
Finally, using \eqref{eq:Gammapsieta} and integrating by parts, we obtain
	\begin{align}\label{eq:3}
		 \circled{3} &\simeq \begin{aligned}[t] \frac{Ze}{mc}\iota_{\rm r}'(\psi-\psi_\mathrm{r})\Biggl[ &\int_0^{{\zeta}} 2\upi M \biggl(\frac{\bb{v}_\mathrm{d}\0\bb{\nabla}\psi}{v_\parallel \bh\0\bb{\nabla}\zeta}\biggr)(\psi_\mathrm{r},\eta,\zeta')\,\rmd\zeta'\\
        &+ \int_0^{2\upi M}(\zeta' - \zeta - \upi M)\biggl(\frac{\bb{v}_\mathrm{d}\0\bb{\nabla}\psi}{v_\parallel \bh\0\bb{\nabla}\zeta}\biggr)(\psi_\mathrm{r},\eta,\zeta')\,\rmd\zeta'\Biggr]\\
        \end{aligned}\nn
		\al +\iota_{\rm r}'(\psi-\psi_\mathrm{r})\int_0^{2\upi M} [\nu(\psi_\mathrm{r},\eta,\zeta') - \nu(\psi_\mathrm{r},\eta,\zeta)]\,\partial_\eta(v_\parallel{\bh}\0\partial_\zeta\bb{x})(\psi_\mathrm{r},\eta,\zeta')\,\rmd\zeta'\nn
		\al + \iota_{\rm r}'(\psi-\psi_\mathrm{r})\int_0^{2\upi M} (\partial_\zeta \nu\, v_\parallel{\bh}\0\partial_\eta\bb{x})(\psi_\mathrm{r},\eta,\zeta')\,\rmd\zeta'\,.
	\end{align}
	Summing \circled{1}, \circled{2}, and \circled{3} gives $\tilde{\mybar{\mathcal{K}}}(\psi,\tilde{\eta},\tilde{\zeta}) = \mybar{\mathcal{I}}(\psi,\eta,\zeta)$, up to terms of order $\rho_\star vL$. Therefore, formula \eqref{eq:higherorderinvariant} for the higher-order transit invariant is gauge invariant, as claimed.

\section{Limiting forms of $\mybarup{\chi}_{\rm r}$}\label{app:limits}

In \cref{subsec:oneres}, we show that passing orbits in the region $\mathfrak{R}_1$ (which contains a single low-order rational surface) conserve $\mybarup{\chi}_{\rm r}$, defined in \eqref{eq:H}. In this appendix, we show that this conservation law reproduces the correct behaviour for orbits close to (\cref{subsec:close}) or far from (\cref{subsec:far}) the rational surface. 

\subsection{Close to the rational surface}\label{subsec:close}

First, we discuss particles whose distance to the rational surface is on the order of a drift-island width. In \cref{sec:higherorder}, we described such particles by ordering $s\sim 1$ and $\psi-\psi_\mathrm{r}\sim \sqrt{\rho_\star}\,\Psi_\mathrm{t}$. Here, we will show that, when this ordering is satisfied, $\mybarup{\chi}_{\rm r}$ becomes (up to a constant factor) the transit adiabatic invariant, including the higher-order corrections derived in \cref{sec:higherorder}.

With $s\sim 1$ and $\psi-\psi_\mathrm{r}\sim \sqrt{\rho_\star}\,\Psi_\mathrm{t}$, formula \eqref{eq:H} for $\mybarup{\chi}_{\rm r}$ becomes
\begin{align}\label{eq:Qclose}
    \mybarup{\chi}_{\rm r}(\psi,\eta,\zeta) &= \int_{\psi_\mathrm{r}}^{\psi}\biggl(\iota(\psi') - \frac{N}{M}\biggr)\,\rmd\psi' - \frac{1}{2\upi M}\frac{mc}{Ze}\oint v_\parallel\bh\bb{\cdot}\biggl(\con{\zeta}+\frac{N}{M}\?\con{\theta}\biggr)\,\rmd\zeta'\nn
    \al + \frac{mc}{Ze}\? \iota_{\rm r}'(\psi-\psi_\mathrm{r})\biggl(v_\parallel\bh\bb{\cdot}\con{\theta} - \frac{1}{2\upi M}\oint_\mathrm{r}\, v_\parallel\bh\bb{\cdot}\con{\theta}\, \rmd\zeta' \biggr)\nn
    \al - \iota_{\rm r}'(\psi-\psi_\mathrm{r})\?\partial_\theta S_\mathrm{nr}^{(1)}+O(\rho_\star^2\?  \Psi_\mathrm{t})\,.
\end{align}
In the second and third lines of \eqref{eq:Qclose}, all functions of position except $\psi-\psi_\mathrm{r}$ are evaluated on the rational surface, at $(\psi_\mathrm{r}, \theta, \zeta)$. 
To make further progress, we need to know $\partial_\theta S^{(1)}_\mathrm{nr}(\psi_\mathrm{r},\theta,\zeta)$. We make a negligible error by restoring, in the definition \eqref{eq:Rnonres} of $S^{(1)}_\mathrm{nr}$, all non-resonant harmonics with $|{p|+|q|\geq K}$. This gives
\begin{equation}\label{eq:Srestored}
    S_\mathrm{nr}^{(1)}(\psi_\mathrm{r},\theta,\zeta) \simeq \mspace{-8mu}\sum_{(p,q) \in \mathbb{I}^{<K}_\mathrm{nr}}\mspace{-8mu}\frac{\mathrm{i}(V_\mathrm{nr}^{<K})_{pq}(\psi_\mathrm{r})}{q-Np/M}\,\mathrm{e}^{\mathrm{i}(p\theta-q\zeta)}\, + \mspace{-8mu}\sum_{(p,q) \in \mathbb{I}^{\geq K}_\mathrm{nr}}\mspace{-8mu}\frac{\mathrm{i}(V_\mathrm{nr}^{\geq K})_{pq}(\psi_\mathrm{r})}{q-Np/M}\,\mathrm{e}^{\mathrm{i}(p\theta-q\zeta)}\,,
\end{equation}
where $\mathbb{I}^{\geq K}_\mathrm{nr} \coloneq \{ (p, q)\in \mathbb{Z}^2: |p|+|q| \geq K \text{ and } (p,q) \neq (kM,kN) \text{ for any }k\in\mathbb{Z}\}$. Restoring these harmonics is allowed because the fact that we are working precisely on the rational surface means their size can be bounded as follows. Since $|Mq-Np|\geq 1$ for non-resonant harmonics, we have
\begin{equation}\label{eq:errorsum}
    \Bigl|\sum_{(p,q) \in \mathbb{I}^{\geq K}_\mathrm{nr}}\mspace{-8mu}\frac{\mathrm{i}(V_\mathrm{nr}^{\geq K})_{pq}(\psi_\mathrm{r})}{q-Np/M}\,\mathrm{e}^{\mathrm{i}(p\theta-q\zeta)}\Bigr| \leq M\mspace{-8mu}\sum_{(p,q) \in \mathbb{I}^{\geq K}_\mathrm{nr}}\mspace{-4mu} \bigl|(V_\mathrm{nr}^{\geq K})_{pq}(\psi_\mathrm{r})\bigr|\,.
\end{equation}
If the Fourier series of $V(\psi_
\mathrm{r},\theta,\zeta)$ is absolutely convergent, then the sum on the right side of \eqref{eq:errorsum} converges and will be small when $K$ is large. Hence, the approximate form of $S^{(1)}_\mathrm{nr}(\psi_\mathrm{r},\theta,\zeta)$ in \eqref{eq:Srestored} satisfies
\begin{equation}
        \biggl(\partial_\zeta + \frac{N}{M}\?\partial_\theta\biggr)\/ S^{(1)}_\mathrm{nr}(\psi_\mathrm{r}, \theta, \zeta) \simeq V(\psi_\mathrm{r}, \theta, \zeta) - \langle V \rangle_\zeta(\psi_\mathrm{r}, \theta, \zeta)\,.
\end{equation}
This magnetic differential equation is useful because we can solve it by integrating along rational-surface field lines to derive a simple integral formula for $S^{(1)}_\mathrm{nr}$. The result is
\begin{equation}
    S^{(1)}_\mathrm{nr}(\psi_\mathrm{r},\theta,\zeta) \simeq \int_{0, \mathrm{r}}^\zeta ( V - \langle V \rangle_\zeta )\,\rmd\zeta' + h(\eta)\,,
\end{equation}
for some function $h(\eta)$. This function can be determined using the fact that $S^{(1)}_\mathrm{nr}$ contains no $e^{\mathrm{i}k(M\theta - N\zeta)}$ harmonics, for any $k\in\mathbb{Z}$, according to definition \eqref{eq:Rnonres}. Therefore, we may impose $\langle S^{(1)}_\mathrm{nr} \rangle_\zeta = 0$. This gives
\begin{align}
\langle S^{(1)}_\mathrm{nr} \rangle_\zeta = 0 &\simeq \frac{1}{2\upi M}\oint_{\rm r} \?\biggl(\?\int_{0, \mathrm{r}}^\zeta  V \,\rmd\zeta'\biggr)\,\rmd\zeta - \frac{1}{2\upi M}\oint_{\rm r}\? \zeta  \langle V \rangle_\zeta \,\rmd\zeta + h(\eta)\nn
&= \upi M  \langle V \rangle_\zeta - \frac{1}{2\upi M}\oint_{\rm r} \?\zeta\? V \,\rmd\zeta  + h(\eta)\,,
\end{align}
where we have integrated by parts. Solving for $h$, we find
\begin{equation}\label{eq:Rnr}
S^{(1)}_\mathrm{nr}(\psi_\mathrm{r},\theta,\zeta) \simeq \int_{0, \mathrm{r}}^\zeta \, V \,\rmd\zeta' + \frac{1}{2\upi M}\oint_\mathrm{r}\,(\zeta' - \zeta - \upi M)\?  V\,\rmd\zeta'\,.
\end{equation}
Now we substitute this result into \eqref{eq:Qclose}. Using $\partial_\theta=\partial_\eta$ and $\con{\zeta} + (N/M)\?\con{\theta}=\con{\zeta}|_\eta$, where we have temporarily introduced the notation $\con{\zeta}|_\eta$ to indicate a $\zeta$ derivative holding $\eta$ fixed, as well as $V = (mc/Ze)\?\vp\bh\0\con{\zeta}|_\eta$ on the rational surface, we find
\begin{align}
    \mybarup{\chi}_{\rm r}(\psi,\eta,\zeta) &\simeq \int_{\psi_\mathrm{r}}^{\psi}\biggl(\iota(\psi') - \frac{N}{M}\biggr)\,\rmd\psi' - \frac{1}{2\upi M}\frac{mc}{Ze}\oint v_\parallel\bh\bb{\cdot}\con{\zeta}|_\eta\?\rmd\zeta'\nn
    \al + \begin{aligned}[t] \frac{mc}{Ze}\?  &\iota_{\rm r}'(\psi-\psi_\mathrm{r})\biggl(\vp\bh\0\con{\eta} - \int_{0, \mathrm{r}}^\zeta \partial_\eta(\vp\bh\0\con{\zeta}|_\eta)\,\rmd\zeta'\\
    & + \frac{1}{2\upi M} \oint_{\rm r}\?\bigl[(\zeta'-\zeta-\upi M)\?\partial_\eta(\vp\bh\0\con{\zeta}|_\eta) - \vp\bh\0\con{\eta}\bigr]\,\rmd\zeta'\biggr)\,.\taghere
    \end{aligned}
\end{align}
From here on, we use $\partial_\zeta$ to denote $\zeta$ derivatives at fixed $\eta$. Integrating by parts allows us to express $\mybarup{\chi}_{\rm r}$ in the form
\begin{align}
    \mybarup{\chi}_{\rm r}(\psi,\eta,\zeta) &\simeq \int_{\psi_\mathrm{r}}^{\psi}\biggl(\iota(\psi') - \frac{N}{M}\biggl)\,\rmd\psi' - \frac{1}{2\upi M}\frac{mc}{Ze}\oint v_\parallel\bh\bb{\cdot}\con{\zeta}\,\rmd\zeta'\nn
    \al - \begin{aligned}[t]&\frac{mc}{Ze}\? \iota_{\rm r}'(\psi-\psi_\mathrm{r})\biggl(\?\int_{0,\mathrm{r}}^\zeta\bigl[ \partial_\eta(v_\parallel \bh\bb{\cdot}\con{\zeta}) - \partial_\zeta(v_\parallel \bh\bb{\cdot}\con{\eta}) \bigr]\,\rmd\zeta'\\
    &+ \frac{1}{2\upi M}\oint_\mathrm{r}\,(\zeta'-\zeta-\upi M)\bigl[ \partial_\eta(v_\parallel \bh\bb{\cdot}\con{\zeta}) - \partial_\zeta(v_\parallel \bh\bb{\cdot}\con{\eta}) \bigr]\,\rmd\zeta'\biggr)\,,\taghere
    \end{aligned}
\end{align}
which is proportional to the higher-order transit adiabatic invariant \eqref{eq:higherorderinvariant}, as claimed.


\subsection{Far from the rational surface}\label{subsec:far}

Next, we discuss particles that are far from the rational surface, by which we mean ${\iota - N/M \sim 1}$. The particle orbits that satisfy this ordering are contained in a subregion of $\mathfrak{R}_1$ that lies much further than a drift-island width from the rational surface $\psi_\mathrm{r}$. This subregion is a permissible choice for the region $\mathfrak{R}_0$ introduced in \cref{subsec:weakresonances}, according to \cref{def:R0}. Therefore, for such particles, we expect conservation of $\mybarup{\chi}_{\rm r}$ to be equivalent to conservation of $\mybar{\psi}$, defined by \eqref{eq:psistar}. We proceed to show that this is the case.

First, since the formula for $\mybar{\psi}$ involves $S^{(1)}$ while the formula for $\mybarup{\chi}_{\rm r}$ involves $S^{(1)}_\mathrm{nr}$, it will be useful to relate these two functions. The difference between $S^{(1)}$ and $S^{(1)}_\mathrm{nr}$ consists entirely of resonant harmonics:
\begin{equation}\label{eq:RRdiff}
    S^{(1)} = S^{(1)}_\mathrm{nr} + \mspace{-12mu}\sum_{\substack{k\in \mathbb{Z}\backslash\{0\}\\
    |kM|+|kN|<K}}\mspace{-8mu} \frac{\mathrm{i}V_{kM\!,\? kN}}{k(N-\iota\?  M)}\,e^{\mathrm{i}k(M\theta-N\zeta)}\,.
\end{equation}
This difference clearly blows up as $\psi\to\psi_\mathrm{r}$; however, we are now working in a subregion of $\mathfrak{R}_1$ that excludes this divergence. Thus, since ${\iota-N/M\sim 1}$, we make a small error by restoring, in the sum, the harmonics with $|p|+|q|\geq K$:
\begin{equation}\label{eq:RRdiff2}
     S^{(1)} \simeq S^{(1)}_\mathrm{nr} + \mspace{-4mu}\sum_{k \in \mathbb{Z}\backslash\{0\}}\frac{\mathrm{i}V_{kM\!,\? kN}}{k(N-\iota\?  M)}\,e^{\mathrm{i}k(M\theta-N\zeta)}\,.
\end{equation}
Only the $\theta$ derivatives of $S^{(1)}$ and $S^{(1)}_\mathrm{nr}$ appear in $\mybar{\psi}$ and $\mybarup{\chi}_{\rm r}$, so we differentiate \eqref{eq:RRdiff2} with respect to $\theta$ and find
\begin{equation}
    \partial_\theta S^{(1)} \simeq \partial_\theta S^{(1)}_\mathrm{nr} + \frac{\langle V \rangle_\zeta - \langle V \rangle}{\iota - N/M}\,.
\end{equation}
 With this approximation, equation \eqref{eq:H} for $\mybarup{\chi}_{\rm r}$ becomes
 \begin{align}
    \mybarup{\chi}_{\rm r}(\psi,\eta,\zeta) \simeq &\int_{\psi_\mathrm{r}}^{\psi}\biggl(\iota(\psi') - \frac{N}{M}\biggr)\,\rmd\psi' + \biggl(\iota(\psi) - \frac{N}{M}\biggr)\Bigl(\frac{mc}{Ze}v_\parallel\bh\bb{\cdot}\con{\theta} -\partial_\theta S^{(1)}\Bigr) - \langle V \rangle\,.
\end{align}
To lowest order, our conserved quantity is $\mybarup{\chi}_{\rm r}(\psi, \eta, \zeta)\simeq \chi_\mathrm{r}(\psi)$, which means $\psi$ is constant. Therefore, it must be possible to re-express conservation of $\mybarup{\chi}_{\rm r}$ in the form
\begin{equation}\label{eq:psiseries}
    \psi + \psi^{[1]}(\psi, \theta, \zeta) + \psi^{[2]}(\psi, \theta, \zeta) + \ldots = \text{const.}\,,
\end{equation}
where $\Psi_\mathrm{t}\gg \psi^{[1]} \gg \psi^{[2]}\gg\ldots$ Note that we are using square brackets in the superscripts to distinguish these corrections from $\psi^{(1)}$ and $\psi^{(2)}$ defined in \eqref{eq:irrationaltransform} and \eqref{eq:nearidentity2}, respectively. Equation \eqref{eq:psiseries} is consistent with $\mybarup{\chi}_{\rm r}(\psi, \theta, \zeta) = \text{const.}$ if we take
\begin{equation}\label{eq:psichoice}
    \psi^{[1]}(\psi, \theta, \zeta) \coloneq \frac{mc}{Ze} v_\parallel\bh\bb{\cdot}\con{\theta} - \partial_\theta S^{(1)}\,.
\end{equation}
Note that we are free to add any function of $\psi$ to $\psi^{[1]}$, by reasoning similar to that of \cref{app:constant}. With the choice \eqref{eq:psichoice}, we recover \eqref{eq:psistar} as claimed.

\section{Higher-order drift surfaces in the presence of a single resonance}\label{app:higherorder}

In this appendix, we show how the higher-order conservation law \eqref{eq:hoconserved} for particles in region $\mathfrak{R}_1$ is obtained. To begin, we must retrace the steps leading up to Lagrangian \eqref{eq:resonantLagrangian} and restore all terms of order $\rho_\star^2$ that we previously suppressed. We find
\begin{align}\label{eq:hoLag1}
    \mathcal{L} &\equiv \Bigl( \mybar{\psi}_\mathrm{nr} + \psi^{(1)}_\mathrm{nr}\Bigl[ \partial_\psi\Bigl(\frac{\vp}{\Omega}\bb{B}\0\con{\theta} \Bigr) - \partial_\theta\Bigl( \frac{\vp}{\Omega}\bb{B}\0\con{\psi} \Bigr)\Bigr] - \theta^{(1)}_\mathrm{nr}\partial_\theta\psi^{(1)}_\mathrm{nr} \Bigr)\frac{\rmd\mybar{\theta}_\mathrm{nr}}{\rmd \zeta}\nn
    \al +\Bigl( \theta^{(1)}_\mathrm{nr}\Bigl[\partial_\theta\Bigl( \frac{\vp}{\Omega}\bb{B}\0\con{\psi} \Bigr) - \partial_\psi\Bigl( \frac{\vp}{\Omega}\bb{B}\0\con{\theta} \Bigr)\Bigr] -\theta^{(1)}_\mathrm{nr}\partial_\psi\psi^{(1)}_\mathrm{nr} \Bigr)\frac{\rmd \mybar{\psi}_\mathrm{nr}}{\rmd \zeta}\nn
    \al - \chi - \frac{1}{2}\iota'\bigl( \psi^{(1)}_\mathrm{nr} \bigr)^2 + \psi^{(1)}_\mathrm{nr}\Bigl[\partial_\psi\Bigl( \frac{\vp}{\Omega}\bb{B}\0\con{\zeta} \Bigr) - \partial_\zeta\Bigl( \frac{\vp}{\Omega}\bb{B}\0\con{\psi} \Bigr)\Bigr]\nn
    \al + \theta^{(1)}_\mathrm{nr}\Bigl[\partial_\theta\Bigl( \frac{\vp}{\Omega}\bb{B}\0\con{\zeta} \Bigr) - \partial_\zeta\Bigl( \frac{\vp}{\Omega}\bb{B}\0\con{\theta} \Bigr)\Bigr] - \theta^{(1)}_\mathrm{nr}\partial_\zeta\psi^{(1)}_\mathrm{nr} + V_{\rm r} + V^{\geq K}_{\rm nr}\,,
\end{align}
where all functions are evaluated at $(\mybar{\psi}_\mathrm{nr}, \mybar{\theta}_\mathrm{nr}, \zeta)$. Here, and for the remainder of this appendix, we drop terms of order $\rho_\star^3$ or smaller. Next, we make the near-identity change of coordinates defined by
\begin{subequations}\label{eq:hocoords}
\begin{align}
    \mybar{\psi}_{\rm nr} &= \mybardown{\mybar{\psi}}_{\rm nr} + \psi^{(2)}_{\rm nr}(\mybardown{\mybar{\psi}}_{\rm nr}, \mybardown{\mybar{\theta}}_{\rm nr}, \zeta)\,,\\
    \mybar{\theta}_{\rm nr} &= \mybardown{\mybar{\theta}}_{\rm nr} + \theta^{(2)}_{\rm nr}(\mybardown{\mybar{\psi}}_{\rm nr}, \mybardown{\mybar{\theta}}_{\rm nr}, \zeta)\,.
\end{align}
\end{subequations}
Substituting \eqref{eq:hocoords} into \eqref{eq:hoLag1} gives
\begin{align}\label{eq:hoLag2}
    \mathcal{L} &\equiv \begin{alignedat}[t]{2} &\Bigl( \mybardown{\mybar{\psi}}_{\rm nr} + \psi^{(2)}_{\rm nr} + \psi^{(1)}_\mathrm{nr}\Bigl[&& \partial_\psi\Bigl(\frac{\vp}{\Omega}\bb{B}\0\con{\theta} \Bigr) \\
    & && \mathbin{-} \partial_\theta\Bigl( \frac{\vp}{\Omega}\bb{B}\0\con{\psi} \Bigr)\Bigr] - \theta^{(1)}_\mathrm{nr}\partial_\theta\psi^{(1)}_\mathrm{nr} \Bigr)\biggl(\frac{\rmd\mybardown{\mybar{\theta}}_{\rm nr}}{\rmd \zeta} + \frac{\rmd \theta^{(2)}_{\rm nr}}{\rmd \zeta}\biggr)
    \end{alignedat}\nn
    \al +\Bigl( \theta^{(1)}_\mathrm{nr}\Bigl[\partial_\theta\Bigl( \frac{\vp}{\Omega}\bb{B}\0\con{\psi} \Bigr) - \partial_\psi\Bigl( \frac{\vp}{\Omega}\bb{B}\0\con{\theta} \Bigr)\Bigr] -\theta^{(1)}_\mathrm{nr}\partial_\psi\psi^{(1)}_\mathrm{nr} \Bigr)\biggl(\frac{\rmd\mybardown{\mybar{\psi}}_{\rm nr}}{\rmd \zeta} + \frac{\rmd \psi^{(2)}_{\rm nr}}{\rmd \zeta}\biggr)\nn
    \al - \chi - \iota\?\psi^{(2)}_{\rm nr} - \frac{1}{2}\iota'\bigl( \psi^{(1)}_\mathrm{nr} \bigr)^2 + \psi^{(1)}_\mathrm{nr}\Bigl[\partial_\psi\Bigl( \frac{\vp}{\Omega}\bb{B}\0\con{\zeta} \Bigr) - \partial_\zeta\Bigl( \frac{\vp}{\Omega}\bb{B}\0\con{\psi} \Bigr)\Bigr] \nn
    \al + \theta^{(1)}_\mathrm{nr}\Bigl[\partial_\theta\Bigl( \frac{\vp}{\Omega}\bb{B}\0\con{\zeta} \Bigr) - \partial_\zeta\Bigl( \frac{\vp}{\Omega}\bb{B}\0\con{\theta} \Bigr)\Bigr] - \theta^{(1)}_\mathrm{nr}\partial_\zeta\psi^{(1)}_\mathrm{nr} + V_{\rm r} + V^{\geq K}_{\rm nr}\,,
\end{align}
where all functions are now evaluated at $(\mybardown{\mybar{\psi}}_{\rm nr}, \mybardown{\mybar{\theta}}_{\rm nr}, \zeta)$. We now subtract the following total $\zeta$ derivative from the Lagrangian:
\begin{align}
    \frac{\rmd}{\rmd\zeta}\biggl\{ &\Bigl( \mybardown{\mybar{\psi}}_{\rm nr} + \psi^{(2)}_{\rm nr} + \psi^{(1)}_\mathrm{nr}\Bigl[ \partial_\psi\Bigl(\frac{\vp}{\Omega}\bb{B}\0\con{\theta} \Bigr) - \partial_\theta\Bigl( \frac{\vp}{\Omega}\bb{B}\0\con{\psi} \Bigr)\Bigr] - \theta^{(1)}_\mathrm{nr}\partial_\theta\psi^{(1)}_\mathrm{nr} \Bigr)\?\theta^{(2)}_{\rm nr}\nn
    &+ \Bigl( \theta^{(1)}_\mathrm{nr}\Bigl[\partial_\theta\Bigl( \frac{\vp}{\Omega}\bb{B}\0\con{\psi} \Bigr) - \partial_\psi\Bigl( \frac{\vp}{\Omega}\bb{B}\0\con{\theta} \Bigr)\Bigr] -\theta^{(1)}_\mathrm{nr}\partial_\psi\psi^{(1)}_\mathrm{nr} \Bigr)\?\psi^{(2)}_{\rm nr} + S^{(2)}_{\rm nr} \biggr\}\,,
\end{align}
where $S^{(2)}_{\rm nr}\sim \rho_\star^2\Psi_{\rm t}$ is an as-yet-undetermined function. The result is
\begin{align}\label{eq:hoLag3}
    \mathcal{L} &\equiv \begin{alignedat}[t]{2} &\Bigl( \mybardown{\mybar{\psi}}_{\rm nr} + \psi^{(2)}_{\rm nr} + \psi^{(1)}_\mathrm{nr}\Bigl[ &&\partial_\psi\Bigl(\frac{\vp}{\Omega}\bb{B}\0\con{\theta} \Bigr)\\
    & &&\mathbin{-} \partial_\theta\Bigl( \frac{\vp}{\Omega}\bb{B}\0\con{\psi} \Bigr)\Bigr] -\theta^{(1)}_\mathrm{nr}\partial_\theta\psi^{(1)}_\mathrm{nr} - \partial_\theta S^{(2)}_{\rm nr} \Bigr)\frac{\rmd\mybardown{\mybar{\theta}}_{\rm nr}}{\rmd \zeta}
    \end{alignedat}\nn
    \al +\Bigl( -\theta^{(2)}_{\rm nr} + \theta^{(1)}_\mathrm{nr}\Bigl[\partial_\theta\Bigl( \frac{\vp}{\Omega}\bb{B}\0\con{\psi} \Bigr) - \partial_\psi\Bigl( \frac{\vp}{\Omega}\bb{B}\0\con{\theta} \Bigr)\Bigr] -\theta^{(1)}_\mathrm{nr}\partial_\psi\psi^{(1)}_\mathrm{nr} - \partial_\psi S^{(2)}_{\rm nr}\Bigr)\frac{\rmd\mybardown{\mybar{\psi}}_{\rm nr}}{\rmd \zeta}\nn
    \al -\chi - \iota\?\psi^{(2)}_{\rm nr} - \frac{1}{2}\iota'\bigl( \psi^{(1)}_\mathrm{nr} \bigr)^2 + \psi^{(1)}_\mathrm{nr}\Bigl[\partial_\psi\Bigl( \frac{\vp}{\Omega}\bb{B}\0\con{\zeta} \Bigr) - \partial_\zeta\Bigl( \frac{\vp}{\Omega}\bb{B}\0\con{\psi} \Bigr)\Bigr] \nn
    \al + \theta^{(1)}_\mathrm{nr}\Bigl[\partial_\theta\Bigl( \frac{\vp}{\Omega}\bb{B}\0\con{\zeta} \Bigr) - \partial_\zeta\Bigl( \frac{\vp}{\Omega}\bb{B}\0\con{\theta} \Bigr)\Bigr] - \theta^{(1)}_\mathrm{nr}\partial_\zeta\psi^{(1)}_\mathrm{nr} + V_{\rm r} + V^{\geq K}_{\rm nr} - \partial_\zeta S^{(2)}_{\rm nr}\,.
\end{align}
In order to eliminate the $\zeta$-dependent terms in the symplectic parts of this Lagrangian, we pick
\begin{subequations}
\begin{align}
    \psi^{(2)}_\mathrm{nr} &\coloneq \psi^{(1)}_\mathrm{nr}\Bigl[\partial_\theta \Bigl( \frac{v_\parallel}{\Omega}\bb{B} \0\con{\psi} \Bigr) - \partial_\psi\Bigl( \frac{v_\parallel}{\Omega}\bb{B} \0\con{\theta} \Bigr) \Bigr] + \theta^{(1)}_\mathrm{nr}\partial_\theta\psi^{(1)}_\mathrm{nr} + \partial_\theta S^{(2)}_\mathrm{nr}\,,\\
    \theta^{(2)}_\mathrm{nr} &\coloneq \theta^{(1)}_\mathrm{nr}\Bigl[\partial_\theta \Bigl( \frac{v_\parallel}{\Omega}\bb{B} \0\con{\psi} \Bigr) - \partial_\psi\Bigl( \frac{v_\parallel}{\Omega}\bb{B} \0\con{\theta} \Bigr) \Bigr] - \theta^{(1)}_\mathrm{nr}\partial_\psi\psi^{(1)}_\mathrm{nr} - \partial_\psi S^{(2)}_\mathrm{nr}\,,
\end{align}
\end{subequations}
for whatever choice of $S^{(2)}_{\rm nr}$ we eventually make. This leaves
\begin{align}\label{eq:hoLag4}
    \mathcal{L} \equiv \mybardown{\mybar{\psi}}_{\rm nr}\frac{\rmd\mybardown{\mybar{\theta}}_{\rm nr}}{\rmd \zeta} - \chi &- \frac{1}{2}\iota'\bigl( \psi^{(1)}_\mathrm{nr} \bigr)^2 + V_{\rm r} + V^{\geq K}_{\rm nr}\nn
    & + \iota\?\psi^{(1)}_\mathrm{nr}\Bigl[\partial_\psi\Bigl( \frac{\vp}{\Omega}\bb{B}\0\con{\theta} \Bigr) - \partial_\theta\Bigl( \frac{\vp}{\Omega}\bb{B}\0\con{\psi} \Bigr)\Bigr] \nn
    & + \psi^{(1)}_\mathrm{nr}\Bigl[\partial_\psi\Bigl( \frac{\vp}{\Omega}\bb{B}\0\con{\zeta} \Bigr) - \partial_\zeta\Bigl( \frac{\vp}{\Omega}\bb{B}\0\con{\psi} \Bigr)\Bigr]\nn
    & + \theta^{(1)}_\mathrm{nr}\Bigl[\partial_\theta\Bigl( \frac{\vp}{\Omega}\bb{B}\0\con{\zeta} \Bigr) - \partial_\zeta\Bigl( \frac{\vp}{\Omega}\bb{B}\0\con{\theta} \Bigr)\Bigr]\nn
    & - \theta^{(1)}_\mathrm{nr}\bigl(\partial_\zeta\psi^{(1)}_\mathrm{nr} +\iota\?\partial_\theta\psi^{(1)}_\mathrm{nr}\bigr) - \bigl(\partial_\zeta S^{(2)}_{\rm nr} + \iota\?\partial_\theta S^{(2)}_{\rm nr}\bigr)\,,
\end{align}
After some algebra, using the relations \eqref{eq:nrtransforms}, \eqref{eq:Rnrdef}, and $(\vp/\Omega)\?\bb{B}\0(\con{\zeta}+\iota\?\con{\theta})\simeq V_{\rm r} + V_{\rm nr}^{<K}$ to rewrite several terms, this Lagrangian simplifies to
\begin{align}
    \mathcal{L} \equiv \mybardown{\mybar{\psi}}_{\rm nr}\frac{\rmd\mybardown{\mybar{\theta}}_{\rm nr}}{\rmd \zeta} - \chi + V_{\rm r} &+ V^{\geq K}_\mathrm{nr} + \bigl(\psi^{(1)}_\mathrm{nr}\partial_\psi V_\mathrm{r} + \theta^{(1)}_\mathrm{nr}\partial_\theta V_\mathrm{r}\bigr) + \frac{1}{2}\iota' \bigl(\psi^{(1)}_{\rm nr}\bigr)^2\nn
     &- \psi^{(1)}_\mathrm{nr}\bigl(\partial_\zeta \theta^{(1)}_\mathrm{nr} + \iota\? \partial_\theta \theta^{(1)}_\mathrm{nr}\bigr) - \bigl(\partial_\zeta S^{(2)}_\mathrm{nr} + \iota\? \partial_\theta S^{(2)}_\mathrm{nr}\bigr)\,.
\end{align}
Let $V_2$ be defined by
\begin{equation}
    V_2 \coloneq V^{\geq K}_\mathrm{nr} + \bigl(\psi^{(1)}_\mathrm{nr}\partial_\psi V_\mathrm{r} + \theta^{(1)}_\mathrm{nr}\partial_\theta V_\mathrm{r}\bigr) + \frac{1}{2}\iota' \bigl(\psi^{(1)}_\mathrm{nr}\bigr)^{\!2} - \psi^{(1)}_\mathrm{nr}\bigl(\partial_\zeta \theta^{(1)}_\mathrm{nr} + \iota\? \partial_\theta \theta^{(1)}_\mathrm{nr}\bigr)\,,
\end{equation}   
as announced in \eqref{eq:hoV1}. Then, the Lagrangian takes the form
\begin{equation}
    \mathcal{L}\equiv \mybardown{\mybar{\psi}}_{\rm nr}\frac{\rmd\mybardown{\mybar{\theta}}_{\rm nr}}{\rmd \zeta} - \chi + V_{\rm r} + V_2 - \bigl(\partial_\zeta S^{(2)}_\mathrm{nr} + \iota\? \partial_\theta S^{(2)}_\mathrm{nr}\bigr).
\end{equation}
Let $V_2 = \langle V_2 \rangle_\zeta + V_{2,\mathrm{nr}}^{<K_2} + V_{2,\mathrm{nr}}^{\geq K_2}$. We choose $S^{(2)}_\mathrm{nr}$ to remove the angle dependence contained in the low-order, non-resonant harmonics $V_{2,\mathrm{nr}}^{<K_2}$; that is, we take
\begin{equation}
    S^{(2)}_\mathrm{nr} \coloneq \mspace{-8mu}\sum_{(p,q) \in \mathbb{I}^{<K_2}_\mathrm{nr}}\mspace{-8mu} \frac{\mathrm{i}(V_{2,\mathrm{nr}}^{<K_2})_{pq}}{q-\iota\? p}\,\mathrm{e}^{\mathrm{i}(p\theta-q\zeta)}\,,
\end{equation}
as in \eqref{eq:hoS1}, so that $\partial_\zeta S^{(2)}_\mathrm{nr} + \iota\? \partial_\theta S^{(2)}_\mathrm{nr} = V_{2,\mathrm{nr}}^{<K_2}$. If we define $\mybardown{\mybar{\eta}}_{\rm nr} = \mybardown{\mybar{\theta}}_{\rm nr} - (N/M)\?\zeta$, then our final result may be written
\begin{equation}
    \mathcal{L}\equiv \mybardown{\mybar{\psi}}_{\rm nr}\frac{\rmd\mybardown{\mybar{\eta}}_{\rm nr}}{\rmd \zeta} - \chi_{\rm r} + \langle V \rangle_\zeta + \langle V_2 \rangle_\zeta\,.
\end{equation}
This Lagrangian does not depend on $\zeta$, so the Hamiltonian
\begin{equation}
\mathcal{H}_{2, \mathrm{r}}(\mybardown{\mybar{\psi}}_\mathrm{nr}, \mybardown{\mybar{\eta}}_\mathrm{nr}) = \chi_\mathrm{r}(\mybardown{\mybar{\psi}}_\mathrm{nr}) - \langle V \rangle_\zeta(\mybardown{\mybar{\psi}}_\mathrm{nr}, \mybardown{\mybar{\eta}}_\mathrm{nr}) - \langle V_2 \rangle_\zeta(\mybardown{\mybar{\psi}}_\mathrm{nr}, \mybardown{\mybar{\eta}}_\mathrm{nr})
\end{equation}
is conserved, as claimed.

\bibliographystyle{jpp}
\bibliography{resonant}

\end{document}